%% file: formalizing-mk.tex
\documentclass[fleqn]{llncs}
\usepackage{latexsym}
\usepackage{amssymb}
\usepackage{stmaryrd}
\usepackage{graphicx}
\usepackage{color}
\usepackage{url}
\usepackage{phonetic}
\usepackage{amsmath}
\usepackage{url}
\usepackage{tikz}
\usetikzlibrary{arrows,decorations,decorations.pathmorphing}
\usepackage[inline]{enumitem}
\usepackage{etoolbox}

\newtoggle{cicm}
\togglefalse{cicm}      % arXiv paper with appendices
\newcommand{\appendixref}[1]
  {\iftoggle{cicm}
    {appendix #1 of \cite{CaretteFarmerArxiv17}}
    {appendix #1}%
  }
\newcommand{\appendicesref}[2]
  {\iftoggle{cicm}
    {appendices #1 and #2 of \cite{CaretteFarmerArxiv17}}
    {appendices #1 and #2}%
  }

\usepackage[references]{agda}
\usepackage{ucs}
\usepackage[utf8x]{inputenc}

\DeclareUnicodeCharacter{8799}{$\stackrel{?}{=}$}
\DeclareUnicodeCharacter{931}{$\Sigma$}
\DeclareUnicodeCharacter{9679}{\ensuremath{\bullet}}
\DeclareUnicodeCharacter{945}{\ensuremath{\alpha}}
\DeclareUnicodeCharacter{964}{\ensuremath{\tau}}
\DeclareUnicodeCharacter{958}{\ensuremath{\xi}}
\DeclareUnicodeCharacter{8759}{\ensuremath{\:\!\!\:}}

\iftoggle{cicm} 
  {\newcommand{\syn}[1]{#1}}
  {\newcommand{\syn}[1]{\textcolor{red}{#1}}}

\input{formalizing-mk-def}

\title{Formalizing Mathematical Knowledge as a Biform Theory Graph: A
  Case Study\thanks{Published without appendices in: H. Geuvers et
    al., eds, \emph{Intelligent Computer Mathematics (CICM 2017)},
    \emph{Lecture Notes in Computer Science}, Vol.~10383, pp.~9--24,
    Springer, 2017. The final publication is available at Springer via
    \texttt{http://dx.doi.org/10.1007/978-3-319-62075-6\_2}.  This
    research was supported by NSERC.}}

\author{Jacques Carette and William M. Farmer}

\institute{%
Computing and Software, McMaster University, Canada\\
\url{http://www.cas.mcmaster.ca/~carette}\\
\url{http://imps.mcmaster.ca/wmfarmer}\\[1.5ex]
\iftoggle{cicm}{}{\today}
}

\begin{document}

\maketitle

\begin{abstract}
A \emph{biform theory} is a combination of an axiomatic theory and an
algorithmic theory that supports the integration of reasoning and
computation.  These are ideal for formalizing algorithms that
manipulate mathematical expressions.  A \emph{theory graph} is a
network of \emph{theories} connected by meaning-preserving
\emph{theory morphisms} that map the formulas of one theory to the
formulas of another theory.  Theory graphs are in turn well suited for
formalizing mathematical knowledge at the most convenient level of
abstraction using the most convenient vocabulary.  We are interested
in the problem of whether a body of mathematical knowledge can be
effectively formalized as a theory graph of biform theories.  As a
test case, we look at the graph of theories encoding natural number
arithmetic.  We used two different formalisms to do this, which we
describe and compare.  The first is realized in {\churchuqe}, a
version of Church's type theory with quotation and evaluation, and the
second is realized in Agda, a dependently typed programming language.
\end{abstract}

\iffalse 


\fi

\section{Introduction}\label{sec:introduction}

There are many methods for encoding mathematical knowledge.  The two
most prevalent are the \emph{axiomatic} and the \emph{algorithmic}.
The axiomatic method, famously employed by Euclid in his
\emph{Elements} circa 300 BCE, encodes a body of knowledge as an
\emph{axiomatic theory} composed of a language and a set of
\emph{axioms} expressed in that language.  The axioms are assumptions
about the \emph{concepts} of the language and the logical consequences
of the axioms are the \emph{facts} about the concepts.  The
algorithmic method in contrast uses an \emph{algorithmic theory},
composed of a language and a set of \emph{algorithms} that perform
symbolic computations over the expressions of the language.  Each
algorithm procedurally encodes its input/output relation.  For
example, an algorithm that symbolically adds expressions that
represent rational numbers encodes the addition function $+ :
\mathbb{Q} \times \mathbb{Q} \tarrow \mathbb{Q}$ over the rational
numbers.

A complex body of mathematical knowledge comprises many different
theories; these can be captured by the
\emph{little theories method}~\cite{FarmerEtAl92b} as a \emph{theory
  graph}~\cite{Kohlhase14} consisting of theories as nodes and theory
morphisms as directed edges.  A theory morphism is a
meaning-preserving mapping from the formulas of one theory to the
formulas of another.  The theories serve as abstract
mathematical models and the morphisms serve as information
conduits that enable definitions and theorems to be transported
from one theory to another~\cite{BarwiseSeligman97}.  A theory graph enables
mathematical knowledge to be formalized at the most convenient level
of abstraction using the most convenient vocabulary.  Moreover, the
structure of a theory graph provides the means to access relevant
concepts and facts (c\&f), reduce the duplication of c\&f, and
enable c\&f to be interpreted in multiple ways.

The axiomatic method is the basis for formalizing mathematical
knowledge in proof assistants and logical frameworks.  Although many
proof assistants support the little theories method to some extent,
very few provide the means to explicitly build theory graphs.  Notable
exceptions are the {\imps} theorem proving system~\cite{FarmerEtAl93}
and the {\mmt} module system for mathematical
theories~\cite{RabeKohlhase13}.

Computer algebra systems on the other hand are based on
algorithmic theories, which are not usually
organized as a graph.  An exception is the
Axiom system~\cite{JenksSutor92} in which a
network of abstract and concrete algorithmic theories are represented
by Axiom categories and domains, respectively.  Algorithmic theories
are challenging to fully formalize because a specification of
a symbolic algorithm that encodes a mathematical function requires
the ability to talk about the relationship between syntax and
semantics.

Axiomatic and algorithmic knowledge complement each other, and both
are needed.  A \emph{biform
  theory}~\cite{CaretteFarmer08,FarmerMohrenschildt03,Farmer07b}
combines both, and furthermore supports the integration of reasoning
and computation.  We argue in~\cite{CaretteFarmer08} that biform
theories are needed to build \emph{high-level theories} analogous to
high-level programming languages.  Biform theories are challenging to
formalize for the same reasons that algorithmic theories are
challenging to formalize.

We are interested in the problem of whether the little theories method
can be applied to biform theories.  That is, can a body of
mathematical knowledge be effectively formalized as a theory graph of
biform theories?  We use a graph (of biform theories) encoding
natural number arithmetic as a test case. We describe
two different formalizations, and compare the
results.  The first formalization is
realized using the global approach in
{\churchuqe}~\cite{FarmerArxiv17}, a variant of
{\churchqe}~\cite{FarmerArxiv16,Farmer16}, a version of Church's type
theory with quotation and evaluation, while the second is realized
using the local approach in Agda~\cite{Norell07,Norell09} , a dependently typed
programming language.  This dual formalization, contrasting the two 
approaches, forms the core of our contribution; each formalization has 
some smaller contributions, some of which may be of independent 
interest.

The rest of the paper is organized as follows.  The notion of a biform
theory is defined and discussed in section~\ref{sec:biform}.  The
theories that encode natural number arithmetic are presented in
section~\ref{sec:test-case}.  The {\churchuqe} formalization is
discussed in section~\ref{sec:cttuqe}, and the Agda version in
section~\ref{sec:agda}.  These two are presented in full in
\appendicesref{A}{B}.  Section~\ref{sec:comparison} compares the two
formalizations.  The paper ends with conclusions and future work in
section~\ref{sec:conclusion}.

The authors are grateful to the reviewers for their comments and
suggestions.

\section{Biform Theories}\label{sec:biform}

Let $\sE$ be a set of expressions and $f : \sE^n \tarrow \sE$ be an
$n$-ary function where $n \ge 1$.  A \emph{transformer for $f$} is an
algorithm that implements $f$.  Transformers manipulate expressions
$e$ in various ways: simple ones build bigger expressions from pieces,
select components of $e$, or check whether $e$ satisfies some
syntactic property.  More sophisticated transformers manipulate
expressions in a mathematically meaningful way.  We call these kinds
of transformers \emph{syntax-based mathematical algorithms
  (SBMAs)}~\cite{Farmer13}.  Examples include algorithms that apply
arithmetic operations to numerals, factor polynomials, transpose
matrices, and symbolically differentiate expressions with variables.
The \emph{computational behavior} of a transformer is the relationship
between its input and output expressions.  When the transformer is an
SBMA, its \emph{mathematical meaning} is the relationship between the
mathematical meanings of its input and output expressions.

A \emph{biform theory} $T$ is a triple $(L,\Pi,\Gamma)$ where $L$ is a
language of some underlying logic, $\Pi$ is a set of transformers for
functions over expressions of $L$, and $\Gamma$ is a set of formulas
of $L$.  $L$ is generated from a set of symbols that include, e.g.,
types and constants.  Each symbol is the name for a concept of $T$.
The transformers in $\Pi$ are for functions represented by symbols of
$L$.  The members of $\Gamma$ are the \emph{axioms} of $T$.  They
specify the concepts of $T$ including the computational behaviors of
transformers and the mathematical meanings of SBMAs.  The underlying
logic provides the semantic foundation for $T$.  We say $T$ is an
\emph{axiomatic theory} if $\Pi$ is empty and an \emph{algorithmic
 theory} if $\Gamma$ is empty.

Expressing a biform theory in the underlying logic requires infrastructure for
reasoning about expressions manipulated by the transformers as syntactic
entities.  The infrastructure provides a basis for \emph{metareasoning
 with reflection}~\cite{FarmerArxiv16}.
There are two main approaches for obtaining this
infrastructure~\cite{Farmer13}.  The \emph{local approach} is to
produce a deep embedding of a sublanguage $L'$ of $L$ that include all
the expressions manipulated by the transformers of $\Pi$.  The deep
embedding consists of 
\begin{enumerate*}[label=(\arabic*)]
\item an inductive type of \emph{syntactic values}
that represent the syntactic structures of the expressions in $L'$,
\item an \emph{informal quotation operator} that maps the expressions in
$L'$ to syntactic values, and 
\item a \emph{formal evaluation operator}
that maps syntactic values to the values of the expressions in $L'$
that they represent.
\end{enumerate*}
The \emph{global approach} is to replace the underlying logic of $L$
with a logic such as that of~\cite{FarmerArxiv16} that has
\begin{enumerate*}[label=(\arabic*)]
\item an inductive type of \emph{syntactic values} for all the
expressions in $L$, 
\item a \emph{global formal quotation operator}, and
\item a \emph{global formal evaluation operator}.
\end{enumerate*}

There are several ways, in a proof assistant, to construct a
transformer $\pi$ for $f : \sE^n \tarrow \sE$.  The simplest is to
define $f$ as a lambda abstraction $A_f$, and then $\pi$ computes the
value $f(e_1,\ldots,e_n)$ by reducing $A_f(e_1,\ldots,e_n)$ using
$\beta$-reduction (and possibly other transformations such as
$\delta$-reduction, etc).  Another method is to specify the
computational behavior of $f$ by axioms, and then $\pi$ can be
implemented as a tactic that applies the axioms to $f(e_1,\ldots,e_n)$
as, e.g., rewrite rules or conditional rewrite rules.  Finally, the
computational behavior or mathematical meaning of $f$ can be specified
by axioms, and then $\pi$ can be a program which satisfies these
axioms; this program can operate on either internal or external data
structures representing the expressions $e_1,\ldots,e_n$.

\section{Natural Number Arithmetic: A Test Case}\label{sec:test-case}

\begin{figure}
\center
\begin{tikzpicture}[scale=3.0]
\node[thy] (bt1) at (0,0) {BT1};
\node[thy] (bt2) at (1,0) {BT2};
\node[thy] (bt3) at (2,0) {BT3};
\node[thy] (bt4) at (3,0) {BT4};
\node[thy] (bt5) at (0,1) {BT5};
\node[thy] (bt6) at (1,1) {BT6};
\node[thy] (bt7) at (2,1) {BT7};
\node[thy] (bt8) at (3,1) {BT8};
\draw[inclusion-arrow] (bt1) to (bt2);
\draw[inclusion-arrow] (bt2) to (bt3);
\draw[inclusion-arrow] (bt3) to (bt4);
\draw[inclusion-arrow] (bt1) to (bt5);
\draw[inclusion-arrow] (bt2) to (bt6);
\draw[inclusion-arrow] (bt3) to (bt7);
\draw[inclusion-arrow] (bt5) to (bt6);
\draw[inclusion-arrow] (bt6) to (bt7);
\draw[view-arrow] (bt4) to (bt7);
\draw[view-arrow] (bt7) to (bt8);

\end{tikzpicture}
\caption{Biform Theory Graph Test Case}  \label{fig:biform-tg}
\end{figure}
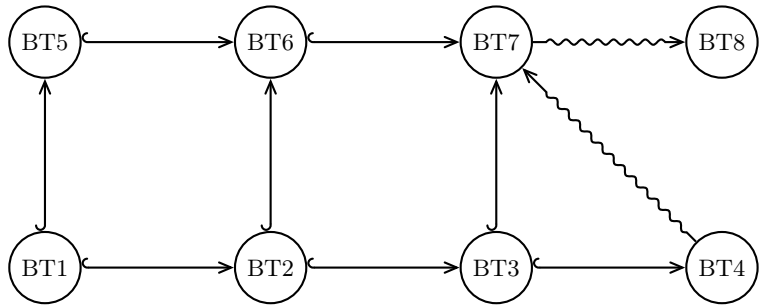

Figure~\ref{fig:biform-tg} shows a theory graph composed of biform
theories encoding natural number arithmetic.  We start with eight
axiomatic theories (seven in first-order logic (FOL) and one in simple
type theory (STT)) and then add a variety of useful transformers in
the appropriate theories.  These eight are chosen because they fit
together closely and have simple axiomatizations.  Of the first-order
theories, BT1 and BT5 are theories of $0$ and $S$ (which denotes the
successor function); BT2 and BT6 are theories of $0$, $S$, and $+$;
and BT3, BT4, and BT7 are theories of $0$, $S$, $+$, and $*$.  Several
other biform theories could be added to this graph, most notably
Skolem arithmetic, the complete theory of $0$, $S$, and $*$, which has
a very complicated axiomatization~\cite{Smorynski91}.  The details of
each theory is given below.

Figure~\ref{fig:biform-tg} shows the morphisms that connect these
theories.  The $\inclusion$ arrows denote strict theory inclusions.
The morphism from BT4 to BT7 is the identity mapping.  It is
meaning-preserving since each axiom of BT4 is a theorem of BT7.  In
particular, A7 follows from the induction schema A10.  The theory
morphism from BT7 to BT8 is interlogical since their logics are
different.  It is defined by the mapping of $0, S, +, *$ to $0_\iota,
S_{\iota \tarrow \iota}, +_{\iota \tarrow \iota \tarrow \iota},
*_{\iota \tarrow \iota \tarrow \iota}$, respectively, where $+_{\iota
  \tarrow \iota \tarrow \iota}$ and $*_{\iota \tarrow \iota \tarrow
  \iota}$ are defined constants in BT8. It is meaning-preserving since
A1--A6 and the instances of the induction schema A10 map to theorems
of BT8.

We have formalized this biform theory graph in two ways: the first in
{\churchuqe} using the global approach and the second in Agda using
the local approach.  These are discussed in the next
two sections, while the full details are given in \appendicesref{A}{B}.
A ``conventional'' mathematical presentation of the theories would be
as follows.

\begin{biformthy}[BT\thebiformthy: Simple Theory of $0$ and $S$]\em\ \\
\emph{Logic}: FOL. \emph{Constants}: $0$ (0-ary), $S$ (unary).\\
\emph{Axioms}:\\
\indent A1. $S(x) \not= 0$.\\
\indent A2. $S(x) = S(y) \Implies x = y$.\\
\emph{Properties}: incomplete, undecidable.\\
\emph{Transformers}: Recognizer for the formulas of the theory.
\end{biformthy}

\begin{biformthy}[BT\thebiformthy: Simple Theory of $0$, $S$, and $+$]\em\ \\
Extends BT1.\\
\emph{Logic}: FOL. \emph{Constants}: $+$ (binary, infix).\\
\emph{Axioms}:\\
\indent A3. $x + 0 = x$.\\
\indent A4. $x + S(y) = S(x + y)$.\\
\emph{Properties}: incomplete, undecidable.\\
\emph{Transformers}: Recognizer for the formulas of the theory and  
algorithm for adding natural numbers as binary numerals.
\end{biformthy}

\begin{biformthy}[BT\thebiformthy: Simple Theory of $0$, $S$, $+$, and $*$]\em\ \\
Extends BT2.\\
\emph{Logic}: FOL. \emph{Constants}: $*$ (binary, infix).\\
\emph{Axioms}:\\
\indent A5. $x * 0 = 0$.\\
\indent A6. $x * S(y) = (x * y) + x$.\\
\emph{Properties}: incomplete, undecidable.\\
\emph{Transformers}: Recognizer for the formulas of the theory and  
algorithm for multiplying natural numbers as binary numerals.
\end{biformthy}

\begin{biformthy}[BT\thebiformthy: Robinson Arithmetic (Q)]\em\ \\
Extends BT3.\\
\emph{Logic}: FOL.\\
\emph{Axioms}:\\
\indent A7. $x = 0 \OR \ForsomeApp y \mdot S(y) = x$.\\
\emph{Properties}: essentially incomplete, essentially undecidable.
\end{biformthy}

\begin{biformthy}[BT\thebiformthy: Complete Theory of $0$ and $S$]\em\ \\
Extends BT1.\\
\emph{Logic}: FOL.\\
\emph{Axioms}:\\
\indent A8. $(A(0) \And \ForallApp x \mdot (A(x) \Implies A(S(x))))
\Implies \ForallApp x \mdot A(x)$\\
\indent
where $A$ is any formula of BT{\thebiformthy} in which $x$ is not bound and $A(t)$
is the result\\
\indent
of replacing each free occurrence of $x$ in $A$ with the term $t$.\\
\emph{Properties}: complete, decidable.\\
\emph{Transformers}: Generator for instances of the theory's induction schema 
and decision procedure for the theory.
\end{biformthy}

\begin{biformthy}[BT\thebiformthy: Presburger Arithmetic]\em\ \\
Extends BT2 and BT5.\\
\emph{Logic}: FOL.\\
\emph{Axioms}:\\
\indent A9. $(A(0) \And \ForallApp x \mdot (A(x) \Implies A(S(x))))
\Implies \ForallApp x \mdot A(x)$\\
\indent 
where $A$ is any formula of BT{\thebiformthy} in which $x$ is not bound and $A(t)$ 
is the result\\
\indent
of replacing each free occurrence of $x$ in $A$ with the term $t$.\\
\emph{Properties}: complete, decidable.\\
\emph{Transformers}: Generator for instances of the theory's induction schema 
and decision procedure for the theory.
\end{biformthy}

\begin{biformthy}[BT\thebiformthy: First-Order Peano Arithmetic]\em\ \\
Extends BT3 and BT6.\\ 
\emph{Logic}: FOL.\\
\emph{Axioms}:\\
\indent A10. $(A(0) \And \ForallApp x \mdot (A(x) \Implies A(S(x))))
\Implies \ForallApp x \mdot A(x)$\\
\indent 
where $A$ is any formula of BT{\thebiformthy} in which $x$ is not bound and $A(t)$
is the result\\
\indent
of replacing each free occurrence of $x$ in $A$ with the term $t$.\\
\emph{Properties}: essentially incomplete, essentially decidable.\\
\emph{Transformers}: Generator for instances of the theory's induction schema.
\end{biformthy}

\begin{biformthy}[BT\thebiformthy: Higher-Order Peano Arithmetic]\em\ \\
\emph{Logic}: STT. \emph{Types}: $\iota$. 
\emph{Constants}: $0_\iota$, $S_{\iota \tarrow \iota}$.\\
\emph{Axioms}:\\
\indent A11. $S_{\iota \tarrow \iota}(x_\iota) \not= 0_\iota$.\\
\indent A12. $S_{\iota \tarrow \iota}(x_\iota) = S_{\iota \tarrow \iota}(y_\iota) 
\Implies x_\iota = y_\iota$.\\
\indent A13. $(p_{\iota \tarrow o}(0) \And 
\ForallApp x_\iota \mdot (p_{\iota \tarrow o}(x_\iota) 
\Implies p_{\iota \tarrow o}(S(x_\iota)))) 
\Implies \ForallApp x_\iota \mdot p_{\iota \tarrow o}(x_\iota)$.\\
\emph{Properties}: essentially incomplete, essentially decidable, categorical for standard models.
\end{biformthy}

It is important to note that axioms A8, A9 and A10 are all different since 
they are over different languages; in particular, {\bf BT6} adds $+$ to the
language of {\bf BT5}, and {\bf BT7} adds $*$ to the language of {\bf BT6}.

\section{Study 1: Test Case Formalized in ${\bf CTT}_{\bf uqe}$}\label{sec:cttuqe}

{\churchuqe} supports the global approach for metareasoning with
reflection.  {\churchuqe} contains%
\renewcommand{\labelenumi}{(\theenumi)}
\begin{enumerate*}
\item a logical base type $\epsilon$ that
is an inductive type of syntactic values called \emph{constructions}
which are expressions of type $\epsilon$,
\item a
global quotation operator $\synbrack{\cdot}$ that maps each expression
$\textbf{A}_\alpha$ of {\churchuqe} to a construction that represents
the syntactic structure of $\textbf{A}_\alpha$, and
\item a typed global evaluation operator $\sembrack{\cdot}_\alpha$
that maps each construction $\textbf{B}_\epsilon$ of {\churchuqe}
representing an expression $\textbf{A}_\alpha$ of type~$\alpha$ to an
expression whose value is the same as $\textbf{A}_\alpha$.
\end{enumerate*}
\renewcommand{\labelenumi}{\theenumi.}
See~\cite{FarmerArxiv17} for details.

A \emph{biform theory} of {\churchuqe} is a triple $(L,\Pi,\Gamma)$
where $L$ is a language generated by a set of base types and constants
of {\churchuqe}, $\Pi$ is a set of transformers over expressions of
$L$, and $\Gamma$ is a set of formulas of $L$.  Each transformer is
for a constant in $L$ whose type has the form $\epsilon \tarrow \cdots
\tarrow \epsilon$.  We present biform theories in {\churchuqe} as a
set of base types, constants, axioms, transformers, and theorems.  The
base types are divided into primitive and defined base types.  A
defined base type is declared by a formula that equates the base type
to a nonempty subset of some type of $L$.  Similarly, the constants
are divided into primitive and defined constants.  A defined constant
$\textbf{c}_\alpha$ is declared by an equation $\textbf{c}_\alpha =
\textbf{A}_\alpha$ where $\textbf{A}_\alpha$ is a defined expression.

The biform theory graph test case given in section~\ref{sec:test-case}
is formalized in {\churchuqe} as a theory graph of eight {\churchuqe}
theories as shown in \appendixref{A}.  Since {\churchuqe} is
not currently implemented, it is not possible to give the transformers
as implemented algorithms.  Instead we described their intended
behavior.

We will concentrate our discussion on BT6 (given below).  We have not
included the following components of BT6 (that should be in BT6
according to its definition in section~\ref{sec:test-case}) that are
redundant or are subsumed by other components: Constants
$\mname{BT5-DEC-PROC}_{\epsilon \tarrow \epsilon}$,
$\mname{IS-FO-BT1}_{\epsilon \tarrow \epsilon}$, and
$\mname{IS-FO-BT1-ABS}_{\epsilon \tarrow \epsilon}$; axioms 27 and 28;
and transformers $\pi_1$, $\pi_2$, $\pi_{11}$, $\pi_{12}$, and
$\pi_{13}$.  See~\cite{FarmerArxiv17} for details.
\iftoggle{cicm}{}{Expressions of type~$\epsilon$, i.e., expressions
  that denote constructions, are colored red to identify where
  reasoning about syntax occurs.}

BT6 has the usual constants ($0_\iota$, $S_{\iota \tarrow \iota}$, and
$+_{\iota \tarrow \iota \tarrow \iota}$) and axioms (axioms 1--4 and
29) of Presburger arithmetic.  Axiom 29 is the direct formalization of
A9, the induction schema for Presburger arithmetic, stated in
section~\ref{sec:test-case}.  It is expressed as a single universal
formula in {\churchuqe} that ranges over constructions representing
function abstractions of the form $\LambdaApp \textbf{x}_\iota \mdot
\textbf{A}_o$.  These constructions are identified by the transformers
$\pi_{15}$ and $\pi_{16}$ for the defined constant
$\mname{IS-FO-BT2-ABS}_{\epsilon \tarrow \epsilon}$.  $\pi_{15}$ works
by accessing data about variables, constants, and other subexpressions
stored in the data structure for an expression, while $\pi_{16}$ works
by expanding the definition of $\mname{IS-FO-BT2-ABS}_{\epsilon
  \tarrow \epsilon}$.  $\pi_{15} $ is sound if the definition
expansion mechanism is sound.  Showing the soundness of $\pi_{14}$
would require a formal verification of the implementation of the data
structure for expressions.  Of course, the results of $\pi_{14}$ could
be checked using $\pi_{16}$.

This biform theory has a defined constant $\mname{bnat}_{\iota \tarrow
  \iota \tarrow \iota}$ with the usual base~2 notation for expressing
natural numbers in a binary form.  There is a constant
$\mname{BPLUS}_{\epsilon \tarrow \epsilon \tarrow \epsilon}$ specified
by axioms 5--15 for adding quotations of these natural numbers in
binary form.  $\mname{BPLUS}_{\epsilon \tarrow \epsilon \tarrow
  \epsilon}$ is implemented by transformers $\pi_3$ and $\pi_4$.
$\pi_3$ is some efficient algorithm implemented outside of
{\churchuqe}, and $\pi_4$ is an algorithm that uses axioms 5--15 as
conditional rewrite rules.  $\pi_4$ is sound if the rewriting
mechanism is sound.  Showing the soundness of $\pi_3$ would require a
formal verification of its program.  The meaning formula for
$\mname{BPLUS}_{\epsilon \tarrow \epsilon \tarrow \epsilon}$, theorem
3, follows from axioms 5--15.

This biform theory also has a transformer $\pi_{14}$ for
$\mname{BT6-DEC-PROC}_{\epsilon \tarrow \epsilon}$ that implements an
efficient decision procedure for the first-order formulas of the
theory that is specified by axiom 30.  The first-order formulas of the
theory are identified by the transformers $\pi_{5}$ and $\pi_{6}$ for
the defined constant $\mname{IS-FO-BT2}_{\epsilon \tarrow \epsilon}$
that are analogous to the transformers $\pi_{15}$ and $\pi_{16}$ for
$\mname{IS-FO-BT2-ABS}_{\epsilon \tarrow \epsilon}$.

\setcounter{biformthy}{5}

\begin{biformthy}[BT\thebiformthy: Presburger Arithmetic]\em
\bi

  \item[] 

  \item[] \textbf{Primitive Base Types}

  \be

    \item $\iota$ (type of natural numbers).

  \ee

  \item[] \textbf{Primitive Constants}

  \be

    \item $0_\iota$.

    \item $S_{\iota \tarrow \iota}$.

    \item $+_{\iota \tarrow \iota \tarrow \iota}$ (infix).

    \item $\mname{BPLUS}_{\epsilon \tarrow \epsilon \tarrow \epsilon}$ (infix).

    \setcounter{enumi}{5}

    \iffalse
    \item $\mname{BT5-DEC-PROC}_{\epsilon \tarrow \epsilon}$.
    \fi

    \item $\mname{BT6-DEC-PROC}_{\epsilon \tarrow \epsilon}$.

  \ee

  \item[] \textbf{Defined Constants (selected)}

  \be

    \item $1_\iota = S \, 0_\iota$.

    \setcounter{enumi}{2}

    \iffalse
    \item $\mname{IS-FO-BT1}_{\epsilon \tarrow \epsilon} = \LambdaApp
      x_\epsilon \mdot \textbf{B}_\epsilon$ {\sglsp} where
      $\textbf{B}_\epsilon$ is a complex expression such that
      $\syn{(\LambdaApp x_\epsilon \mdot \textbf{B}_\epsilon) \,
        \synbrack{\textbf{A}_\alpha}}$ equals $\syn{\synbrack{T_o}}$
      [$\syn{\synbrack{F_o}}$] if $\textbf{A}_\alpha$ is [not] a term
      or formula of first-order logic with equality whose variables
      are of type $\iota$ and whose nonlogical constants are members
      of $\set{0_\iota,S_{\iota \tarrow \iota}}$.  \fi

    \item $\mname{bnat}_{\iota \tarrow \iota \tarrow \iota} =
      \LambdaApp x_\iota \mdot \LambdaApp y_\iota \mdot ((x_\iota +
      x_\iota) + y_\iota)$.
      
    Notational definition:

    \bi

      \item[] $(0)_2 = \mname{bnat} \, 0_\iota \, 0_\iota$.
  
      \item[] $(1)_2 = \mname{bnat} \, 0_\iota \, 1_\iota$.
  
      \item[] $(a_1 \cdots a_n0)_2 = \mname{bnat} \, (a_1 \cdots
        a_n)_2 \, 0_\iota$ {\sglsp} where each $a_i$ is 0 or 1.
  
      \item[] $(a_1 \cdots a_n1)_2 = \mname{bnat} \, (a_1 \cdots
        a_n)_2 \, 1_\iota$ {\sglsp} where each $a_i$ is 0 or 1.
  
    \ei

    \item $\mname{is-bnum}_{\epsilon \tarrow o} = 
      \IotaApp f_{\epsilon \tarrow o} \mdot
      \ForallApp \syn{u_\epsilon} \mdot
      (f_{\epsilon \tarrow \epsilon} \, \syn{u_\epsilon} \Iff {}\\
      \hspace*{2ex}\ForsomeApp \syn{v_\epsilon} \mdot 
      \ForsomeApp \syn{w_\epsilon} \mdot
      (\syn{u_\epsilon} = \syn{\synbrack{\mname{bnat} \, 
      \commabrack{v_\epsilon} \, \commabrack{w_\epsilon}}} \And {}\\
      \hspace*{4ex}(\syn{v_\epsilon} = \syn{\synbrack{0_\iota}} \OR
      f_{\epsilon \tarrow \epsilon} \, \syn{v_\epsilon}) \And
      (\syn{w_\epsilon} = \syn{\synbrack{0_\iota}} \OR
      \syn{w_\epsilon} =
      \syn{\synbrack{1_\iota}})))$.\footnote{Notation of the form
        $\synbrack{\cdots\commabrack{\cdot}\cdots}$ represents a
        quasiquotation; see~\cite{FarmerArxiv16} for details.}

    \item $\mname{IS-FO-BT2}_{\epsilon \tarrow \epsilon} = \LambdaApp
      x_\epsilon \mdot \textbf{B}_\epsilon$ {\sglsp} where
      $\textbf{B}_\epsilon$ is a complex expression such that
      $\syn{(\LambdaApp x_\epsilon \mdot \textbf{B}_\epsilon) \,
        \synbrack{\textbf{A}_\alpha}}$ equals $\syn{\synbrack{T_o}}$
      [$\syn{\synbrack{F_o}}$] if $\textbf{A}_\alpha$ is [not] a term
      or formula of first-order logic with equality whose variables
      are of type $\iota$ and whose nonlogical constants are members
      of $\set{0_\iota,S_{\iota \tarrow \iota},+_{\iota \tarrow \iota
          \tarrow \iota}}$.

    \setcounter{enumi}{6}

    \iffalse

    \item $\mname{IS-FO-BT1-ABS}_{\epsilon \tarrow \epsilon} = {}\\
    \hspace*{2ex}\LambdaApp \syn{x_\epsilon} \mdot 
    (\If \; (\mname{is-abs}_{\epsilon \tarrow o} \, \syn{x_\epsilon}) \;
    \syn{(\mname{IS-FO-BT1}_{\epsilon \tarrow \epsilon} \,
    (\mname{abs-body}_{\epsilon \tarrow \epsilon} \, x_\epsilon))} \;
    \syn{\synbrack{F_o}})$.

    \fi

    \item $\mname{IS-FO-BT2-ABS}_{\epsilon \tarrow \epsilon} = {}\\
    \hspace*{2ex}\LambdaApp \syn{x_\epsilon} \mdot 
    (\If \; (\mname{is-abs}_{\epsilon \tarrow o} \, \syn{x_\epsilon}) \;
    \syn{(\mname{IS-FO-BT2}_{\epsilon \tarrow \epsilon} \,
    (\mname{abs-body}_{\epsilon \tarrow \epsilon} \, x_\epsilon))} \;
    \syn{\synbrack{F_o}})$.

  \ee

  \item[] \textbf{Axioms}

  \be

    \item $S \, x_\iota \not= 0_\iota$.

    \item $S \, x_\iota = S \, y_\iota \Implies x_\iota =
      y_\iota$.

    \item $x_\iota + 0_\iota = x_\iota$.

    \item $x_\iota + S \, y_\iota = S \, (x_\iota + y_\iota)$.

    \item $\mname{is-bnum} \, \syn{u_\epsilon} \Implies
      \syn{u_\epsilon \; \mname{BPLUS} \; \synbrack{(0)_2}} =
      \syn{u_\epsilon}$.

    \item[] $\vdots$

    \setcounter{enumi}{14}

    \item $(\mname{is-bnum} \, \syn{u_\epsilon} \And \mname{is-bnum}
      \, \syn{v_\epsilon}) \Implies {}\\
        \hspace*{2ex} \syn{\synbrack{\mname{bnat} \,
            \commabrack{u_\epsilon} \, 1_\iota} \; \mname{BPLUS} \;
          \synbrack{\mname{bnat} \, \commabrack{v_\epsilon} \, 1_\iota}} = {}\\
        \hspace*{2ex}\syn{\synbrack{\mname{bnat} \,
            \commabrack{(u_\epsilon \; \mname{BPLUS} \; v_\epsilon)
              \; \mname{BPLUS} \; \synbrack{(1)_2}} \, 0_\iota}}$.

    \iffalse
    \item Induction Schema for $S$

    $\ForallApp \syn{f_\epsilon} \mdot 
    ((\mname{is-expr}_{\epsilon \tarrow o}^{\iota \tarrow o} \, \syn{f_\epsilon} \And
    \sembrack{\syn{\mname{IS-FO-BT1-ABS}_{\epsilon \tarrow \epsilon} \, 
    f_\epsilon}}_o) \Implies {} \\
    \hspace*{2ex}((\sembrack{\syn{f_\epsilon}}_{\iota \tarrow o} \, 0_\iota \And
    (\ForallApp x_\iota \mdot 
    \sembrack{\syn{f_\epsilon}}_{\iota \tarrow o} \, x_\iota \Implies
    \sembrack{\syn{f_\epsilon}}_{\iota \tarrow o} \, 
    (\mname{S} \, x_\iota))) \Implies 
    \ForallApp x_\iota \mdot 
    \sembrack{\syn{f_\epsilon}}_{\iota \tarrow o} \, x_\iota))$.

    \item Meaning Formula for $\mname{BT5-DEC-PROC}_{\epsilon \tarrow \epsilon}$

    $\ForallApp \syn{u_\epsilon} \mdot 
    ((\mname{is-expr}_{\epsilon \tarrow o}^{o} \, 
    \syn{u_\epsilon} \And
    \mname{is-closed}_{\epsilon \tarrow o} \, \syn{u_\epsilon} \And 
    \sembrack{\syn{\mname{IS-FO-BT1}_{\epsilon \tarrow \epsilon} \, 
    u_\epsilon}}_o) \Implies {}\\
    \hspace*{2ex}((\syn{\mname{BT5-DEC-PROC}_{\epsilon \tarrow \epsilon} \, u_\epsilon} = 
    \syn{\synbrack{T_o}} \OR 
    \syn{\mname{BT5-DEC-PROC}_{\epsilon \tarrow \epsilon} \, u_\epsilon} = 
    \syn{\synbrack{F_o}}) \And {}\\
    \hspace*{3.1ex}
    \sembrack{\syn{\mname{BT5-DEC-PROC}_{\epsilon \tarrow \epsilon} \, u_\epsilon}}_o =
    \sembrack{\syn{u_\epsilon}}_o))$.
    \fi

    \setcounter{enumi}{28}

    \item Induction Schema for $S$ and $+$

    $\ForallApp \syn{f_\epsilon} \mdot 
    ((\mname{is-expr}_{\epsilon \tarrow o}^{\iota \tarrow o} \, \syn{f_\epsilon} \And
    \sembrack{\syn{\mname{IS-FO-BT2-ABS}_{\epsilon \tarrow \epsilon} \, 
    f_\epsilon}}_o) \Implies {} \\
    \hspace*{2ex}((\sembrack{\syn{f_\epsilon}}_{\iota \tarrow o} \, 0_\iota \And
    (\ForallApp x_\iota \mdot 
    \sembrack{\syn{f_\epsilon}}_{\iota \tarrow o} \, x_\iota \Implies
    \sembrack{\syn{f_\epsilon}}_{\iota \tarrow o} \, 
    (\mname{S} \, x_\iota))) \Implies 
    \ForallApp x_\iota \mdot 
    \sembrack{\syn{f_\epsilon}}_{\iota \tarrow o} \, x_\iota))$.

    \item Meaning formula for $\mname{BT6-DEC-PROC}_{\epsilon \tarrow \epsilon}$.

    $\ForallApp \syn{u_\epsilon} \mdot 
    ((\mname{is-expr}_{\epsilon \tarrow o}^{o} \, 
    \syn{u_\epsilon}  \And
    \mname{is-closed}_{\epsilon \tarrow o} \, \syn{u_\epsilon} \And 
    \sembrack{\syn{\mname{IS-FO-BT2}_{\epsilon \tarrow \epsilon} \, 
    u_\epsilon}}_o) \Implies {}\\
    \hspace*{2ex}((\syn{\mname{BT6-DEC-PROC}_{\epsilon \tarrow \epsilon} \, u_\epsilon} = 
    \syn{\synbrack{T_o}} \OR 
    \syn{\mname{BT6-DEC-PROC}_{\epsilon \tarrow \epsilon} \, u_\epsilon} = 
    \syn{\synbrack{F_o}}) \And {}\\
    \hspace*{3.1ex}
    \sembrack{\syn{\mname{BT6-DEC-PROC}_{\epsilon \tarrow \epsilon} \, u_\epsilon}}_o =
    \sembrack{\syn{u_\epsilon}}_o))$.

  \ee

  \item[] \textbf{Transformers}

  \be

    \iffalse

    \item $\pi_1$ for $\mname{IS-FO-BT1}_{\epsilon \tarrow \epsilon}$
      is an efficient program that accesses the data stored in the
      data structures that represent expressions.

    \item $\pi_2$ for $\mname{IS-FO-BT1}_{\epsilon \tarrow \epsilon}$
      uses the definition of $\mname{IS-FO-BT1}_{\epsilon \tarrow
        \epsilon}$.  

    \fi

    \setcounter{enumi}{2}

    \item $\pi_3$ for $\mname{BPLUS}_{\epsilon \tarrow \epsilon
      \tarrow \epsilon}$ is an efficient program that satisfies Axioms
      5--15.

    \item $\pi_4$ for $\mname{BPLUS}_{\epsilon \tarrow \epsilon
      \tarrow \epsilon}$ uses Axioms 5--15 as conditional rewrite
      rules.

    \item $\pi_5$ for $\mname{IS-FO-BT2}_{\epsilon \tarrow
      \epsilon}$ is an efficient program that accesses the data
      stored in the data structures that represent expressions.

    \item $\pi_6$ for $\mname{IS-FO-BT2}_{\epsilon \tarrow \epsilon}$
      uses the definition of $\mname{IS-FO-BT2}_{\epsilon \tarrow
        \epsilon}$.

    \iffalse

    \item $\pi_{11}$ for $\mname{BT5-DEC-PROC}_{\epsilon \tarrow
      \epsilon \tarrow \epsilon}$ is an efficient decision procedure
      that satisfies Axiom 28.

    \item $\pi_{12}$ for $\mname{IS-FO-BT1-ABS}_{\epsilon \tarrow
      \epsilon}$ is an efficient program that accesses the data stored
      in the data structures that represent expressions.

    \item $\pi_{13}$ for $\mname{IS-FO-BT1-ABS}_{\epsilon \tarrow
      \epsilon}$ uses the definition of
      $\mname{IS-FO-BT1-ABS}_{\epsilon \tarrow \epsilon}$.  

    \fi

    \setcounter{enumi}{13}

    \item $\pi_{14}$ for $\mname{BT6-DEC-PROC}_{\epsilon \tarrow
      \epsilon \tarrow \epsilon}$ is an efficient decision procedure
      that satisfies Axiom 30.

    \item $\pi_{15}$ for $\mname{IS-FO-BT2-ABS}_{\epsilon \tarrow
      \epsilon}$ is an efficient program that accesses the data stored
      in the data structures that represent expressions.

    \item $\pi_{16}$ for $\mname{IS-FO-BT2-ABS}_{\epsilon \tarrow
      \epsilon}$ uses the definition of
      $\mname{IS-FO-BT2-ABS}_{\epsilon \tarrow \epsilon}$.

  \ee

  \item[] \textbf{Theorems (selected)}

    \be

      \setcounter{enumi}{2}

      \item Meaning formula for 
      $\mname{BPLUS}_{\epsilon \tarrow \epsilon \tarrow \epsilon}$

      $\ForallApp \syn{u_\epsilon} \mdot \ForallApp \syn{v_\epsilon} \mdot
      ((\mname{is-bnum} \, \syn{u_\epsilon} \And \mname{is-bnum} \, \syn{v_\epsilon}) 
      \Implies {}\\
      \hspace*{2ex}(\mname{is-bnum} \, 
      \syn{(u_\epsilon \; \mname{BPLUS} \; v_\epsilon)} \And {}
      (\sembrack{\syn{u_\epsilon \; \mname{BPLUS} \; 
      v_\epsilon}}_\iota = 
      \sembrack{\syn{u_\epsilon}}_\iota + \sembrack{\syn{v_\epsilon}}_\iota)))$.

    \ee

\ei
\end{biformthy}

\section{Study 2: Test Case Formalized in Agda}\label{sec:agda}

As our goal is to, in part, compare the global approach and the local
approach, the formalization in Agda~\cite{Norell09,AgdaWiki} eschews the use of
its reflection capabilities\footnote{As of early $2017$, there is no official
publication describing these features outside of the Agda
documentation, but see~\cite{VanDerWalt12,VanDerWaltSwierstra12}}.
Thus this formalization replaces the global type $\epsilon$ (of
{\churchuqe}) by a \emph{set} of inductive types, one for each of the
biform theories.  This is still reflection, just hand-rolled.  We also
need to express formulas in FOL (as syntax), so we need a type for
that as well.  To be more modular, this is done as a type for
first-order logic (with equality) over any ground language.
We display some illustrative samples here; the full code is available
in~\appendixref{B}.

\iffalse
and at~\cite{agda-code}.
\fi

An abstract theory is modeled as a \emph{record}.  For example, we have
\AgdaRecord{BT₁}:
\input{T1.tex}
One of the languages needed is an extension of the naturals which
allows variables:

\input{NatVar.tex}

But where the informal description in section~\ref{sec:test-case} can
get away with saying ``Logic: FOL'' and ``Transformers: Recognizers for
the formulas of the theory'', here we need to be very explicit.  To do
so, we need to define some language infrastructure.

\input{Language.tex}
\input{T6.tex}

\input{Numerals.tex}
\input{T2a.tex}

\section{Comparison of the Two Formalizations}\label{sec:comparison}

As expected, we were able to formalize this network of theories using
both methods.  Neither are fully complete; both are missing the actual
decision procedures (which would be large undertakings).  In particular,
\begin{itemize}
\item The {\churchuqe} formalization is missing the definition of the language
recognizers, as well as the full assurance of being mechanically checked.
It has no ``implementation'' of any transformers.
\item The Agda version implements evaluation but not substitution --- which means
that the induction statement in BT5--BT7 are not quite the same as in
{\churchuqe}; the models will be the same however.  It also does not implement
any theory morphisms, as record definitions are not first-class in Agda. 
\end{itemize}

More importantly, because of our (explicit) choice to contrast the
global and local approaches, each version uses different infrastructure to
reason about syntax.
\begin{itemize}
\item {\churchuqe} has a built-in inductive type of ``all syntax'',
along with quotation and evaluation operators for the entire language
of expressions.
\item In the local approach, a new inductive type for each new
``language'' (the numerals, the numerals with plus, the numerals with
plus and times, all three of these augmented with variables, first-order
logic, binary digits, binary numerals) has to be created.  For many of 
these, a variety of traversals (folds) have to be implemented ``by hand''
even though the recursion patterns are obvious, at least to humans.
Some of these are evaluation operators (one per language).  There is no
formal quotation operator.
\end{itemize}

The Agda version has a number of extra features: some transformers (such as
for \AgdaFunction{bplus} and \AgdaFunction{btimes}) are implemented.
Furthermore, the \emph{meaning formula} for \AgdaFunction{bplus} is
shown to be a theorem.  A variety of coherence theorems are also shown,
to gain confidence that the theories really are the ones we want.

It is worth remarking that defining the language of first-order formulas
is complicated in \emph{both} versions.  This has been noticed before by
people doing programming language meta-theory with proof assistants: 
encoding languages, especially languages with binders (such as FOL) along
with traversals and basic reasoning can be very tedious~\cite{poplmark}.

The most notable differences in the two formalizations are:
\begin{enumerate}
\item Because FOL is classical, but Agda's host logic is constructive,
a double-negation embedding was needed.
\item The use of \emph{type equivalence} instead of boolean equality for
verifying that the interpretation of a formula of FOL and the results
of the decision procedure are ``the same'',
\item Borrowing the notion of \emph{contractibility} from 
HoTT~\cite{hottbook}, to encode \emph{definite description}.
\item Extending the decision procedure to \emph{closeable} terms (by
providing an explicit, total valuation) instead of restricting to
closed terms.
\end{enumerate}
The first is basically forced upon us by Agda: it has no Prop type (unlike
Coq), and so we do not know a priori that all interpretations of
first-order formulas are actually $0$-types.  The second is an active
design decision: the infrastructure required to define the meaning of
\emph{closed} which is useful in a constructive setting is quite complex%
\footnote{It would require us to define \emph{paths} in terms, bound and
free variables along paths, quantification over paths, etc.}.  We believe
the third is novel.  The fourth point requires deeper investigating.

\section{Conclusion}\label{sec:conclusion}

\iffalse
\begin{verbatim}
Outline:

  1. Both CTT_uqe and Agda can be used to formalize the biform theories
     involved in the network in Fig.1.

  2. CTT_uqe provides a built-in infrastructure for reasoning about
     syntax, but it is a nonstandard language with an unfamiliar
     semantics that has not been implemented.

  3. Agda is implemented, but by using the local approach,
     (1) the size of its infrastructure for
     reasoning about syntax depends on the size of the biform theory
     graph that is formalized and (2) it does not have an implemented
     notion of theory morphism (Is this true?).  This is not inherently
     a fault of Agda's.

  4. Recommendation: CTT_uqe is implemented, and then the test study
     is fully formalized in this implementation.

  5. Recommendation: implementing the global approach using Agda's
     reflection mechanism, to see how that compares to the global
     approach in CTT_uqe.

  6. To better understand the local approach, some automation to 
     derive languages related to each theory should be implemented.
\end{verbatim}
\fi

We have proposed a biform theory graph test case composed of eight
theories that encode natural number arithmetic and include a variety
of useful transformers.  We have formalized this test case (as a set
of biform theories and theory morphisms) in {\churchuqe} using the
global approach (for metareasoning with reflection) and in Agda using
the local approach.  In both cases, we have produced substantial
partial formalizations that indicate that full formalizations could be
obtained with additional work.

Our results show that, by providing a built-in global infrastructure,
the global approach has a significant advantage over the local
approach.  The local approach is burdened by the necessity to define
an infrastructure --- consisting of an inductive type and an
evaluation operator for the type --- for every set of expressions
manipulated by a transformer.  In general, new local infrastructures
must be created each time a new theory is added to the theory graph.
On the other hand, the global approach employs an infrastructure ---
consisting of an inductive type, a quotation operator, and an
evaluation operator --- for the entire set of expressions in the
logic.  This single infrastructure is used for every theory in the
theory graph.

We recommend that future research is directed toward making the global
approach for metareasoning with reflection into a practical method for
formalizing biform theories.  This can be done by developing and
implementing logics such as {\churchqe}~\cite{FarmerArxiv16,Farmer16}
and {\churchuqe}~\cite{FarmerArxiv17} and by adding global
infrastructures to proof systems such as Agda and Coq
(see~\cite{VanDerWalt12,VanDerWaltSwierstra12} for work in this
direction).

\bibliography{imps}
\bibliographystyle{plain}

\iftoggle{cicm}{}{%
\appendix
\input{app-ctt.tex}
\section{Agda Formalization}\label{app:agda}
First, some infrastructure, then the theories themselves.

\long\def\AgdaHide#1{#1}
\subsection{Definite Description}
\input{DefiniteDescr.tex}
\subsection{Equivalences of Types}
\input{Equiv.tex}
\subsection{Numerals}
\input{Numerals.tex}
\subsection{NumPlus}
\input{NumPlus.tex}
\subsection{NumPlusTimes}
\input{NumPlusTimes.tex}
\subsection{Naturals as variables, and with variables}
\input{NatVar.tex}
\subsection{Language infrastructure}
Note that much of this code is in the main paper already.

\input{Language.tex}
\subsection{Some languages of variables}
\input{Variables.tex}

\subsection{T1}

\input{T1.tex}
\subsection{T2}
\input{T2.tex}

\subsection{T2a}

\input{T2a.tex}
\subsection{T3}
\input{T3.tex}

\subsection{T4}
\input{T4.tex}

\subsection{T5}
\input{T5.tex}

\subsection{T6}

\input{T6.tex}
\subsection{T7}
\input{T7.tex}

\subsection{T8}
\input{T8.tex}

}

\end{document}

%% file: formalizing-mk-def.tex
\newcommand{\be}{\begin{enumerate}}
\newcommand{\ee}{\end{enumerate}}
\newcommand{\bi}{\begin{itemize}}
\newcommand{\ei}{\end{itemize}}
\newcommand{\bc}{\begin{center}}
\newcommand{\ec}{\end{center}}
\newcommand{\bsp}{\begin{sloppypar}}
\newcommand{\esp}{\end{sloppypar}}

\newtheorem{biformthy}{Biform Theory}
\newtheorem{thymorphism}{Theory Morphism}

\newcommand{\sglsp}{\ }

\newcommand{\sE}{\mbox{$\cal E$}}

\renewcommand{\phi}{\varphi}

\newcommand{\churchqe}{$\mbox{\sc ctt}_{\rm qe}$}
\newcommand{\churchuqe}{$\mbox{\sc ctt}_{\rm uqe}$}

\newcommand{\set}[1]{{\{ #1 \}}}
\newcommand{\sembrack}[1]{\llbracket#1\rrbracket}
\newcommand{\synbrack}[1]{\ulcorner#1\urcorner}
\newcommand{\commabrack}[1]{\lfloor#1\rfloor}
\newcommand{\mname}[1]{\mbox{\sf #1}}

\newcommand{\mdot}{\mathrel.}
\newcommand{\tarrow}{\rightarrow}
\newcommand{\LambdaApp}{\lambda\,}

\renewcommand{\And}{\wedge}
\newcommand{\Implies}{\supset}
\newcommand{\OR}{\vee} % \Or is used elsewhere
\newcommand{\Iff}{\equiv}

\newcommand{\ForallApp}{\forall\,}

\newcommand{\ForsomeApp}{\exists\,}

\newcommand{\IotaApp}{\mbox{\rm I}\,}

\newcommand{\If}{\mname{if}}

\newcommand{\imps}{\mbox{\sc imps}}
\newcommand{\mmt}{\mbox{\sc Mmt}}

\newcommand{\inclusion}{\hookrightarrow}

\tikzstyle{thy}=[draw, circle, thick]
\tikzstyle{inclusion-arrow}= [right hook-angle 45, thick]
\tikzstyle{preview}=[decorate,
                     decoration={coil,aspect=0,amplitude=1pt,
                                 segment length=6pt,
                                 pre=lineto,pre length=5pt,
                                 post=lineto,post length=7pt},
                     thick]
\tikzstyle{view-arrow}= [preview,-angle 45, thick]

%% file: T1.tex
\AgdaHide{
\begin{code}%
\>\AgdaComment{-- The encoding uses the 'local method'.}\<%
\\
\>\AgdaKeyword{module} \AgdaModule{T1} \AgdaKeyword{where}\<%
\\
\\
\>\AgdaKeyword{open} \AgdaKeyword{import} \AgdaModule{Relation.Binary} \AgdaKeyword{using} \AgdaSymbol{(}\AgdaRecord{DecSetoid}\AgdaSymbol{)}\<%
\\
\>\AgdaKeyword{open} \AgdaModule{DecSetoid} \AgdaKeyword{using} \AgdaSymbol{(}\AgdaField{Carrier}\AgdaSymbol{)}\<%
\\
\>\AgdaKeyword{open} \AgdaKeyword{import} \AgdaModule{Level} \AgdaKeyword{using} \AgdaSymbol{()} \AgdaKeyword{renaming} \AgdaSymbol{(}\AgdaPrimitive{zero} \AgdaSymbol{to} \AgdaPrimitive{lzero}\AgdaSymbol{)}\<%
\\
\\
\>\AgdaComment{-- we use ⊥, ¬ and ≡ from the 'meta' logic}\<%
\\
\>\AgdaKeyword{open} \AgdaKeyword{import} \AgdaModule{Data.Empty} \AgdaKeyword{using} \AgdaSymbol{(}\AgdaDatatype{⊥}\AgdaSymbol{)}\<%
\\
\>\AgdaKeyword{open} \AgdaKeyword{import} \AgdaModule{Relation.Nullary} \AgdaKeyword{using} \AgdaSymbol{(}\AgdaFunction{¬\_}\AgdaSymbol{)}\<%
\\
\>\AgdaKeyword{open} \AgdaKeyword{import} \AgdaModule{Relation.Binary.PropositionalEquality}\<%
\\
\>[0]\AgdaIndent{2}{}\<[2]%
\>[2]\AgdaKeyword{using} \AgdaSymbol{(}\AgdaDatatype{\_≡\_}\AgdaSymbol{;} \AgdaInductiveConstructor{refl}\AgdaSymbol{)}\<%
\\
\>\AgdaKeyword{open} \AgdaKeyword{import} \AgdaModule{Data.Nat} \AgdaKeyword{using} \AgdaSymbol{(}\AgdaDatatype{ℕ}\AgdaSymbol{;} \AgdaInductiveConstructor{suc}\AgdaSymbol{)} \AgdaComment{-- instead of defining our own}\<%
\\
\>[0]\AgdaIndent{2}{}\<[2]%
\>[2]\AgdaComment{-- isomorphic copy}\<%
\\
\>\AgdaKeyword{open} \AgdaKeyword{import} \AgdaModule{Data.Product} \AgdaKeyword{using} \AgdaSymbol{(}\AgdaRecord{Σ}\AgdaSymbol{;} \AgdaInductiveConstructor{\_,\_}\AgdaSymbol{;} \AgdaField{proj₁}\AgdaSymbol{;} \AgdaField{proj₂}\AgdaSymbol{)}\<%
\\
\>\AgdaKeyword{open} \AgdaKeyword{import} \AgdaModule{Data.List} \AgdaKeyword{using} \AgdaSymbol{(}\AgdaFunction{[\_]}\AgdaSymbol{)}\<%
\\
\>\AgdaKeyword{open} \AgdaKeyword{import} \AgdaModule{Data.Bool} \AgdaKeyword{using} \AgdaSymbol{(}\AgdaInductiveConstructor{false}\AgdaSymbol{)}\<%
\\
\\
\>\AgdaComment{-- we will eventually need this}\<%
\\
\>\AgdaKeyword{open} \AgdaKeyword{import} \AgdaModule{Language}\<%
\\
\>\AgdaKeyword{open} \AgdaKeyword{import} \AgdaModule{NatVar}\<%
\\
\\
\>\AgdaKeyword{private}\<%
\\
\>[0]\AgdaIndent{2}{}\<[2]%
\>[2]\AgdaFunction{DT} \AgdaSymbol{=} \AgdaRecord{DecSetoid} \AgdaPrimitive{lzero} \AgdaPrimitive{lzero}\<%
\end{code}
}

\begin{code}%
\>\<%
\\
\>\AgdaKeyword{record} \AgdaRecord{BT₁} \AgdaSymbol{:} \AgdaPrimitiveType{Set₁} \AgdaKeyword{where}\<%
\\
\>[0]\AgdaIndent{2}{}\<[2]%
\>[2]\AgdaKeyword{field}\<%
\\
\>[2]\AgdaIndent{4}{}\<[4]%
\>[4]\AgdaField{nat} \AgdaSymbol{:} \AgdaPrimitiveType{Set₀}\<%
\\
\>[2]\AgdaIndent{4}{}\<[4]%
\>[4]\AgdaField{Z} \AgdaSymbol{:} \AgdaField{nat}\<%
\\
\>[2]\AgdaIndent{4}{}\<[4]%
\>[4]\AgdaField{S} \AgdaSymbol{:} \AgdaField{nat} \AgdaSymbol{→} \AgdaField{nat}\<%
\\
\>[2]\AgdaIndent{4}{}\<[4]%
\>[4]\AgdaField{S≠Z} \AgdaSymbol{:} \AgdaSymbol{∀} \AgdaBound{x} \AgdaSymbol{→} \AgdaFunction{¬} \AgdaSymbol{(}\AgdaField{S} \AgdaBound{x} \AgdaDatatype{≡} \AgdaField{Z}\AgdaSymbol{)}\<%
\\
\>[2]\AgdaIndent{4}{}\<[4]%
\>[4]\AgdaField{inj} \AgdaSymbol{:} \AgdaSymbol{∀} \AgdaBound{x} \AgdaBound{y} \AgdaSymbol{→} \AgdaField{S} \AgdaBound{x} \AgdaDatatype{≡} \AgdaField{S} \AgdaBound{y} \AgdaSymbol{→} \AgdaBound{x} \AgdaDatatype{≡} \AgdaBound{y}\<%
\\
\\
\>[0]\AgdaIndent{2}{}\<[2]%
\>[2]\AgdaFunction{One} \AgdaSymbol{:} \AgdaField{nat}\<%
\\
\>[0]\AgdaIndent{2}{}\<[2]%
\>[2]\AgdaFunction{One} \AgdaSymbol{=} \AgdaField{S} \AgdaField{Z}\<%
\end{code} 

\noindent where we see a field \AgdaField{nat} for the
new type (pronounced \AgdaPrimitiveType{Set} in Agda), the two constants,
and the two axioms.  The host logic is dependently typed, and so the
axioms refer to the constants just defined.  \AgdaFunction{One} is not
a field, but a defined constant.

For simplicity, we will take the built-in type $\mathbb{N}$,
defined as an inductive type, as the \emph{syntax} for natural
numbers, which is also the syntax associated to the theory
\AgdaRecord{BT₁}.  Whereas in {\churchuqe} there is a global evaluation, here
we also need to define evaluation explicitly (a subscript is used to
indicate which theory it belongs to).\\

\begin{code}%
\>[0]\AgdaIndent{2}{}\<[2]%
\>[2]\AgdaFunction{⟦\_⟧₁} \AgdaSymbol{:} \AgdaDatatype{ℕ} \AgdaSymbol{→} \AgdaField{nat}\<%
\\
\>[0]\AgdaIndent{2}{}\<[2]%
\>[2]\AgdaFunction{⟦} \AgdaNumber{0} \AgdaFunction{⟧₁} \AgdaSymbol{=} \AgdaField{Z}\<%
\\
\>[0]\AgdaIndent{2}{}\<[2]%
\>[2]\AgdaFunction{⟦} \AgdaInductiveConstructor{suc} \AgdaBound{x} \AgdaFunction{⟧₁} \AgdaSymbol{=} \AgdaField{S} \AgdaFunction{⟦} \AgdaBound{x} \AgdaFunction{⟧₁}\<%
\end{code}
\AgdaHide{
\begin{code}%
\>[0]\AgdaIndent{2}{}\<[2]%
\>[2]\AgdaComment{-- and some coherence theorems:}\<%
\\
\>[0]\AgdaIndent{2}{}\<[2]%
\>[2]\AgdaFunction{pres-S≠Z} \AgdaSymbol{:} \AgdaSymbol{(}\AgdaBound{x} \AgdaSymbol{:} \AgdaDatatype{ℕ}\AgdaSymbol{)} \AgdaSymbol{→} \AgdaFunction{¬} \AgdaFunction{⟦} \AgdaInductiveConstructor{suc} \AgdaBound{x} \AgdaFunction{⟧₁} \AgdaDatatype{≡} \AgdaFunction{⟦} \AgdaNumber{0} \AgdaFunction{⟧₁}\<%
\\
\>[0]\AgdaIndent{2}{}\<[2]%
\>[2]\AgdaFunction{pres-S≠Z} \AgdaBound{x} \AgdaSymbol{=} \AgdaField{S≠Z} \AgdaFunction{⟦} \AgdaBound{x} \AgdaFunction{⟧₁}\<%
\\
\\
\>[0]\AgdaIndent{2}{}\<[2]%
\>[2]\AgdaFunction{pres-inj} \AgdaSymbol{:} \AgdaSymbol{(}\AgdaBound{x} \AgdaBound{y} \AgdaSymbol{:} \AgdaDatatype{ℕ}\AgdaSymbol{)} \AgdaSymbol{→} \AgdaField{S} \AgdaFunction{⟦} \AgdaBound{x} \AgdaFunction{⟧₁} \AgdaDatatype{≡} \AgdaField{S} \AgdaFunction{⟦} \AgdaBound{y} \AgdaFunction{⟧₁} \AgdaSymbol{→} \AgdaFunction{⟦} \AgdaBound{x} \AgdaFunction{⟧₁} \AgdaDatatype{≡} \AgdaFunction{⟦} \AgdaBound{y} \AgdaFunction{⟧₁}\<%
\\
\>[0]\AgdaIndent{2}{}\<[2]%
\>[2]\AgdaFunction{pres-inj} \AgdaBound{x} \AgdaBound{y} \AgdaBound{pf} \AgdaSymbol{=} \AgdaField{inj} \AgdaFunction{⟦} \AgdaBound{x} \AgdaFunction{⟧₁} \AgdaFunction{⟦} \AgdaBound{y} \AgdaFunction{⟧₁} \AgdaBound{pf}\<%
\end{code}
}
The accompanying code furthermore proves some basic coherence
theorems which are elided here.  We make two further definitions
(\AgdaRecord{GroundLanguage} describing some language features,
and \AgdaModule{FOL} as our definition of first order logic)
which will be explained in more detail on the next page.

\begin{code}%
\>[0]\AgdaIndent{2}{}\<[2]%
\>[2]\AgdaFunction{nat-lang} \AgdaSymbol{:} \AgdaRecord{GroundLanguage} \AgdaField{nat}\<%
\\
\>[0]\AgdaIndent{2}{}\<[2]%
\>[2]\AgdaFunction{nat-lang} \AgdaSymbol{=} \AgdaKeyword{record} \AgdaSymbol{\{} \AgdaField{Lang} \AgdaSymbol{=} \AgdaSymbol{λ} \AgdaBound{X} \AgdaSymbol{→} \AgdaDatatype{ℕX} \AgdaSymbol{(}\AgdaField{Carrier} \AgdaBound{X}\AgdaSymbol{)}\<%
\\
\>[2]\AgdaIndent{20}{}\<[20]%
\>[20]\AgdaSymbol{;} \AgdaField{value} \AgdaSymbol{=} \AgdaSymbol{λ} \AgdaSymbol{\{}\AgdaBound{V}\AgdaSymbol{\}} \AgdaSymbol{→} \AgdaFunction{val} \AgdaSymbol{\{}\AgdaBound{V}\AgdaSymbol{\}} \AgdaSymbol{\}}\<%
\\
\>[0]\AgdaIndent{4}{}\<[4]%
\>[4]\AgdaKeyword{where}\<%
\\
\>[4]\AgdaIndent{6}{}\<[6]%
\>[6]\AgdaFunction{val} \AgdaSymbol{:} \AgdaSymbol{\{}\AgdaBound{V} \AgdaSymbol{:} \AgdaFunction{DT}\AgdaSymbol{\}} \AgdaSymbol{→} \AgdaDatatype{ℕX} \AgdaSymbol{(}\AgdaField{Carrier} \AgdaBound{V}\AgdaSymbol{)} \AgdaSymbol{→} \AgdaSymbol{(}\AgdaField{Carrier} \AgdaBound{V} \AgdaSymbol{→} \AgdaField{nat}\AgdaSymbol{)} \AgdaSymbol{→} \AgdaField{nat}\<%
\\
\>[4]\AgdaIndent{6}{}\<[6]%
\>[6]\AgdaFunction{val} \<[15]%
\>[15]\AgdaInductiveConstructor{z} \<[20]%
\>[20]\AgdaBound{env} \AgdaSymbol{=} \AgdaField{Z}\<%
\\
\>[4]\AgdaIndent{6}{}\<[6]%
\>[6]\AgdaFunction{val} \AgdaSymbol{\{}\AgdaBound{V}\AgdaSymbol{\}} \AgdaSymbol{(}\AgdaInductiveConstructor{s} \AgdaBound{e}\AgdaSymbol{)} \AgdaBound{env} \AgdaSymbol{=} \AgdaField{S} \AgdaSymbol{(}\AgdaFunction{val} \AgdaSymbol{\{}\AgdaBound{V}\AgdaSymbol{\}} \AgdaBound{e} \AgdaBound{env}\AgdaSymbol{)}\<%
\\
\>[4]\AgdaIndent{6}{}\<[6]%
\>[6]\AgdaFunction{val} \<[14]%
\>[14]\AgdaSymbol{(}\AgdaInductiveConstructor{v} \AgdaBound{x}\AgdaSymbol{)} \AgdaBound{env} \AgdaSymbol{=} \AgdaBound{env} \AgdaBound{x}\<%
\\
\>[4]\AgdaIndent{6}{}\<[6]%
\>[6]\<%
\\
\>[0]\AgdaIndent{2}{}\<[2]%
\>[2]\AgdaKeyword{module} \AgdaModule{fo₁} \AgdaSymbol{=} \AgdaModule{FOL} \AgdaFunction{nat-lang}\<%
\end{code}

We can also demonstrate that the natural numbers are a model:\\

\begin{code}%
\>\AgdaFunction{ℕ-is-T1} \AgdaSymbol{:} \AgdaRecord{BT₁}\<%
\\
\>\AgdaFunction{ℕ-is-T1} \AgdaSymbol{=} \AgdaKeyword{record} \AgdaSymbol{\{} \AgdaField{nat} \AgdaSymbol{=} \AgdaDatatype{ℕ} \AgdaSymbol{;} \AgdaField{Z} \AgdaSymbol{=} \AgdaNumber{0} \AgdaSymbol{;} \AgdaField{S} \AgdaSymbol{=} \AgdaInductiveConstructor{suc}\<%
\\
\>[0]\AgdaIndent{2}{}\<[2]%
\>[2]\AgdaSymbol{;} \AgdaField{S≠Z} \AgdaSymbol{=} \AgdaSymbol{λ} \AgdaBound{x} \AgdaSymbol{→} \AgdaSymbol{λ} \AgdaSymbol{()} \AgdaSymbol{;} \AgdaField{inj} \AgdaSymbol{=} \AgdaSymbol{λ} \AgdaSymbol{\{} \AgdaBound{x} \AgdaSymbol{.}\AgdaBound{x} \AgdaInductiveConstructor{refl} \AgdaSymbol{→} \AgdaInductiveConstructor{refl} \AgdaSymbol{\}} \AgdaSymbol{\}}\<%
\end{code}

\AgdaHide{
\begin{code}%
\>\<%
\\
\>\AgdaComment{-- inverse of the type of sub-term of ℕX}\<%
\\
\>\AgdaFunction{SubTermType} \AgdaSymbol{:} \AgdaSymbol{\{}\AgdaBound{V} \AgdaSymbol{:} \AgdaFunction{DT}\AgdaSymbol{\}} \AgdaSymbol{→} \AgdaDatatype{ℕX} \AgdaSymbol{(}\AgdaField{Carrier} \AgdaBound{V}\AgdaSymbol{)} \AgdaSymbol{→} \AgdaPrimitiveType{Set₀}\<%
\\
\>\AgdaFunction{SubTermType} \AgdaSymbol{\{\_\}} \AgdaInductiveConstructor{z} \AgdaSymbol{=} \AgdaDatatype{⊥}\<%
\\
\>\AgdaFunction{SubTermType} \AgdaSymbol{\{}\AgdaBound{V}\AgdaSymbol{\}} \AgdaSymbol{(}\AgdaInductiveConstructor{s} \AgdaBound{x}\AgdaSymbol{)} \AgdaSymbol{=} \AgdaDatatype{ℕX} \AgdaSymbol{(}\AgdaField{Carrier} \AgdaBound{V}\AgdaSymbol{)}\<%
\\
\>\AgdaFunction{SubTermType} \AgdaSymbol{\{}\AgdaBound{V}\AgdaSymbol{\}} \AgdaSymbol{(}\AgdaInductiveConstructor{v} \AgdaBound{x}\AgdaSymbol{)} \AgdaSymbol{=} \AgdaField{Carrier} \AgdaBound{V}\<%
\\
\\
\>\AgdaComment{-- paths in a ℕX}\<%
\\
\>\AgdaKeyword{data} \AgdaDatatype{Path} \AgdaSymbol{\{}\AgdaBound{V} \AgdaSymbol{:} \AgdaFunction{DT}\AgdaSymbol{\}} \AgdaSymbol{:} \AgdaSymbol{(}\AgdaBound{e} \AgdaSymbol{:} \AgdaDatatype{ℕX} \AgdaSymbol{(}\AgdaField{Carrier} \AgdaBound{V}\AgdaSymbol{))} \AgdaSymbol{→} \AgdaFunction{SubTermType} \AgdaSymbol{\{}\AgdaBound{V}\AgdaSymbol{\}} \AgdaBound{e} \AgdaSymbol{→} \AgdaPrimitiveType{Set₀} \AgdaKeyword{where}\<%
\\
\>[0]\AgdaIndent{2}{}\<[2]%
\>[2]\AgdaInductiveConstructor{sp} \AgdaSymbol{:} \AgdaSymbol{(}\AgdaBound{x} \AgdaSymbol{:} \AgdaDatatype{ℕX} \AgdaSymbol{(}\AgdaField{Carrier} \AgdaBound{V}\AgdaSymbol{))} \AgdaSymbol{→} \AgdaDatatype{Path} \AgdaSymbol{(}\AgdaInductiveConstructor{s} \AgdaBound{x}\AgdaSymbol{)} \AgdaBound{x}\<%
\\
\>[0]\AgdaIndent{2}{}\<[2]%
\>[2]\AgdaInductiveConstructor{vp} \AgdaSymbol{:} \AgdaSymbol{(}\AgdaBound{x} \AgdaSymbol{:} \AgdaField{Carrier} \AgdaBound{V}\AgdaSymbol{)} \AgdaSymbol{→} \AgdaDatatype{Path} \AgdaSymbol{(}\AgdaInductiveConstructor{v} \AgdaBound{x}\AgdaSymbol{)} \AgdaBound{x}\<%
\end{code}
}

%% file: NatVar.tex
\AgdaHide{
Showing that ℕ has decidable equality (and thus can be used as a set
of variables).  Also build ℕX which is ℕ augmented with variables.

\begin{code}%
\>\AgdaKeyword{module} \AgdaModule{NatVar} \AgdaKeyword{where}\<%
\\
\\
\>\AgdaKeyword{open} \AgdaKeyword{import} \AgdaModule{Level} \AgdaKeyword{renaming} \AgdaSymbol{(}\AgdaPrimitive{zero} \AgdaSymbol{to} \AgdaPrimitive{lzero}\AgdaSymbol{)} \AgdaKeyword{hiding} \AgdaSymbol{(}\AgdaPrimitive{suc}\AgdaSymbol{)}\<%
\\
\>\AgdaKeyword{open} \AgdaKeyword{import} \AgdaModule{Data.Nat} \AgdaKeyword{using} \AgdaSymbol{(}\AgdaDatatype{ℕ}\AgdaSymbol{;} \AgdaInductiveConstructor{zero}\AgdaSymbol{;} \AgdaInductiveConstructor{suc}\AgdaSymbol{;} \AgdaFunction{\_≟\_}\AgdaSymbol{)}\<%
\\
\>\AgdaKeyword{open} \AgdaKeyword{import} \AgdaModule{Equiv}\<%
\\
\>\AgdaKeyword{open} \AgdaKeyword{import} \AgdaModule{Data.Empty} \AgdaKeyword{using} \AgdaSymbol{(}\AgdaDatatype{⊥}\AgdaSymbol{)}\<%
\\
\>\AgdaKeyword{open} \AgdaKeyword{import} \AgdaModule{Function} \AgdaKeyword{using} \AgdaSymbol{(}\AgdaFunction{\_∘\_}\AgdaSymbol{;} \AgdaFunction{id}\AgdaSymbol{)}\<%
\\
\>\AgdaKeyword{open} \AgdaKeyword{import} \AgdaModule{Relation.Binary.PropositionalEquality}\<%
\\
\>[0]\AgdaIndent{2}{}\<[2]%
\>[2]\AgdaKeyword{using} \AgdaSymbol{(}\AgdaDatatype{\_≡\_}\AgdaSymbol{;} \AgdaInductiveConstructor{refl}\AgdaSymbol{;} \AgdaFunction{cong}\AgdaSymbol{;} \AgdaFunction{isEquivalence}\AgdaSymbol{)}\<%
\\
\>\AgdaKeyword{open} \AgdaKeyword{import} \AgdaModule{Relation.Binary} \AgdaKeyword{using} \AgdaSymbol{(}\AgdaRecord{DecSetoid}\AgdaSymbol{)}\<%
\\
\\
\>\AgdaKeyword{private}\<%
\\
\>[0]\AgdaIndent{2}{}\<[2]%
\>[2]\AgdaFunction{DT} \AgdaSymbol{:} \AgdaPrimitiveType{Set₁}\<%
\\
\>[0]\AgdaIndent{2}{}\<[2]%
\>[2]\AgdaFunction{DT} \AgdaSymbol{=} \AgdaRecord{DecSetoid} \AgdaPrimitive{lzero} \AgdaPrimitive{lzero}\<%
\\
\>[0]\AgdaIndent{2}{}\<[2]%
\>[2]\<%
\\
\>\AgdaFunction{ℕS} \AgdaSymbol{:} \AgdaFunction{DT}\<%
\\
\>\AgdaFunction{ℕS} \AgdaSymbol{=} \AgdaKeyword{record}\<%
\\
\>[0]\AgdaIndent{2}{}\<[2]%
\>[2]\AgdaSymbol{\{} \AgdaField{Carrier} \AgdaSymbol{=} \AgdaDatatype{ℕ}\<%
\\
\>[0]\AgdaIndent{2}{}\<[2]%
\>[2]\AgdaSymbol{;} \AgdaField{\_≈\_} \AgdaSymbol{=} \AgdaDatatype{\_≡\_}\<%
\\
\>[0]\AgdaIndent{2}{}\<[2]%
\>[2]\AgdaSymbol{;} \AgdaField{isDecEquivalence} \AgdaSymbol{=} \AgdaKeyword{record}\<%
\\
\>[2]\AgdaIndent{4}{}\<[4]%
\>[4]\AgdaSymbol{\{} \AgdaField{isEquivalence} \AgdaSymbol{=} \AgdaFunction{isEquivalence}\<%
\\
\>[2]\AgdaIndent{4}{}\<[4]%
\>[4]\AgdaSymbol{;} \AgdaField{\_≟\_} \AgdaSymbol{=} \AgdaFunction{\_≟\_} \AgdaSymbol{\}} \AgdaSymbol{\}}\<%
\end{code}
}

\begin{code}%
\>\AgdaKeyword{data} \AgdaDatatype{ℕX} \AgdaSymbol{(}\AgdaBound{Var} \AgdaSymbol{:} \AgdaPrimitiveType{Set₀}\AgdaSymbol{)} \AgdaSymbol{:} \AgdaPrimitiveType{Set₀} \AgdaKeyword{where}\<%
\\
\>[0]\AgdaIndent{2}{}\<[2]%
\>[2]\AgdaInductiveConstructor{z} \AgdaSymbol{:} \AgdaDatatype{ℕX} \AgdaBound{Var}\<%
\\
\>[0]\AgdaIndent{2}{}\<[2]%
\>[2]\AgdaInductiveConstructor{s} \AgdaSymbol{:} \AgdaDatatype{ℕX} \AgdaBound{Var} \AgdaSymbol{→} \AgdaDatatype{ℕX} \AgdaBound{Var}\<%
\\
\>[0]\AgdaIndent{2}{}\<[2]%
\>[2]\AgdaInductiveConstructor{v} \AgdaSymbol{:} \AgdaBound{Var} \AgdaSymbol{→} \AgdaDatatype{ℕX} \AgdaBound{Var}\<%
\end{code}

\AgdaHide{
\begin{code}%
\>\AgdaFunction{ℕ⊥≃ℕ} \AgdaSymbol{:} \AgdaDatatype{ℕX} \AgdaDatatype{⊥} \AgdaRecord{≃} \AgdaDatatype{ℕ}\<%
\\
\>\AgdaFunction{ℕ⊥≃ℕ} \AgdaSymbol{=} \AgdaInductiveConstructor{qeq} \AgdaFunction{f} \AgdaFunction{g} \AgdaFunction{f∘g∼id} \AgdaFunction{g∘f∼id}\<%
\\
\>[0]\AgdaIndent{2}{}\<[2]%
\>[2]\AgdaKeyword{where}\<%
\\
\>[2]\AgdaIndent{4}{}\<[4]%
\>[4]\AgdaFunction{f} \AgdaSymbol{:} \AgdaDatatype{ℕX} \AgdaDatatype{⊥} \AgdaSymbol{→} \AgdaDatatype{ℕ}\<%
\\
\>[2]\AgdaIndent{4}{}\<[4]%
\>[4]\AgdaFunction{f} \AgdaInductiveConstructor{z} \AgdaSymbol{=} \AgdaNumber{0}\<%
\\
\>[2]\AgdaIndent{4}{}\<[4]%
\>[4]\AgdaFunction{f} \AgdaSymbol{(}\AgdaInductiveConstructor{s} \AgdaBound{x}\AgdaSymbol{)} \AgdaSymbol{=} \AgdaInductiveConstructor{suc} \AgdaSymbol{(}\AgdaFunction{f} \AgdaBound{x}\AgdaSymbol{)}\<%
\\
\>[2]\AgdaIndent{4}{}\<[4]%
\>[4]\AgdaFunction{f} \AgdaSymbol{(}\AgdaInductiveConstructor{v} \AgdaSymbol{())}\<%
\\
\>[2]\AgdaIndent{4}{}\<[4]%
\>[4]\AgdaFunction{g} \AgdaSymbol{:} \AgdaDatatype{ℕ} \AgdaSymbol{→} \AgdaDatatype{ℕX} \AgdaDatatype{⊥}\<%
\\
\>[2]\AgdaIndent{4}{}\<[4]%
\>[4]\AgdaFunction{g} \AgdaInductiveConstructor{zero} \AgdaSymbol{=} \AgdaInductiveConstructor{z}\<%
\\
\>[2]\AgdaIndent{4}{}\<[4]%
\>[4]\AgdaFunction{g} \AgdaSymbol{(}\AgdaInductiveConstructor{suc} \AgdaBound{n}\AgdaSymbol{)} \AgdaSymbol{=} \AgdaInductiveConstructor{s} \AgdaSymbol{(}\AgdaFunction{g} \AgdaBound{n}\AgdaSymbol{)}\<%
\\
\>[2]\AgdaIndent{4}{}\<[4]%
\>[4]\AgdaFunction{f∘g∼id} \AgdaSymbol{:} \AgdaFunction{f} \AgdaFunction{∘} \AgdaFunction{g} \AgdaFunction{∼} \AgdaFunction{id}\<%
\\
\>[2]\AgdaIndent{4}{}\<[4]%
\>[4]\AgdaFunction{f∘g∼id} \AgdaInductiveConstructor{zero} \AgdaSymbol{=} \AgdaInductiveConstructor{refl}\<%
\\
\>[2]\AgdaIndent{4}{}\<[4]%
\>[4]\AgdaFunction{f∘g∼id} \AgdaSymbol{(}\AgdaInductiveConstructor{suc} \AgdaBound{x}\AgdaSymbol{)} \AgdaSymbol{=} \AgdaFunction{cong} \AgdaInductiveConstructor{suc} \AgdaSymbol{(}\AgdaFunction{f∘g∼id} \AgdaBound{x}\AgdaSymbol{)}\<%
\\
\>[2]\AgdaIndent{4}{}\<[4]%
\>[4]\AgdaFunction{g∘f∼id} \AgdaSymbol{:} \AgdaFunction{g} \AgdaFunction{∘} \AgdaFunction{f} \AgdaFunction{∼} \AgdaFunction{id}\<%
\\
\>[2]\AgdaIndent{4}{}\<[4]%
\>[4]\AgdaFunction{g∘f∼id} \AgdaInductiveConstructor{z} \AgdaSymbol{=} \AgdaInductiveConstructor{refl}\<%
\\
\>[2]\AgdaIndent{4}{}\<[4]%
\>[4]\AgdaFunction{g∘f∼id} \AgdaSymbol{(}\AgdaInductiveConstructor{s} \AgdaBound{x}\AgdaSymbol{)} \AgdaSymbol{=} \AgdaFunction{cong} \AgdaInductiveConstructor{s} \AgdaSymbol{(}\AgdaFunction{g∘f∼id} \AgdaBound{x}\AgdaSymbol{)}\<%
\\
\>[2]\AgdaIndent{4}{}\<[4]%
\>[4]\AgdaFunction{g∘f∼id} \AgdaSymbol{(}\AgdaInductiveConstructor{v} \AgdaSymbol{())}\<%
\end{code}
}

%% file: Language.tex
\AgdaHide{
\begin{code}%
\>\AgdaKeyword{module} \AgdaModule{Language} \AgdaKeyword{where}\<%
\\
\\
\>\AgdaKeyword{open} \AgdaKeyword{import} \AgdaModule{Level} \AgdaKeyword{using} \AgdaSymbol{(}\AgdaPostulate{Level}\AgdaSymbol{;} \AgdaPrimitive{zero}\AgdaSymbol{;} \AgdaPrimitive{suc}\AgdaSymbol{;} \AgdaPrimitive{\_⊔\_}\AgdaSymbol{)}\<%
\\
\>\AgdaKeyword{open} \AgdaKeyword{import} \AgdaModule{Relation.Binary} \AgdaKeyword{using} \AgdaSymbol{(}\AgdaRecord{DecSetoid}\AgdaSymbol{)}\<%
\\
\>\AgdaKeyword{open} \AgdaKeyword{import} \AgdaModule{Relation.Nullary} \AgdaKeyword{using} \AgdaSymbol{(}\AgdaDatatype{Dec}\AgdaSymbol{;} \AgdaInductiveConstructor{yes}\AgdaSymbol{;} \AgdaInductiveConstructor{no}\AgdaSymbol{;} \AgdaFunction{¬\_}\AgdaSymbol{)}\<%
\\
\>\AgdaKeyword{open} \AgdaKeyword{import} \AgdaModule{Data.Bool} \AgdaKeyword{using} \AgdaSymbol{(}\AgdaDatatype{Bool}\AgdaSymbol{;} \AgdaInductiveConstructor{true}\AgdaSymbol{;} \AgdaInductiveConstructor{false}\AgdaSymbol{;} \AgdaFunction{\_∧\_}\AgdaSymbol{;} \AgdaFunction{\_∨\_}\AgdaSymbol{;} \AgdaFunction{\_xor\_}\AgdaSymbol{)}\<%
\\
\>[0]\AgdaIndent{2}{}\<[2]%
\>[2]\AgdaKeyword{renaming} \AgdaSymbol{(}\AgdaFunction{not} \AgdaSymbol{to} \AgdaFunction{bnot}\AgdaSymbol{;} \AgdaFunction{\_≟\_} \AgdaSymbol{to} \AgdaFunction{\_=𝔹\_}\AgdaSymbol{)}\<%
\\
\>\AgdaKeyword{open} \AgdaKeyword{import} \AgdaModule{Relation.Binary.PropositionalEquality}\<%
\\
\>[0]\AgdaIndent{2}{}\<[2]%
\>[2]\AgdaKeyword{using} \AgdaSymbol{(}\AgdaDatatype{\_≡\_}\AgdaSymbol{;} \AgdaInductiveConstructor{refl}\AgdaSymbol{;} \AgdaFunction{sym}\AgdaSymbol{;} \AgdaFunction{trans}\AgdaSymbol{;} \AgdaFunction{cong₂}\AgdaSymbol{)}\<%
\\
\>\AgdaKeyword{open} \AgdaKeyword{import} \AgdaModule{Data.Empty} \AgdaKeyword{using} \AgdaSymbol{(}\AgdaDatatype{⊥}\AgdaSymbol{)}\<%
\\
\>\AgdaKeyword{open} \AgdaKeyword{import} \AgdaModule{Data.Unit} \AgdaKeyword{using} \AgdaSymbol{(}\AgdaRecord{⊤}\AgdaSymbol{)}\<%
\\
\>\AgdaKeyword{open} \AgdaKeyword{import} \AgdaModule{Data.Product} \AgdaKeyword{using} \AgdaSymbol{(}\AgdaRecord{Σ}\AgdaSymbol{;} \AgdaFunction{\_×\_}\AgdaSymbol{;} \AgdaField{proj₁}\AgdaSymbol{;} \AgdaField{proj₂}\AgdaSymbol{;} \AgdaInductiveConstructor{\_,\_}\AgdaSymbol{)}\<%
\\
\>\AgdaKeyword{open} \AgdaKeyword{import} \AgdaModule{Data.Sum} \AgdaKeyword{using} \AgdaSymbol{(}\AgdaDatatype{\_⊎\_}\AgdaSymbol{)}\<%
\\
\>\AgdaKeyword{open} \AgdaKeyword{import} \AgdaModule{Data.List} \AgdaKeyword{using} \AgdaSymbol{(}\AgdaDatatype{List}\AgdaSymbol{;} \AgdaInductiveConstructor{[]}\AgdaSymbol{;} \AgdaInductiveConstructor{\_∷\_}\AgdaSymbol{;} \AgdaFunction{[\_]}\AgdaSymbol{)}\<%
\\
\\
\>\AgdaKeyword{open} \AgdaKeyword{import} \AgdaModule{Variables}\<%
\\
\\
\>\AgdaKeyword{private}\<%
\\
\>[0]\AgdaIndent{2}{}\<[2]%
\>[2]\AgdaFunction{DT} \AgdaSymbol{:} \AgdaPrimitiveType{Set} \AgdaSymbol{(}\AgdaPrimitive{suc} \AgdaPrimitive{zero}\AgdaSymbol{)}\<%
\\
\>[0]\AgdaIndent{2}{}\<[2]%
\>[2]\AgdaFunction{DT} \AgdaSymbol{=} \AgdaRecord{DecSetoid} \AgdaPrimitive{zero} \AgdaPrimitive{zero}\<%
\end{code}
}

One of the important concepts is that of a \emph{language with variables},
in other words a language with a reasonable definition of substitution.
This requires \emph{variables} to come from a
type which has the structure of a decidable setoid (from the
Agda library \AgdaModule{DecSetoid}, and denoted
\AgdaFunction{DT} below).

A language, expressed as an inductive type, is closed, i.e., cannot be
extended.  If a language does not have variables, we cannot add them.
One solution is to deal with \emph{contexts} as first-class citizens.
While that is likely the best long-term solution, here we have gone with
something simpler:  create another language which does, and show that its
variable-free fragment is equivalent to the original.  As that aspect of
our development is straightforward, albeit tedious, we elide it.

As we are concerned with statements in first-order logic over a
variety of languages, it makes sense to modularize this aspect somewhat.
Note that, as we are building syntax via inductive types, we can either
build these as functors and then use a fixpoint combinator to tie the
knot, or we can just bite the bullet and make one large definition.
For now, we chose the latter.  We do parametrize over a
\emph{ground language with variables}.  In turn, this is defined
as a type parametrized by a decidable setoid along with an evaluation
function into some type \AgdaSymbol{T}.

\begin{code}%
\>\AgdaKeyword{record} \AgdaRecord{GroundLanguage} \AgdaSymbol{(}\AgdaBound{T} \AgdaSymbol{:} \AgdaPrimitiveType{Set₀}\AgdaSymbol{)} \AgdaSymbol{:} \AgdaPrimitiveType{Set₁} \AgdaKeyword{where}\<%
\\
\>[0]\AgdaIndent{2}{}\<[2]%
\>[2]\AgdaKeyword{open} \AgdaModule{DecSetoid} \AgdaKeyword{using} \AgdaSymbol{(}\AgdaField{Carrier}\AgdaSymbol{)}\<%
\\
\>[0]\AgdaIndent{2}{}\<[2]%
\>[2]\AgdaKeyword{field}\<%
\\
\>[2]\AgdaIndent{4}{}\<[4]%
\>[4]\AgdaField{Lang} \AgdaSymbol{:} \AgdaFunction{DT} \AgdaSymbol{→} \AgdaPrimitiveType{Set₀}\<%
\\
\>[2]\AgdaIndent{4}{}\<[4]%
\>[4]\AgdaField{value} \AgdaSymbol{:} \AgdaSymbol{\{}\AgdaBound{V} \AgdaSymbol{:} \AgdaFunction{DT}\AgdaSymbol{\}} \AgdaSymbol{→} \AgdaField{Lang} \AgdaBound{V} \AgdaSymbol{→} \AgdaSymbol{(}\AgdaField{Carrier} \AgdaBound{V} \AgdaSymbol{→} \AgdaBound{T}\AgdaSymbol{)} \AgdaSymbol{→} \AgdaBound{T}\<%
\end{code}

A logic over a language (with variables), is then also a parametrized
type as well as a parametrized interpretation into types.  The
definition is almost the same, except that a ground language
interprets into \AgdaSymbol{T} and a logic into \AgdaPrimitiveType{Set₀}.

\begin{code}%
\>\AgdaKeyword{record} \AgdaRecord{LogicOverL} \AgdaSymbol{(}\AgdaBound{T} \AgdaSymbol{:} \AgdaPrimitiveType{Set₀}\AgdaSymbol{)} \AgdaSymbol{(}\AgdaBound{L} \AgdaSymbol{:} \AgdaRecord{GroundLanguage} \AgdaBound{T}\AgdaSymbol{)} \AgdaSymbol{:} \AgdaPrimitiveType{Set₁} \AgdaKeyword{where}\<%
\\
\>[0]\AgdaIndent{2}{}\<[2]%
\>[2]\AgdaKeyword{open} \AgdaModule{DecSetoid} \AgdaKeyword{using} \AgdaSymbol{(}\AgdaField{Carrier}\AgdaSymbol{)}\<%
\\
\>[0]\AgdaIndent{2}{}\<[2]%
\>[2]\AgdaKeyword{field}\<%
\\
\>[2]\AgdaIndent{4}{}\<[4]%
\>[4]\AgdaField{Logic} \AgdaSymbol{:} \AgdaFunction{DT} \AgdaSymbol{→} \AgdaPrimitiveType{Set₀}\<%
\\
\>[2]\AgdaIndent{4}{}\<[4]%
\>[4]\AgdaField{⟦\_⟧\_} \AgdaSymbol{:} \AgdaSymbol{∀} \AgdaSymbol{\{}\AgdaBound{V}\AgdaSymbol{\}} \AgdaSymbol{→} \AgdaField{Logic} \AgdaBound{V} \AgdaSymbol{→} \AgdaSymbol{(}\AgdaField{Carrier} \AgdaBound{V} \AgdaSymbol{→} \AgdaBound{T}\AgdaSymbol{)} \AgdaSymbol{→} \AgdaPrimitiveType{Set₀}\<%
\end{code}

The definition of first-order logic is then straightforward.

\begin{code}%
\>\AgdaKeyword{module} \AgdaModule{FOL} \AgdaSymbol{\{}\AgdaBound{T} \AgdaSymbol{:} \AgdaPrimitiveType{Set₀}\AgdaSymbol{\}} \AgdaSymbol{(}\AgdaBound{L} \AgdaSymbol{:} \AgdaRecord{GroundLanguage} \AgdaBound{T}\AgdaSymbol{)} \AgdaKeyword{where}\<%
\\
\>[0]\AgdaIndent{2}{}\<[2]%
\>[2]\AgdaKeyword{open} \AgdaModule{DecSetoid} \AgdaKeyword{using} \AgdaSymbol{(}\AgdaField{Carrier}\AgdaSymbol{)}\<%
\\
\>[0]\AgdaIndent{2}{}\<[2]%
\>[2]\AgdaKeyword{open} \AgdaModule{GroundLanguage} \AgdaBound{L}\<%
\\
\>[0]\AgdaIndent{2}{}\<[2]%
\>[2]\<%
\\
\>[0]\AgdaIndent{2}{}\<[2]%
\>[2]\AgdaKeyword{data} \AgdaDatatype{FOL} \AgdaSymbol{(}\AgdaBound{V} \AgdaSymbol{:} \AgdaFunction{DT}\AgdaSymbol{)} \AgdaSymbol{:} \AgdaPrimitiveType{Set} \AgdaKeyword{where}\<%
\\
\>[2]\AgdaIndent{4}{}\<[4]%
\>[4]\AgdaInductiveConstructor{tt} \AgdaSymbol{:} \AgdaDatatype{FOL} \AgdaBound{V}\<%
\\
\>[2]\AgdaIndent{4}{}\<[4]%
\>[4]\AgdaInductiveConstructor{ff} \AgdaSymbol{:} \AgdaDatatype{FOL} \AgdaBound{V}\<%
\\
\>[2]\AgdaIndent{4}{}\<[4]%
\>[4]\AgdaInductiveConstructor{\_and\_} \AgdaSymbol{:} \AgdaDatatype{FOL} \AgdaBound{V} \AgdaSymbol{→} \AgdaDatatype{FOL} \AgdaBound{V} \AgdaSymbol{→} \AgdaDatatype{FOL} \AgdaBound{V}\<%
\\
\>[2]\AgdaIndent{4}{}\<[4]%
\>[4]\AgdaInductiveConstructor{\_or\_} \AgdaSymbol{:} \AgdaDatatype{FOL} \AgdaBound{V} \AgdaSymbol{→} \AgdaDatatype{FOL} \AgdaBound{V} \AgdaSymbol{→} \AgdaDatatype{FOL} \AgdaBound{V}\<%
\\
\>[2]\AgdaIndent{4}{}\<[4]%
\>[4]\AgdaInductiveConstructor{not} \AgdaSymbol{:} \AgdaDatatype{FOL} \AgdaBound{V} \AgdaSymbol{→} \AgdaDatatype{FOL} \AgdaBound{V}\<%
\\
\>[2]\AgdaIndent{4}{}\<[4]%
\>[4]\AgdaInductiveConstructor{\_⊃\_} \AgdaSymbol{:} \AgdaDatatype{FOL} \AgdaBound{V} \AgdaSymbol{→} \AgdaDatatype{FOL} \AgdaBound{V} \AgdaSymbol{→} \AgdaDatatype{FOL} \AgdaBound{V}\<%
\\
\>[2]\AgdaIndent{4}{}\<[4]%
\>[4]\AgdaInductiveConstructor{\_==\_} \AgdaSymbol{:} \AgdaField{Lang} \AgdaBound{V} \AgdaSymbol{→} \AgdaField{Lang} \AgdaBound{V} \AgdaSymbol{→} \AgdaDatatype{FOL} \AgdaBound{V}\<%
\\
\>[2]\AgdaIndent{4}{}\<[4]%
\>[4]\AgdaInductiveConstructor{all} \AgdaSymbol{:} \AgdaField{Carrier} \AgdaBound{V} \AgdaSymbol{→} \AgdaDatatype{FOL} \AgdaBound{V} \AgdaSymbol{→} \AgdaDatatype{FOL} \AgdaBound{V}\<%
\\
\>[2]\AgdaIndent{4}{}\<[4]%
\>[4]\AgdaInductiveConstructor{exist} \AgdaSymbol{:} \AgdaField{Carrier} \AgdaBound{V} \AgdaSymbol{→} \AgdaDatatype{FOL} \AgdaBound{V} \AgdaSymbol{→} \AgdaDatatype{FOL} \AgdaBound{V}\<%
\\
\\
\>[0]\AgdaIndent{2}{}\<[2]%
\>[2]\AgdaFunction{override} \AgdaSymbol{:} \AgdaSymbol{\{}\AgdaBound{V} \AgdaSymbol{:} \AgdaFunction{DT}\AgdaSymbol{\}} \AgdaSymbol{→} \AgdaSymbol{(}\AgdaField{Carrier} \AgdaBound{V} \AgdaSymbol{→} \AgdaBound{T}\AgdaSymbol{)} \AgdaSymbol{→} \AgdaField{Carrier} \AgdaBound{V} \AgdaSymbol{→} \AgdaBound{T} \AgdaSymbol{→} \AgdaSymbol{(}\AgdaField{Carrier} \AgdaBound{V} \AgdaSymbol{→} \AgdaBound{T}\AgdaSymbol{)}\<%
\\
\>[0]\AgdaIndent{2}{}\<[2]%
\>[2]\AgdaFunction{override} \AgdaSymbol{\{}\AgdaBound{V}\AgdaSymbol{\}} \AgdaBound{f} \AgdaBound{x} \AgdaBound{t} \AgdaBound{y} \AgdaKeyword{with} \AgdaFunction{DecSetoid.\_≟\_} \AgdaBound{V} \AgdaBound{y} \AgdaBound{x}\<%
\\
\>[0]\AgdaIndent{2}{}\<[2]%
\>[2]\AgdaSymbol{...} \AgdaSymbol{|} \AgdaInductiveConstructor{yes} \AgdaSymbol{\_} \AgdaSymbol{=} \AgdaBound{t}\<%
\\
\>[0]\AgdaIndent{2}{}\<[2]%
\>[2]\AgdaSymbol{...} \AgdaSymbol{|} \AgdaInductiveConstructor{no} \AgdaSymbol{\_} \<[14]%
\>[14]\AgdaSymbol{=} \AgdaBound{f} \AgdaBound{y}\<%
\end{code}

\noindent We can also prove that \AgdaSymbol{FOL} is a logic
over \AgdaSymbol{L} by providing an interpretation.  Of course,
as we are modeling classical logic into a constructive logic,
we have to use a double-negation embedding.  We also choose to
interpret the logic's equality $\AgdaInductiveConstructor{\_==\_}$ as
\emph{propositional equality}, but we could make that choice a
parameter as well.

\begin{code}%
\>[0]\AgdaIndent{2}{}\<[2]%
\>[2]\AgdaFunction{LoL-FOL} \AgdaSymbol{:} \AgdaRecord{LogicOverL} \AgdaBound{T} \AgdaBound{L}\<%
\\
\>[0]\AgdaIndent{2}{}\<[2]%
\>[2]\AgdaFunction{LoL-FOL} \AgdaSymbol{=} \AgdaKeyword{record} \AgdaSymbol{\{} \AgdaField{Logic} \AgdaSymbol{=} \AgdaDatatype{FOL} \AgdaSymbol{;} \AgdaField{⟦\_⟧\_} \AgdaSymbol{=} \AgdaFunction{interp} \AgdaSymbol{\}}\<%
\\
\>[2]\AgdaIndent{3}{}\<[3]%
\>[3]\AgdaKeyword{where}\<%
\\
\>[3]\AgdaIndent{4}{}\<[4]%
\>[4]\AgdaFunction{interp} \AgdaSymbol{:} \AgdaSymbol{\{}\AgdaBound{Var} \AgdaSymbol{:} \AgdaFunction{DT}\AgdaSymbol{\}} \AgdaSymbol{→} \AgdaDatatype{FOL} \AgdaBound{Var} \AgdaSymbol{→} \AgdaSymbol{(}\AgdaField{Carrier} \AgdaBound{Var} \AgdaSymbol{→} \AgdaBound{T}\AgdaSymbol{)} \AgdaSymbol{→} \AgdaPrimitiveType{Set₀}\<%
\\
\>[3]\AgdaIndent{4}{}\<[4]%
\>[4]\AgdaFunction{interp} \AgdaInductiveConstructor{tt} \AgdaBound{env} \AgdaSymbol{=} \AgdaRecord{⊤}\<%
\\
\>[3]\AgdaIndent{4}{}\<[4]%
\>[4]\AgdaFunction{interp} \AgdaInductiveConstructor{ff} \AgdaBound{env} \AgdaSymbol{=} \AgdaDatatype{⊥}\<%
\\
\>[3]\AgdaIndent{4}{}\<[4]%
\>[4]\AgdaFunction{interp} \AgdaSymbol{(}\AgdaBound{e} \AgdaInductiveConstructor{and} \AgdaBound{f}\AgdaSymbol{)} \AgdaBound{env} \AgdaSymbol{=} \AgdaFunction{interp} \AgdaBound{e} \AgdaBound{env} \AgdaFunction{×} \AgdaFunction{interp} \AgdaBound{f} \AgdaBound{env}\<%
\\
\>[3]\AgdaIndent{4}{}\<[4]%
\>[4]\AgdaFunction{interp} \AgdaSymbol{(}\AgdaBound{e} \AgdaInductiveConstructor{or} \AgdaBound{f}\AgdaSymbol{)} \AgdaBound{env} \AgdaSymbol{=} \AgdaFunction{¬} \AgdaFunction{¬} \AgdaSymbol{(}\AgdaFunction{interp} \AgdaBound{e} \AgdaBound{env} \AgdaDatatype{⊎} \AgdaFunction{interp} \AgdaBound{f} \AgdaBound{env}\AgdaSymbol{)}\<%
\\
\>[3]\AgdaIndent{4}{}\<[4]%
\>[4]\AgdaFunction{interp} \AgdaSymbol{(}\AgdaInductiveConstructor{not} \AgdaBound{e}\AgdaSymbol{)} \AgdaBound{env} \AgdaSymbol{=} \AgdaFunction{¬} \AgdaSymbol{(}\AgdaFunction{interp} \AgdaBound{e} \AgdaBound{env}\AgdaSymbol{)}\<%
\\
\>[3]\AgdaIndent{4}{}\<[4]%
\>[4]\AgdaFunction{interp} \AgdaSymbol{(}\AgdaBound{e} \AgdaInductiveConstructor{⊃} \AgdaBound{f}\AgdaSymbol{)} \AgdaBound{env} \AgdaSymbol{=} \AgdaSymbol{(}\AgdaFunction{interp} \AgdaBound{e} \AgdaBound{env}\AgdaSymbol{)} \AgdaSymbol{→} \AgdaSymbol{(}\AgdaFunction{interp} \AgdaBound{f} \AgdaBound{env}\AgdaSymbol{)}\<%
\\
\>[3]\AgdaIndent{4}{}\<[4]%
\>[4]\AgdaFunction{interp} \AgdaSymbol{(}\AgdaBound{x} \AgdaInductiveConstructor{==} \AgdaBound{y}\AgdaSymbol{)} \AgdaBound{env} \AgdaSymbol{=} \AgdaField{value} \AgdaBound{x} \AgdaBound{env} \AgdaDatatype{≡} \AgdaField{value} \AgdaBound{y} \AgdaBound{env}\<%
\\
\>[3]\AgdaIndent{4}{}\<[4]%
\>[4]\AgdaFunction{interp} \AgdaSymbol{\{}\AgdaBound{V}\AgdaSymbol{\}} \AgdaSymbol{(}\AgdaInductiveConstructor{all} \AgdaBound{x} \AgdaBound{p}\AgdaSymbol{)} \AgdaBound{env} \<[31]%
\>[31]\AgdaSymbol{=} \AgdaSymbol{∀} \AgdaBound{z} \AgdaSymbol{→} \AgdaFunction{interp} \AgdaBound{p} \AgdaSymbol{(}\AgdaFunction{override} \AgdaSymbol{\{}\AgdaBound{V}\AgdaSymbol{\}} \AgdaBound{env} \AgdaBound{x} \AgdaBound{z}\AgdaSymbol{)}\<%
\\
\>[3]\AgdaIndent{4}{}\<[4]%
\>[4]\AgdaFunction{interp} \AgdaSymbol{\{}\AgdaBound{V}\AgdaSymbol{\}} \AgdaSymbol{(}\AgdaInductiveConstructor{exist} \AgdaBound{x} \AgdaBound{p}\AgdaSymbol{)} \AgdaBound{env} \AgdaSymbol{=} \AgdaFunction{¬} \AgdaFunction{¬} \AgdaSymbol{(}\AgdaRecord{Σ} \AgdaBound{T} \AgdaSymbol{(λ} \AgdaBound{t} \AgdaSymbol{→} \AgdaFunction{interp} \AgdaBound{p} \AgdaSymbol{(}\AgdaFunction{override} \AgdaSymbol{\{}\AgdaBound{V}\AgdaSymbol{\}} \AgdaBound{env} \AgdaBound{x} \AgdaBound{t}\AgdaSymbol{)))}\<%
\end{code}

%% file: T6.tex
\AgdaHide{
\begin{code}%
\>\AgdaKeyword{module} \AgdaModule{T6} \AgdaKeyword{where}\<%
\\
\>\AgdaKeyword{open} \AgdaKeyword{import} \AgdaModule{Relation.Binary} \AgdaKeyword{using} \AgdaSymbol{(}\AgdaRecord{DecSetoid}\AgdaSymbol{)}\<%
\\
\>\AgdaKeyword{open} \AgdaKeyword{import} \AgdaModule{Level} \AgdaKeyword{using} \AgdaSymbol{()} \AgdaKeyword{renaming} \AgdaSymbol{(}\AgdaPrimitive{zero} \AgdaSymbol{to} \AgdaPrimitive{lzero}\AgdaSymbol{)}\<%
\\
\\
\>\AgdaFunction{DT} \AgdaSymbol{:} \AgdaPrimitiveType{Set₁}\<%
\\
\>\AgdaFunction{DT} \AgdaSymbol{=} \AgdaRecord{DecSetoid} \AgdaPrimitive{lzero} \AgdaPrimitive{lzero}\<%
\\
\\
\>\AgdaKeyword{open} \AgdaKeyword{import} \AgdaModule{T1} \AgdaKeyword{using} \AgdaSymbol{(}\AgdaRecord{BT₁}\AgdaSymbol{)}\<%
\\
\>\AgdaKeyword{open} \AgdaKeyword{import} \AgdaModule{T2} \AgdaKeyword{using} \AgdaSymbol{(}\AgdaRecord{BT₂}\AgdaSymbol{)}\<%
\\
\>\AgdaKeyword{open} \AgdaKeyword{import} \AgdaModule{T5} \AgdaKeyword{using} \AgdaSymbol{(}\AgdaRecord{BT₅}\AgdaSymbol{)}\<%
\\
\\
\>\AgdaKeyword{open} \AgdaKeyword{import} \AgdaModule{Relation.Binary.PropositionalEquality} \AgdaKeyword{using} \AgdaSymbol{(}\AgdaDatatype{\_≡\_}\AgdaSymbol{)}\<%
\\
\>\AgdaKeyword{open} \AgdaKeyword{import} \AgdaModule{Data.Empty} \AgdaKeyword{using} \AgdaSymbol{(}\AgdaDatatype{⊥}\AgdaSymbol{)}\<%
\\
\>\AgdaKeyword{open} \AgdaKeyword{import} \AgdaModule{Data.Sum} \AgdaKeyword{using} \AgdaSymbol{(}\AgdaDatatype{\_⊎\_}\AgdaSymbol{)}\<%
\\
\>\AgdaKeyword{open} \AgdaKeyword{import} \AgdaModule{Data.Product} \AgdaKeyword{using} \AgdaSymbol{(}\AgdaRecord{Σ}\AgdaSymbol{;}\AgdaFunction{\_×\_}\AgdaSymbol{;}\AgdaInductiveConstructor{\_,\_}\AgdaSymbol{)}\<%
\\
\>\AgdaKeyword{open} \AgdaKeyword{import} \AgdaModule{Data.Bool} \AgdaKeyword{using} \AgdaSymbol{(}\AgdaDatatype{Bool}\AgdaSymbol{)}\<%
\\
\>\AgdaKeyword{open} \AgdaKeyword{import} \AgdaModule{Equiv} \AgdaKeyword{using} \AgdaSymbol{(}\AgdaRecord{\_≃\_}\AgdaSymbol{)}\<%
\\
\\
\>\AgdaKeyword{open} \AgdaKeyword{import} \AgdaModule{Variables} \AgdaKeyword{using} \AgdaSymbol{(}\AgdaKeyword{module} \AgdaModule{VarLangs}\AgdaSymbol{;} \AgdaFunction{NoVars}\AgdaSymbol{;} \AgdaFunction{DBool}\AgdaSymbol{)}\<%
\\
\>\AgdaKeyword{open} \AgdaKeyword{import} \AgdaModule{Language}\<%
\\
\>[0]\AgdaIndent{2}{}\<[2]%
\>[2]\AgdaKeyword{using} \AgdaSymbol{(}\AgdaKeyword{module} \AgdaModule{FOL}\AgdaSymbol{;}\<%
\\
\>[2]\AgdaIndent{9}{}\<[9]%
\>[9]\AgdaKeyword{module} \AgdaModule{LogicOverL}\AgdaSymbol{)}\<%
\end{code}
}

With the appropriate infrastructure in place, it is now possible
to define \AgdaRecord{BT₆} from the theories it extends.

\begin{code}%
\>\AgdaKeyword{record} \AgdaRecord{BT₆} \AgdaSymbol{\{}\AgdaBound{t₁} \AgdaSymbol{:} \AgdaRecord{BT₁}\AgdaSymbol{\}} \AgdaSymbol{(}\AgdaBound{t₂} \AgdaSymbol{:} \AgdaRecord{BT₂} \AgdaBound{t₁}\AgdaSymbol{)} \AgdaSymbol{(}\AgdaBound{t₅} \AgdaSymbol{:} \AgdaRecord{BT₅} \AgdaBound{t₁}\AgdaSymbol{)} \AgdaSymbol{:} \AgdaPrimitiveType{Set₁} \AgdaKeyword{where}\<%
\\
\>[0]\AgdaIndent{2}{}\<[2]%
\>[2]\AgdaKeyword{open} \AgdaModule{VarLangs} \AgdaKeyword{using} \AgdaSymbol{(}\AgdaFunction{XV}\AgdaSymbol{;} \AgdaInductiveConstructor{x}\AgdaSymbol{)}\<%
\\
\>[0]\AgdaIndent{2}{}\<[2]%
\>[2]\AgdaKeyword{open} \AgdaModule{DecSetoid} \AgdaKeyword{using} \AgdaSymbol{(}\AgdaField{Carrier}\AgdaSymbol{)}\<%
\\
\>[0]\AgdaIndent{2}{}\<[2]%
\>[2]\AgdaKeyword{open} \AgdaModule{BT₂} \AgdaBound{t₂} \AgdaKeyword{public}\<%
\\
\>[0]\AgdaIndent{2}{}\<[2]%
\>[2]\AgdaKeyword{open} \AgdaModule{fo₂} \AgdaKeyword{using} \AgdaSymbol{(}\AgdaDatatype{FOL}\AgdaSymbol{;} \AgdaInductiveConstructor{tt}\AgdaSymbol{;} \AgdaInductiveConstructor{ff}\AgdaSymbol{;} \AgdaFunction{LoL-FOL}\AgdaSymbol{;} \AgdaInductiveConstructor{\_and\_}\AgdaSymbol{;} \AgdaInductiveConstructor{all}\AgdaSymbol{)}\<%
\\
\>[0]\AgdaIndent{2}{}\<[2]%
\>[2]\AgdaKeyword{open} \AgdaModule{LogicOverL} \AgdaFunction{LoL-FOL}\<%
\\
\\
\>[0]\AgdaIndent{2}{}\<[2]%
\>[2]\AgdaKeyword{field}\<%
\\
\>[2]\AgdaIndent{4}{}\<[4]%
\>[4]\AgdaField{induct} \AgdaSymbol{:} \AgdaSymbol{(}\AgdaBound{e} \AgdaSymbol{:} \AgdaDatatype{FOL} \AgdaFunction{XV}\AgdaSymbol{)} \AgdaSymbol{→}\<%
\\
\>[4]\AgdaIndent{6}{}\<[6]%
\>[6]\AgdaFunction{⟦} \AgdaBound{e} \AgdaFunction{⟧} \AgdaSymbol{(λ} \AgdaSymbol{\{} \AgdaInductiveConstructor{x} \AgdaSymbol{→} \AgdaFunction{⟦} \AgdaNumber{0} \AgdaFunction{⟧₁} \AgdaSymbol{\})} \AgdaSymbol{→}\<%
\\
\>[4]\AgdaIndent{6}{}\<[6]%
\>[6]\AgdaSymbol{(∀} \AgdaBound{y} \AgdaSymbol{→} \AgdaFunction{⟦} \AgdaBound{e} \AgdaFunction{⟧} \AgdaSymbol{(λ} \AgdaSymbol{\{}\AgdaInductiveConstructor{x} \AgdaSymbol{→} \AgdaBound{y}\AgdaSymbol{\})} \AgdaSymbol{→} \AgdaFunction{⟦} \AgdaBound{e} \AgdaFunction{⟧} \AgdaSymbol{(λ} \AgdaSymbol{\{}\AgdaInductiveConstructor{x} \AgdaSymbol{→} \AgdaFunction{S} \AgdaBound{y}\AgdaSymbol{\}))} \AgdaSymbol{→}\<%
\\
\>[4]\AgdaIndent{6}{}\<[6]%
\>[6]\AgdaSymbol{∀} \AgdaBound{y} \AgdaSymbol{→} \AgdaFunction{⟦} \AgdaBound{e} \AgdaFunction{⟧} \AgdaSymbol{(λ} \AgdaSymbol{\{}\AgdaInductiveConstructor{x} \AgdaSymbol{→} \AgdaBound{y}\AgdaSymbol{\})}\<%
\\
\>[0]\AgdaIndent{2}{}\<[2]%
\>[2]\AgdaKeyword{postulate}\<%
\\
\>[2]\AgdaIndent{4}{}\<[4]%
\>[4]\AgdaPostulate{decide} \AgdaSymbol{:} \AgdaSymbol{∀} \AgdaSymbol{\{}\AgdaBound{W}\AgdaSymbol{\}} \AgdaSymbol{→} \AgdaSymbol{(}\AgdaField{Carrier} \AgdaBound{W} \AgdaSymbol{→} \AgdaFunction{nat}\AgdaSymbol{)} \AgdaSymbol{→} \AgdaDatatype{FOL} \AgdaBound{W} \AgdaSymbol{→} \AgdaDatatype{FOL} \AgdaFunction{NoVars}\<%
\\
\>[2]\AgdaIndent{4}{}\<[4]%
\>[4]\AgdaPostulate{meaning-decide} \AgdaSymbol{:} \AgdaSymbol{\{}\AgdaBound{W} \AgdaSymbol{:} \AgdaFunction{DT}\AgdaSymbol{\}} \AgdaSymbol{(}\AgdaBound{env} \AgdaSymbol{:} \AgdaField{Carrier} \AgdaBound{W} \AgdaSymbol{→} \AgdaFunction{nat}\AgdaSymbol{)} \AgdaSymbol{→} \AgdaSymbol{(}\AgdaBound{env′} \AgdaSymbol{:} \AgdaDatatype{⊥} \AgdaSymbol{→} \AgdaFunction{nat}\AgdaSymbol{)} \AgdaSymbol{→}\<%
\\
\>[4]\AgdaIndent{6}{}\<[6]%
\>[6]\AgdaSymbol{(}\AgdaBound{e} \AgdaSymbol{:} \AgdaDatatype{FOL} \AgdaBound{W}\AgdaSymbol{)} \AgdaSymbol{→}\<%
\\
\>[4]\AgdaIndent{6}{}\<[6]%
\>[6]\AgdaKeyword{let} \AgdaBound{res} \AgdaSymbol{=} \AgdaPostulate{decide} \AgdaBound{env} \AgdaBound{e} \AgdaKeyword{in}\<%
\\
\>[4]\AgdaIndent{6}{}\<[6]%
\>[6]\AgdaSymbol{(}\AgdaBound{res} \AgdaDatatype{≡} \AgdaInductiveConstructor{tt} \AgdaDatatype{⊎} \AgdaBound{res} \AgdaDatatype{≡} \AgdaInductiveConstructor{ff}\AgdaSymbol{)} \AgdaFunction{×} \AgdaSymbol{(}\AgdaFunction{⟦} \AgdaBound{e} \AgdaFunction{⟧} \AgdaBound{env}\AgdaSymbol{)} \AgdaRecord{≃} \AgdaSymbol{(}\AgdaFunction{⟦} \AgdaBound{res} \AgdaFunction{⟧} \AgdaBound{env′}\AgdaSymbol{)}\<%
\end{code}
While section~\ref{sec:cttuqe} presents the \emph{flattened} theory,
here we need only define what is new over the extended theory, namely
an induction schema, a decision procedure and its meaning formula.

Here is a guide to understanding the above definition:
\renewcommand{\labelenumi}{(\theenumi)}
\begin{enumerate*}
\item \AgdaDatatype{XV} is a (decidable) type with a single
inhabitant, \AgdaInductiveConstructor{x}.
\item All fields of \AgdaModule{BT₂} are made publicly visible for
\AgdaModule{BT₆}.
\item The language of first-order logic \AgdaDatatype{FOL} over
\AgdaBound{t₂} (and some of its constructors) is also made visible.
\item \AgdaSymbol{(λ} \AgdaSymbol{\{}\AgdaInductiveConstructor{x} \AgdaSymbol{→} \AgdaBound{y}\AgdaSymbol{\})} denotes a substitution for the single
variable \AgdaInductiveConstructor{x}.
\item \AgdaRecord{≃} denotes \emph{type equivalence}.
\end{enumerate*}
\renewcommand{\labelenumi}{\theenumi.}

%% file: Numerals.tex
\AgdaHide{
\begin{code}%
\>\AgdaKeyword{module} \AgdaModule{Numerals} \AgdaKeyword{where}\<%
\end{code}
\noindent Rather that defining our own isomorphic copy, re-use
ℕ and Vec.

\begin{code}%
\>\AgdaKeyword{open} \AgdaKeyword{import} \AgdaModule{Data.Nat} \AgdaKeyword{using} \AgdaSymbol{(}\AgdaDatatype{ℕ}\AgdaSymbol{;} \AgdaInductiveConstructor{suc}\AgdaSymbol{)}\<%
\\
\>\AgdaKeyword{open} \AgdaKeyword{import} \AgdaModule{Data.Vec} \AgdaKeyword{using} \AgdaSymbol{(}\AgdaInductiveConstructor{\_∷\_}\AgdaSymbol{;} \AgdaInductiveConstructor{[]}\AgdaSymbol{;} \AgdaDatatype{Vec}\AgdaSymbol{)}\<%
\end{code}

plus is a \emph{function} on ℕ, which will (of course)
implement addition.  Note that this is not a 'good'
function, in the sense that it is extremely inefficient.
it does have the advantage of being simple and facilitate
proofs.  Note that it is defined (on purpose) by recursion
on the left argument, while the properties in T2
turn out to be "recursive" on the right.

\begin{code}%
\>\AgdaFunction{plus} \AgdaSymbol{:} \AgdaDatatype{ℕ} \AgdaSymbol{→} \AgdaDatatype{ℕ} \AgdaSymbol{→} \AgdaDatatype{ℕ}\<%
\\
\>\AgdaFunction{plus} \AgdaNumber{0} \AgdaBound{y} \AgdaSymbol{=} \AgdaBound{y}\<%
\\
\>\AgdaFunction{plus} \AgdaSymbol{(}\AgdaInductiveConstructor{suc} \AgdaBound{x}\AgdaSymbol{)} \AgdaBound{y} \AgdaSymbol{=} \AgdaInductiveConstructor{suc} \AgdaSymbol{(}\AgdaFunction{plus} \AgdaBound{x} \AgdaBound{y}\AgdaSymbol{)}\<%
\end{code}
}

We represent numerals as vectors (of length at least $1$) of
binary digits.

\begin{code}%
\>\AgdaKeyword{data} \AgdaDatatype{BinDigit} \AgdaSymbol{:} \AgdaPrimitiveType{Set} \AgdaKeyword{where} \AgdaInductiveConstructor{zero} \AgdaInductiveConstructor{one} \AgdaSymbol{:} \AgdaDatatype{BinDigit}\<%
\\
\>\AgdaKeyword{data} \AgdaDatatype{BNum} \AgdaSymbol{:} \AgdaPrimitiveType{Set} \AgdaKeyword{where}\<%
\\
\>[0]\AgdaIndent{2}{}\<[2]%
\>[2]\AgdaInductiveConstructor{bn} \AgdaSymbol{:} \AgdaSymbol{\{}\AgdaBound{n} \AgdaSymbol{:} \AgdaDatatype{ℕ}\AgdaSymbol{\}} \AgdaSymbol{→} \AgdaDatatype{Vec} \AgdaDatatype{BinDigit} \AgdaSymbol{(}\AgdaInductiveConstructor{suc} \AgdaBound{n}\AgdaSymbol{)} \AgdaSymbol{→} \AgdaDatatype{BNum}\<%
\end{code}

\noindent This then allows a straightforward implementation of
\AgdaFunction{bplus} to add numerals.  It is then
possible to \emph{prove} that the meaning function for
\AgdaFunction{bplus} is a theorem.

\AgdaHide{
\noindent Convenient abbreviations

\begin{code}%
\>\AgdaFunction{0b} \AgdaFunction{1b} \AgdaFunction{2b} \AgdaSymbol{:} \AgdaDatatype{BNum}\<%
\\
\>\AgdaFunction{0b} \AgdaSymbol{=} \AgdaInductiveConstructor{bn} \AgdaSymbol{(}\AgdaInductiveConstructor{zero} \AgdaInductiveConstructor{∷} \AgdaInductiveConstructor{[]}\AgdaSymbol{)}\<%
\\
\>\AgdaFunction{1b} \AgdaSymbol{=} \AgdaInductiveConstructor{bn} \AgdaSymbol{(}\AgdaInductiveConstructor{one} \AgdaInductiveConstructor{∷} \AgdaInductiveConstructor{[]}\AgdaSymbol{)}\<%
\\
\>\AgdaFunction{2b} \AgdaSymbol{=} \AgdaInductiveConstructor{bn} \AgdaSymbol{(}\AgdaInductiveConstructor{zero} \AgdaInductiveConstructor{∷} \AgdaInductiveConstructor{one} \AgdaInductiveConstructor{∷} \AgdaInductiveConstructor{[]}\AgdaSymbol{)}\<%
\\
\\
\>\AgdaFunction{<<} \AgdaSymbol{:} \AgdaDatatype{BNum} \AgdaSymbol{→} \AgdaDatatype{BNum}\<%
\\
\>\AgdaFunction{<<} \AgdaSymbol{(}\AgdaInductiveConstructor{bn} \AgdaBound{l}\AgdaSymbol{)} \AgdaSymbol{=} \AgdaInductiveConstructor{bn} \AgdaSymbol{(}\AgdaInductiveConstructor{zero} \AgdaInductiveConstructor{∷} \AgdaBound{l}\AgdaSymbol{)}\<%
\end{code}

\noindent Note how +1 is defined by induction on BNum.

\begin{code}%
\>\AgdaFunction{+1} \AgdaSymbol{:} \AgdaDatatype{BNum} \AgdaSymbol{→} \AgdaDatatype{BNum}\<%
\\
\>\AgdaFunction{+1} \AgdaSymbol{(}\AgdaInductiveConstructor{bn} \AgdaSymbol{(}\AgdaInductiveConstructor{zero} \AgdaInductiveConstructor{∷} \AgdaBound{l}\AgdaSymbol{))} \AgdaSymbol{=} \AgdaInductiveConstructor{bn} \AgdaSymbol{(}\AgdaInductiveConstructor{one} \AgdaInductiveConstructor{∷} \AgdaBound{l}\AgdaSymbol{)}\<%
\\
\>\AgdaFunction{+1} \AgdaSymbol{(}\AgdaInductiveConstructor{bn} \AgdaSymbol{(}\AgdaInductiveConstructor{one} \AgdaInductiveConstructor{∷} \AgdaInductiveConstructor{[]}\AgdaSymbol{))} \AgdaSymbol{=} \AgdaInductiveConstructor{bn} \AgdaSymbol{(}\AgdaInductiveConstructor{zero} \AgdaInductiveConstructor{∷} \AgdaInductiveConstructor{one} \AgdaInductiveConstructor{∷} \AgdaInductiveConstructor{[]}\AgdaSymbol{)}\<%
\\
\>\AgdaFunction{+1} \AgdaSymbol{(}\AgdaInductiveConstructor{bn} \AgdaSymbol{(}\AgdaInductiveConstructor{one} \AgdaInductiveConstructor{∷} \AgdaBound{x} \AgdaInductiveConstructor{∷} \AgdaBound{l}\AgdaSymbol{))} \AgdaSymbol{=} \AgdaFunction{<<} \AgdaSymbol{(}\AgdaFunction{+1} \AgdaSymbol{(}\AgdaInductiveConstructor{bn} \AgdaSymbol{(}\AgdaBound{x} \AgdaInductiveConstructor{∷} \AgdaBound{l}\AgdaSymbol{)))}\<%
\end{code}

Now we want to define a transformer on BNum
with a "meaning formula" that says that it is addition.
this too is defined by induction on BNum
bplus is essentially ξ₃.  It is not the only
such transformer, as different representations could
be even more efficient.
It is also kind of ξ₄ !  What happens it that all the
conditions are all vacuously true.  So the axioms are
then just a bunch of pattern-match.  Turns out that
the rules below are more 'syntax directed' than the
ones given in the axioms.

\begin{code}%
\>\AgdaFunction{bplus} \AgdaSymbol{:} \AgdaDatatype{BNum} \AgdaSymbol{→} \AgdaDatatype{BNum} \AgdaSymbol{→} \AgdaDatatype{BNum}\<%
\\
\>\AgdaFunction{bplus} \AgdaSymbol{(}\AgdaInductiveConstructor{bn} \AgdaSymbol{\{}\AgdaNumber{0}\AgdaSymbol{\}} \AgdaSymbol{(}\AgdaInductiveConstructor{zero} \AgdaInductiveConstructor{∷} \AgdaInductiveConstructor{[]}\AgdaSymbol{))} \AgdaBound{y} \AgdaSymbol{=} \AgdaBound{y}\<%
\\
\>\AgdaFunction{bplus} \AgdaSymbol{(}\AgdaInductiveConstructor{bn} \AgdaSymbol{\{}\AgdaNumber{0}\AgdaSymbol{\}} \AgdaSymbol{(}\AgdaInductiveConstructor{one} \AgdaInductiveConstructor{∷} \AgdaInductiveConstructor{[]}\AgdaSymbol{))} \AgdaBound{y} \AgdaSymbol{=} \AgdaFunction{+1} \AgdaBound{y}\<%
\\
\>\AgdaFunction{bplus} \AgdaSymbol{(}\AgdaInductiveConstructor{bn} \AgdaSymbol{\{}\AgdaInductiveConstructor{suc} \AgdaBound{n}\AgdaSymbol{\}} \AgdaSymbol{(}\AgdaBound{d₀} \AgdaInductiveConstructor{∷} \AgdaBound{l₀}\AgdaSymbol{))} \AgdaSymbol{(}\AgdaInductiveConstructor{bn} \AgdaSymbol{\{}\AgdaInductiveConstructor{ℕ.zero}\AgdaSymbol{\}} \AgdaSymbol{(}\AgdaInductiveConstructor{zero} \AgdaInductiveConstructor{∷} \AgdaInductiveConstructor{[]}\AgdaSymbol{))} \AgdaSymbol{=} \AgdaInductiveConstructor{bn} \AgdaSymbol{(}\AgdaBound{d₀} \AgdaInductiveConstructor{∷} \AgdaBound{l₀}\AgdaSymbol{)}\<%
\\
\>\AgdaFunction{bplus} \AgdaSymbol{(}\AgdaInductiveConstructor{bn} \AgdaSymbol{\{}\AgdaInductiveConstructor{suc} \AgdaBound{n}\AgdaSymbol{\}} \AgdaSymbol{(}\AgdaBound{d₀} \AgdaInductiveConstructor{∷} \AgdaBound{l₀}\AgdaSymbol{))} \AgdaSymbol{(}\AgdaInductiveConstructor{bn} \AgdaSymbol{\{}\AgdaInductiveConstructor{ℕ.zero}\AgdaSymbol{\}} \AgdaSymbol{(}\AgdaInductiveConstructor{one} \AgdaInductiveConstructor{∷} \AgdaInductiveConstructor{[]}\AgdaSymbol{))} \AgdaSymbol{=} \AgdaFunction{+1} \AgdaSymbol{(}\AgdaInductiveConstructor{bn} \AgdaSymbol{(}\AgdaBound{d₀} \AgdaInductiveConstructor{∷} \AgdaBound{l₀}\AgdaSymbol{))}\<%
\\
\>\AgdaFunction{bplus} \AgdaSymbol{(}\AgdaInductiveConstructor{bn} \AgdaSymbol{\{}\AgdaInductiveConstructor{suc} \AgdaBound{n}\AgdaSymbol{\}} \AgdaSymbol{(}\AgdaInductiveConstructor{zero} \AgdaInductiveConstructor{∷} \AgdaBound{l₀}\AgdaSymbol{))} \AgdaSymbol{(}\AgdaInductiveConstructor{bn} \AgdaSymbol{\{}\AgdaInductiveConstructor{suc} \AgdaBound{m}\AgdaSymbol{\}} \AgdaSymbol{(}\AgdaInductiveConstructor{zero} \AgdaInductiveConstructor{∷} \AgdaBound{l₁}\AgdaSymbol{))} \AgdaSymbol{=} \<[58]%
\>[58]\<%
\\
\>[2]\AgdaIndent{10}{}\<[10]%
\>[10]\AgdaFunction{<<} \AgdaSymbol{(}\AgdaFunction{bplus} \AgdaSymbol{(}\AgdaInductiveConstructor{bn} \AgdaBound{l₀}\AgdaSymbol{)} \AgdaSymbol{(}\AgdaInductiveConstructor{bn} \AgdaBound{l₁}\AgdaSymbol{))}\<%
\\
\>\AgdaFunction{bplus} \AgdaSymbol{(}\AgdaInductiveConstructor{bn} \AgdaSymbol{\{}\AgdaInductiveConstructor{suc} \AgdaBound{n}\AgdaSymbol{\}} \AgdaSymbol{(}\AgdaInductiveConstructor{one} \AgdaInductiveConstructor{∷} \AgdaBound{l₀}\AgdaSymbol{))} \AgdaSymbol{(}\AgdaInductiveConstructor{bn} \AgdaSymbol{\{}\AgdaInductiveConstructor{suc} \AgdaBound{m}\AgdaSymbol{\}} \AgdaSymbol{(}\AgdaInductiveConstructor{zero} \AgdaInductiveConstructor{∷} \AgdaBound{l₁}\AgdaSymbol{))} \AgdaSymbol{=}\<%
\\
\>[0]\AgdaIndent{6}{}\<[6]%
\>[6]\AgdaFunction{+1} \AgdaSymbol{(}\AgdaFunction{<<} \AgdaSymbol{(}\AgdaFunction{bplus} \AgdaSymbol{(}\AgdaInductiveConstructor{bn} \AgdaBound{l₀}\AgdaSymbol{)} \AgdaSymbol{(}\AgdaInductiveConstructor{bn} \AgdaBound{l₁}\AgdaSymbol{)))}\<%
\\
\>\AgdaFunction{bplus} \AgdaSymbol{(}\AgdaInductiveConstructor{bn} \AgdaSymbol{\{}\AgdaInductiveConstructor{suc} \AgdaBound{n}\AgdaSymbol{\}} \AgdaSymbol{(}\AgdaInductiveConstructor{zero} \AgdaInductiveConstructor{∷} \AgdaBound{l₀}\AgdaSymbol{))} \AgdaSymbol{(}\AgdaInductiveConstructor{bn} \AgdaSymbol{\{}\AgdaInductiveConstructor{suc} \AgdaBound{m}\AgdaSymbol{\}} \AgdaSymbol{(}\AgdaInductiveConstructor{one} \AgdaInductiveConstructor{∷} \AgdaBound{l₁}\AgdaSymbol{))} \AgdaSymbol{=}\<%
\\
\>[0]\AgdaIndent{6}{}\<[6]%
\>[6]\AgdaFunction{+1} \AgdaSymbol{(}\AgdaFunction{<<} \AgdaSymbol{(}\AgdaFunction{bplus} \AgdaSymbol{(}\AgdaInductiveConstructor{bn} \AgdaBound{l₀}\AgdaSymbol{)} \AgdaSymbol{(}\AgdaInductiveConstructor{bn} \AgdaBound{l₁}\AgdaSymbol{)))}\<%
\\
\>\AgdaFunction{bplus} \AgdaSymbol{(}\AgdaInductiveConstructor{bn} \AgdaSymbol{\{}\AgdaInductiveConstructor{suc} \AgdaBound{n}\AgdaSymbol{\}} \AgdaSymbol{(}\AgdaInductiveConstructor{one} \AgdaInductiveConstructor{∷} \AgdaBound{l₀}\AgdaSymbol{))} \AgdaSymbol{(}\AgdaInductiveConstructor{bn} \AgdaSymbol{\{}\AgdaInductiveConstructor{suc} \AgdaBound{m}\AgdaSymbol{\}} \AgdaSymbol{(}\AgdaInductiveConstructor{one} \AgdaInductiveConstructor{∷} \AgdaBound{l₁}\AgdaSymbol{))} \AgdaSymbol{=} \<[56]%
\>[56]\<%
\\
\>[0]\AgdaIndent{2}{}\<[2]%
\>[2]\AgdaFunction{+1} \AgdaSymbol{(}\AgdaFunction{+1} \AgdaSymbol{(}\AgdaFunction{<<} \AgdaSymbol{(}\AgdaFunction{bplus} \AgdaSymbol{(}\AgdaInductiveConstructor{bn} \AgdaBound{l₀}\AgdaSymbol{)} \AgdaSymbol{(}\AgdaInductiveConstructor{bn} \AgdaBound{l₁}\AgdaSymbol{))))}\<%
\end{code}

Important: because BNum is a type, there is no need for
is-bnum, as it is simply true by construction.  However,
here BNum is an explicit representation (as a Vector of digits)
whereas in {\churchuqe} it is done as a 'recognizer of expressions'
which picksout expressions which are made up of sequences of
digits.  The sequencing is buried (as a right-leaning tree) in
a bunch of 'bnat' calls.  If we're going to go through codes
to represent things, may as well use codes which are specially
built for the task!

Of course, we will need an interpretation of BNum in nat,
but that will be done inside T2.

For btimes, rather than be axiomatic, we go directly to a transformer.

\begin{code}%
\>\AgdaFunction{\_btimes\_} \AgdaSymbol{:} \AgdaDatatype{BNum} \AgdaSymbol{→} \AgdaDatatype{BNum} \AgdaSymbol{→} \AgdaDatatype{BNum}\<%
\\
\>\AgdaBound{x} \AgdaFunction{btimes} \AgdaInductiveConstructor{bn} \AgdaSymbol{(}\AgdaInductiveConstructor{zero} \AgdaInductiveConstructor{∷} \AgdaInductiveConstructor{[]}\AgdaSymbol{)} \AgdaSymbol{=} \AgdaFunction{0b}\<%
\\
\>\AgdaBound{x} \AgdaFunction{btimes} \AgdaInductiveConstructor{bn} \AgdaSymbol{(}\AgdaInductiveConstructor{one} \AgdaInductiveConstructor{∷} \AgdaInductiveConstructor{[]}\AgdaSymbol{)} \AgdaSymbol{=} \AgdaBound{x}\<%
\\
\>\AgdaInductiveConstructor{bn} \AgdaSymbol{(}\AgdaInductiveConstructor{zero} \AgdaInductiveConstructor{∷} \AgdaInductiveConstructor{[]}\AgdaSymbol{)} \AgdaFunction{btimes} \AgdaInductiveConstructor{bn} \AgdaSymbol{(}\AgdaBound{x₂} \AgdaInductiveConstructor{∷} \AgdaBound{x₃} \AgdaInductiveConstructor{∷} \AgdaBound{x₄}\AgdaSymbol{)} \AgdaSymbol{=} \AgdaFunction{0b}\<%
\\
\>\AgdaInductiveConstructor{bn} \AgdaSymbol{(}\AgdaInductiveConstructor{one} \AgdaInductiveConstructor{∷} \AgdaInductiveConstructor{[]}\AgdaSymbol{)} \AgdaFunction{btimes} \AgdaInductiveConstructor{bn} \AgdaSymbol{(}\AgdaBound{x₂} \AgdaInductiveConstructor{∷} \AgdaBound{x₃} \AgdaInductiveConstructor{∷} \AgdaBound{x₄}\AgdaSymbol{)} \AgdaSymbol{=} \AgdaInductiveConstructor{bn} \AgdaSymbol{(}\AgdaBound{x₂} \AgdaInductiveConstructor{∷} \AgdaBound{x₃} \AgdaInductiveConstructor{∷} \AgdaBound{x₄}\AgdaSymbol{)}\<%
\\
\>\AgdaInductiveConstructor{bn} \AgdaSymbol{(}\AgdaInductiveConstructor{zero} \AgdaInductiveConstructor{∷} \AgdaBound{x₁} \AgdaInductiveConstructor{∷} \AgdaBound{x₂}\AgdaSymbol{)} \AgdaFunction{btimes} \AgdaInductiveConstructor{bn} \AgdaSymbol{(}\AgdaBound{x₃} \AgdaInductiveConstructor{∷} \AgdaBound{x₄} \AgdaInductiveConstructor{∷} \AgdaBound{x₅}\AgdaSymbol{)} \AgdaSymbol{=}\<%
\\
\>[2]\AgdaIndent{9}{}\<[9]%
\>[9]\AgdaFunction{<<} \AgdaSymbol{(}\AgdaInductiveConstructor{bn} \AgdaSymbol{(}\AgdaBound{x₁} \AgdaInductiveConstructor{∷} \AgdaBound{x₂}\AgdaSymbol{)} \AgdaFunction{btimes} \AgdaInductiveConstructor{bn} \AgdaSymbol{(}\AgdaBound{x₃} \AgdaInductiveConstructor{∷} \AgdaBound{x₄} \AgdaInductiveConstructor{∷} \AgdaBound{x₅}\AgdaSymbol{))}\<%
\\
\>\AgdaInductiveConstructor{bn} \AgdaSymbol{(}\AgdaInductiveConstructor{one} \AgdaInductiveConstructor{∷} \AgdaBound{x₁} \AgdaInductiveConstructor{∷} \AgdaBound{x₂}\AgdaSymbol{)} \AgdaFunction{btimes} \AgdaInductiveConstructor{bn} \AgdaSymbol{(}\AgdaBound{x₃} \AgdaInductiveConstructor{∷} \AgdaBound{x₄} \AgdaInductiveConstructor{∷} \AgdaBound{x₅}\AgdaSymbol{)} \AgdaSymbol{=}\<%
\\
\>[0]\AgdaIndent{2}{}\<[2]%
\>[2]\AgdaKeyword{let} \AgdaBound{y} \AgdaSymbol{=} \AgdaInductiveConstructor{bn} \AgdaSymbol{(}\AgdaBound{x₃} \AgdaInductiveConstructor{∷} \AgdaBound{x₄} \AgdaInductiveConstructor{∷} \AgdaBound{x₅}\AgdaSymbol{)} \AgdaKeyword{in}\<%
\\
\>[0]\AgdaIndent{2}{}\<[2]%
\>[2]\AgdaFunction{bplus} \AgdaSymbol{(}\AgdaFunction{<<} \AgdaSymbol{(}\AgdaInductiveConstructor{bn} \AgdaSymbol{(}\AgdaBound{x₁} \AgdaInductiveConstructor{∷} \AgdaBound{x₂}\AgdaSymbol{)} \AgdaFunction{btimes} \AgdaBound{y}\AgdaSymbol{))} \AgdaBound{y}\<%
\end{code}
}

%% file: T2a.tex
\AgdaHide{
\begin{code}%
\>\AgdaKeyword{module} \AgdaModule{T2a} \AgdaKeyword{where}\<%
\\
\>\AgdaKeyword{open} \AgdaKeyword{import} \AgdaModule{T1} \AgdaKeyword{using} \AgdaSymbol{(}\AgdaRecord{BT₁}\AgdaSymbol{)}\<%
\\
\>\AgdaKeyword{open} \AgdaKeyword{import} \AgdaModule{T2} \AgdaKeyword{using} \AgdaSymbol{(}\AgdaRecord{BT₂}\AgdaSymbol{)}\<%
\\
\>\AgdaKeyword{open} \AgdaKeyword{import} \AgdaModule{Numerals}\<%
\\
\\
\>\AgdaKeyword{open} \AgdaKeyword{import} \AgdaModule{Relation.Binary.PropositionalEquality}\<%
\\
\>[0]\AgdaIndent{2}{}\<[2]%
\>[2]\AgdaKeyword{using} \AgdaSymbol{(}\AgdaDatatype{\_≡\_}\AgdaSymbol{;} \AgdaInductiveConstructor{refl}\AgdaSymbol{;} \AgdaFunction{trans}\AgdaSymbol{;} \AgdaFunction{cong}\AgdaSymbol{;} \AgdaFunction{sym}\AgdaSymbol{;} \AgdaFunction{cong₂}\AgdaSymbol{)}\<%
\\
\>\AgdaKeyword{open} \AgdaKeyword{import} \AgdaModule{Data.Nat} \AgdaKeyword{using} \AgdaSymbol{(}\AgdaDatatype{ℕ}\AgdaSymbol{;} \AgdaInductiveConstructor{suc}\AgdaSymbol{)} \AgdaComment{-- instead of defining our own}\<%
\\
\>[0]\AgdaIndent{2}{}\<[2]%
\>[2]\AgdaComment{-- isomorphic copy}\<%
\\
\>\AgdaKeyword{open} \AgdaKeyword{import} \AgdaModule{Data.Vec} \AgdaKeyword{using} \AgdaSymbol{(}\AgdaInductiveConstructor{\_∷\_}\AgdaSymbol{;} \AgdaInductiveConstructor{[]}\AgdaSymbol{;} \AgdaDatatype{Vec}\AgdaSymbol{)}\<%
\\
\\
\>\AgdaComment{-------------------------------------------------------------------}\<%
\\
\>\AgdaComment{-- an extension of BT₂ that assumes commutativity and associativity}\<%
\\
\>\AgdaKeyword{record} \AgdaRecord{BT₂ext} \AgdaSymbol{\{}\AgdaBound{t1} \AgdaSymbol{:} \AgdaRecord{BT₁}\AgdaSymbol{\}} \AgdaSymbol{(}\AgdaBound{t2} \AgdaSymbol{:} \AgdaRecord{BT₂} \AgdaBound{t1}\AgdaSymbol{)} \AgdaSymbol{:} \AgdaPrimitiveType{Set₀} \AgdaKeyword{where}\<%
\\
\>[0]\AgdaIndent{2}{}\<[2]%
\>[2]\AgdaKeyword{open} \AgdaModule{BT₂} \AgdaBound{t2} \AgdaKeyword{public}\<%
\\
\>[0]\AgdaIndent{2}{}\<[2]%
\>[2]\AgdaKeyword{field}\<%
\\
\>[2]\AgdaIndent{4}{}\<[4]%
\>[4]\AgdaComment{-- commutativity is needed for some proofs}\<%
\\
\>[2]\AgdaIndent{4}{}\<[4]%
\>[4]\AgdaComment{-- and is not provable; neither is associativity,}\<%
\\
\>[2]\AgdaIndent{4}{}\<[4]%
\>[4]\AgdaComment{-- in general.  So while this is more than Q,}\<%
\\
\>[2]\AgdaIndent{4}{}\<[4]%
\>[4]\AgdaComment{-- it is still 'ok' in the sense that these are}\<%
\\
\>[2]\AgdaIndent{4}{}\<[4]%
\>[4]\AgdaComment{-- equational axioms and not schemas.}\<%
\\
\>[2]\AgdaIndent{4}{}\<[4]%
\>[4]\AgdaField{comm-+} \AgdaSymbol{:} \AgdaSymbol{∀} \AgdaBound{x} \AgdaBound{y} \AgdaSymbol{→} \AgdaBound{x} \AgdaFunction{+} \AgdaBound{y} \AgdaDatatype{≡} \AgdaBound{y} \AgdaFunction{+} \AgdaBound{x}\<%
\\
\>[2]\AgdaIndent{4}{}\<[4]%
\>[4]\AgdaField{assoc-+} \AgdaSymbol{:} \AgdaSymbol{∀} \AgdaBound{x} \AgdaBound{y} \AgdaBound{z} \AgdaSymbol{→} \AgdaSymbol{(}\AgdaBound{x} \AgdaFunction{+} \AgdaBound{y}\AgdaSymbol{)} \AgdaFunction{+} \AgdaBound{z} \AgdaDatatype{≡} \AgdaBound{x} \AgdaFunction{+} \AgdaSymbol{(}\AgdaBound{y} \AgdaFunction{+} \AgdaBound{z}\AgdaSymbol{)}\<%
\\
\\
\>[0]\AgdaIndent{2}{}\<[2]%
\>[2]\AgdaComment{-- useful below. }\<%
\\
\>[0]\AgdaIndent{2}{}\<[2]%
\>[2]\AgdaFunction{left-0} \AgdaSymbol{:} \AgdaSymbol{∀} \AgdaBound{x} \AgdaSymbol{→} \AgdaFunction{Z} \AgdaFunction{+} \AgdaBound{x} \AgdaDatatype{≡} \AgdaBound{x}\<%
\\
\>[0]\AgdaIndent{2}{}\<[2]%
\>[2]\AgdaFunction{left-0} \AgdaBound{x} \AgdaSymbol{=} \AgdaFunction{trans} \AgdaSymbol{(}\AgdaField{comm-+} \AgdaFunction{Z} \AgdaBound{x}\AgdaSymbol{)} \AgdaSymbol{(}\AgdaFunction{right-0} \AgdaBound{x}\AgdaSymbol{)}\<%
\\
\\
\>[0]\AgdaIndent{2}{}\<[2]%
\>[2]\AgdaComment{-- to show that this definition is correct, we need a number}\<%
\\
\>[0]\AgdaIndent{2}{}\<[2]%
\>[2]\AgdaComment{-- of properties}\<%
\\
\>[0]\AgdaIndent{2}{}\<[2]%
\>[2]\AgdaFunction{shift-S} \AgdaSymbol{:} \AgdaSymbol{∀} \AgdaBound{x} \AgdaBound{y} \AgdaSymbol{→} \AgdaBound{x} \AgdaFunction{+} \AgdaFunction{S} \AgdaBound{y} \AgdaDatatype{≡} \AgdaFunction{S} \AgdaBound{x} \AgdaFunction{+} \AgdaBound{y}\<%
\\
\>[0]\AgdaIndent{2}{}\<[2]%
\>[2]\AgdaFunction{shift-S} \AgdaBound{x} \AgdaBound{y} \AgdaSymbol{=} \AgdaFunction{trans} \AgdaSymbol{(}\AgdaFunction{x+Sy≡Sx+y} \AgdaBound{x} \AgdaBound{y}\AgdaSymbol{)} \AgdaSymbol{(}\<%
\\
\>[2]\AgdaIndent{16}{}\<[16]%
\>[16]\AgdaFunction{trans} \AgdaSymbol{(}\AgdaFunction{cong} \AgdaFunction{S} \AgdaSymbol{(}\AgdaField{comm-+} \AgdaBound{x} \AgdaBound{y}\AgdaSymbol{))} \AgdaSymbol{(}\<%
\\
\>[2]\AgdaIndent{16}{}\<[16]%
\>[16]\AgdaFunction{trans} \AgdaSymbol{(}\AgdaFunction{sym} \AgdaSymbol{(}\AgdaFunction{x+Sy≡Sx+y} \AgdaBound{y} \AgdaBound{x}\AgdaSymbol{))} \AgdaSymbol{(}\<%
\\
\>[16]\AgdaIndent{22}{}\<[22]%
\>[22]\AgdaSymbol{(}\AgdaField{comm-+} \AgdaBound{y} \AgdaSymbol{(}\AgdaFunction{S} \AgdaBound{x}\AgdaSymbol{))))} \AgdaSymbol{)}\<%
\\
\\
\>[0]\AgdaIndent{2}{}\<[2]%
\>[2]\AgdaComment{-- two different ways of writing 2x+2}\<%
\\
\>[0]\AgdaIndent{2}{}\<[2]%
\>[2]\AgdaFunction{2x+2} \AgdaSymbol{:} \AgdaSymbol{∀} \AgdaBound{x} \AgdaSymbol{→} \AgdaFunction{S} \AgdaBound{x} \AgdaFunction{+} \AgdaFunction{S} \AgdaBound{x} \AgdaDatatype{≡} \AgdaFunction{S} \AgdaSymbol{((}\AgdaBound{x} \AgdaFunction{+} \AgdaBound{x}\AgdaSymbol{)} \AgdaFunction{+} \AgdaFunction{S} \AgdaFunction{Z}\AgdaSymbol{)}\<%
\\
\>[0]\AgdaIndent{2}{}\<[2]%
\>[2]\AgdaFunction{2x+2} \AgdaBound{x} \AgdaSymbol{=} \AgdaFunction{trans} \AgdaSymbol{(}\AgdaFunction{x+Sy≡Sx+y} \AgdaSymbol{(}\AgdaFunction{S} \AgdaBound{x}\AgdaSymbol{)} \AgdaBound{x}\AgdaSymbol{)}\<%
\\
\>[2]\AgdaIndent{17}{}\<[17]%
\>[17]\AgdaSymbol{(}\AgdaFunction{cong} \AgdaFunction{S} \AgdaSymbol{(}\AgdaFunction{trans} \AgdaSymbol{(}\AgdaField{comm-+} \AgdaSymbol{(}\AgdaFunction{S} \AgdaBound{x}\AgdaSymbol{)} \AgdaBound{x}\AgdaSymbol{)} \AgdaSymbol{(}\<%
\\
\>[17]\AgdaIndent{26}{}\<[26]%
\>[26]\AgdaFunction{trans} \AgdaSymbol{(}\AgdaFunction{x+Sy≡Sx+y} \AgdaBound{x} \AgdaBound{x}\AgdaSymbol{)} \AgdaSymbol{(}\<%
\\
\>[17]\AgdaIndent{26}{}\<[26]%
\>[26]\AgdaFunction{trans} \AgdaSymbol{(}\AgdaFunction{cong} \AgdaFunction{S} \AgdaSymbol{(}\AgdaFunction{sym} \AgdaSymbol{(}\AgdaFunction{right-0} \AgdaSymbol{(}\AgdaBound{x} \AgdaFunction{+} \AgdaBound{x}\AgdaSymbol{))))} \AgdaSymbol{(}\<%
\\
\>[26]\AgdaIndent{32}{}\<[32]%
\>[32]\AgdaSymbol{(}\AgdaFunction{sym} \AgdaSymbol{(}\AgdaFunction{x+Sy≡Sx+y} \AgdaSymbol{(}\AgdaBound{x} \AgdaFunction{+} \AgdaBound{x}\AgdaSymbol{)} \AgdaFunction{Z}\AgdaSymbol{)))))))}\<%
\\
\\
\>[0]\AgdaIndent{2}{}\<[2]%
\>[2]\AgdaFunction{shuffle} \AgdaSymbol{:} \AgdaSymbol{∀} \AgdaBound{x} \AgdaBound{y} \AgdaSymbol{→} \AgdaSymbol{(}\AgdaBound{x} \AgdaFunction{+} \AgdaBound{y}\AgdaSymbol{)} \AgdaFunction{+} \AgdaSymbol{(}\AgdaBound{x} \AgdaFunction{+} \AgdaBound{y}\AgdaSymbol{)} \AgdaDatatype{≡} \AgdaSymbol{(}\AgdaBound{x} \AgdaFunction{+} \AgdaBound{x}\AgdaSymbol{)} \AgdaFunction{+} \AgdaSymbol{(}\AgdaBound{y} \AgdaFunction{+} \AgdaBound{y}\AgdaSymbol{)}\<%
\\
\>[0]\AgdaIndent{2}{}\<[2]%
\>[2]\AgdaFunction{shuffle} \AgdaBound{x} \AgdaBound{y} \AgdaSymbol{=} \AgdaFunction{trans} \AgdaSymbol{(}\AgdaField{assoc-+} \AgdaBound{x} \AgdaBound{y} \AgdaSymbol{(}\AgdaBound{x} \AgdaFunction{+} \AgdaBound{y}\AgdaSymbol{))}\<%
\\
\>[2]\AgdaIndent{16}{}\<[16]%
\>[16]\AgdaSymbol{(}\AgdaFunction{trans} \AgdaSymbol{(}\AgdaFunction{cong} \AgdaSymbol{(λ} \AgdaBound{z} \AgdaSymbol{→} \AgdaBound{x} \AgdaFunction{+} \AgdaBound{z}\AgdaSymbol{)} \AgdaSymbol{(}\AgdaFunction{trans} \AgdaSymbol{(}\AgdaFunction{sym} \AgdaSymbol{(}\AgdaField{assoc-+} \AgdaBound{y} \AgdaBound{x} \AgdaBound{y}\AgdaSymbol{))}\<%
\\
\>[16]\AgdaIndent{43}{}\<[43]%
\>[43]\AgdaSymbol{(}\AgdaFunction{trans} \AgdaSymbol{(}\AgdaFunction{cong} \AgdaSymbol{(λ} \AgdaBound{z} \AgdaSymbol{→} \AgdaBound{z} \AgdaFunction{+} \AgdaBound{y}\AgdaSymbol{)} \AgdaSymbol{(}\AgdaField{comm-+} \AgdaBound{y} \AgdaBound{x}\AgdaSymbol{))}\<%
\\
\>[43]\AgdaIndent{50}{}\<[50]%
\>[50]\AgdaSymbol{(}\AgdaField{assoc-+} \AgdaBound{x} \AgdaBound{y} \AgdaBound{y}\AgdaSymbol{))))}\<%
\\
\>[0]\AgdaIndent{16}{}\<[16]%
\>[16]\AgdaSymbol{(}\AgdaFunction{sym} \AgdaSymbol{(}\AgdaField{assoc-+} \AgdaBound{x} \AgdaBound{x} \AgdaSymbol{\_)))}\<%
\\
\\
\>[0]\AgdaIndent{2}{}\<[2]%
\>[2]\AgdaFunction{add1-is-S} \AgdaSymbol{:} \AgdaSymbol{∀} \AgdaBound{x} \AgdaSymbol{→} \AgdaFunction{⟦} \AgdaFunction{1b} \AgdaFunction{⟧₂} \AgdaFunction{+} \AgdaBound{x} \AgdaDatatype{≡} \AgdaFunction{S} \AgdaBound{x}\<%
\\
\>[0]\AgdaIndent{2}{}\<[2]%
\>[2]\AgdaFunction{add1-is-S} \AgdaBound{x} \AgdaSymbol{=} \AgdaFunction{trans} \AgdaSymbol{(}\AgdaFunction{cong} \AgdaSymbol{(λ} \AgdaBound{z} \AgdaSymbol{→} \AgdaBound{z} \AgdaFunction{+} \AgdaBound{x}\AgdaSymbol{)} \AgdaSymbol{(}\AgdaFunction{lemma₂}\AgdaSymbol{))} \AgdaSymbol{(}\<%
\\
\>[2]\AgdaIndent{16}{}\<[16]%
\>[16]\AgdaFunction{trans} \AgdaSymbol{(}\AgdaField{comm-+} \AgdaSymbol{(}\AgdaFunction{S} \AgdaFunction{Z}\AgdaSymbol{)} \AgdaBound{x}\AgdaSymbol{)} \AgdaSymbol{(}\<%
\\
\>[2]\AgdaIndent{16}{}\<[16]%
\>[16]\AgdaFunction{trans} \AgdaSymbol{(}\AgdaFunction{x+Sy≡Sx+y} \AgdaBound{x} \AgdaFunction{Z}\AgdaSymbol{)}\<%
\\
\>[16]\AgdaIndent{22}{}\<[22]%
\>[22]\AgdaSymbol{(}\AgdaFunction{cong} \AgdaFunction{S} \AgdaSymbol{(}\AgdaFunction{right-0} \AgdaBound{x}\AgdaSymbol{))))}\<%
\\
\\
\>[0]\AgdaIndent{2}{}\<[2]%
\>[2]\AgdaFunction{+1-is-S} \AgdaSymbol{:} \AgdaSymbol{∀} \AgdaBound{x} \AgdaSymbol{→} \AgdaFunction{⟦} \AgdaFunction{+1} \AgdaBound{x} \AgdaFunction{⟧₂} \AgdaDatatype{≡} \AgdaFunction{S} \AgdaFunction{⟦} \AgdaBound{x} \AgdaFunction{⟧₂}\<%
\\
\>[0]\AgdaIndent{2}{}\<[2]%
\>[2]\AgdaFunction{+1-is-S} \AgdaSymbol{(}\AgdaInductiveConstructor{bn} \AgdaSymbol{(}\AgdaInductiveConstructor{zero} \AgdaInductiveConstructor{∷} \AgdaBound{l}\AgdaSymbol{))} \AgdaSymbol{=} \AgdaFunction{x+Sy≡Sx+y} \AgdaSymbol{(}\AgdaFunction{unroll} \AgdaBound{l} \AgdaFunction{+} \AgdaFunction{unroll} \AgdaBound{l}\AgdaSymbol{)} \AgdaFunction{Z}\<%
\\
\>[0]\AgdaIndent{2}{}\<[2]%
\>[2]\AgdaFunction{+1-is-S} \AgdaSymbol{(}\AgdaInductiveConstructor{bn} \AgdaSymbol{(}\AgdaInductiveConstructor{one} \AgdaInductiveConstructor{∷} \AgdaInductiveConstructor{[]}\AgdaSymbol{))} \AgdaSymbol{=}\<%
\\
\>[2]\AgdaIndent{4}{}\<[4]%
\>[4]\AgdaFunction{trans} \AgdaSymbol{(}\AgdaFunction{right-0} \AgdaSymbol{\_)} \AgdaSymbol{(}\<%
\\
\>[2]\AgdaIndent{4}{}\<[4]%
\>[4]\AgdaFunction{trans} \AgdaSymbol{(}\AgdaFunction{cong} \AgdaSymbol{(λ} \AgdaBound{x} \AgdaSymbol{→} \AgdaFunction{⟦} \AgdaFunction{1b} \AgdaFunction{⟧₂} \AgdaFunction{+} \AgdaBound{x}\AgdaSymbol{)} \AgdaFunction{lemma₂}\AgdaSymbol{)} \AgdaSymbol{(}\<%
\\
\>[2]\AgdaIndent{4}{}\<[4]%
\>[4]\AgdaFunction{trans} \AgdaSymbol{(}\AgdaFunction{x+Sy≡Sx+y} \AgdaSymbol{\_} \AgdaFunction{Z}\AgdaSymbol{)}\<%
\\
\>[4]\AgdaIndent{10}{}\<[10]%
\>[10]\AgdaSymbol{(}\AgdaFunction{cong} \AgdaFunction{S} \AgdaSymbol{(}\AgdaFunction{right-0} \AgdaSymbol{\_))))}\<%
\\
\>[0]\AgdaIndent{2}{}\<[2]%
\>[2]\AgdaFunction{+1-is-S} \AgdaSymbol{(}\AgdaInductiveConstructor{bn} \AgdaSymbol{(}\AgdaInductiveConstructor{one} \AgdaInductiveConstructor{∷} \AgdaBound{x} \AgdaInductiveConstructor{∷} \AgdaBound{x₁}\AgdaSymbol{))} \AgdaSymbol{=}\<%
\\
\>[2]\AgdaIndent{4}{}\<[4]%
\>[4]\AgdaFunction{trans} \AgdaSymbol{(}\AgdaFunction{<<-is-*2} \AgdaSymbol{(}\AgdaFunction{+1} \AgdaSymbol{(}\AgdaInductiveConstructor{bn} \AgdaSymbol{(}\AgdaBound{x} \AgdaInductiveConstructor{∷} \AgdaBound{x₁}\AgdaSymbol{))))} \AgdaSymbol{(}\<%
\\
\>[2]\AgdaIndent{4}{}\<[4]%
\>[4]\AgdaFunction{trans} \AgdaSymbol{(}\AgdaFunction{cong₂} \AgdaFunction{\_+\_} \AgdaSymbol{(}\AgdaFunction{+1-is-S} \AgdaSymbol{(}\AgdaInductiveConstructor{bn} \AgdaSymbol{(}\AgdaBound{x} \AgdaInductiveConstructor{∷} \AgdaBound{x₁}\AgdaSymbol{)))} \AgdaSymbol{(}\AgdaFunction{+1-is-S} \AgdaSymbol{(}\AgdaInductiveConstructor{bn} \AgdaSymbol{(}\AgdaBound{x} \AgdaInductiveConstructor{∷} \AgdaBound{x₁}\AgdaSymbol{))))}\<%
\\
\>[4]\AgdaIndent{10}{}\<[10]%
\>[10]\AgdaSymbol{(}\AgdaFunction{2x+2} \AgdaSymbol{\_))}\<%
\\
\\
\>[0]\AgdaIndent{2}{}\<[2]%
\>[2]\AgdaComment{-- because of the invariants that we keep in the types, and the}\<%
\\
\>[0]\AgdaIndent{2}{}\<[2]%
\>[2]\AgdaComment{-- way that bplus is defined \_as a function\_, we can actually}\<%
\\
\>[0]\AgdaIndent{2}{}\<[2]%
\>[2]\AgdaComment{-- prove the meaning function 'internally'.  Only needs 8 cases}\<%
\\
\>[0]\AgdaIndent{2}{}\<[2]%
\>[2]\AgdaComment{-- whereas the paper proof needs 10 (?).}\<%
\\
\>[0]\AgdaIndent{2}{}\<[2]%
\>[2]\AgdaComment{-- Plus I have no idea if the paper spec is complete.}\<%
\end{code}
}
\begin{code}%
\>[0]\AgdaIndent{2}{}\<[2]%
\>[2]\AgdaFunction{bplus-is-+} \AgdaSymbol{:} \AgdaSymbol{∀} \AgdaBound{x} \AgdaBound{y} \AgdaSymbol{→} \AgdaFunction{⟦} \AgdaFunction{bplus} \AgdaBound{x} \AgdaBound{y} \AgdaFunction{⟧₂} \AgdaDatatype{≡} \AgdaFunction{⟦} \AgdaBound{x} \AgdaFunction{⟧₂} \AgdaFunction{+} \AgdaFunction{⟦} \AgdaBound{y} \AgdaFunction{⟧₂}\<%
\end{code}
\AgdaHide{
\begin{code}%
\>[0]\AgdaIndent{2}{}\<[2]%
\>[2]\AgdaFunction{bplus-is-+} \AgdaSymbol{(}\AgdaInductiveConstructor{bn} \AgdaSymbol{\{}\AgdaNumber{0}\AgdaSymbol{\}} \AgdaSymbol{(}\AgdaInductiveConstructor{zero} \AgdaInductiveConstructor{∷} \AgdaInductiveConstructor{[]}\AgdaSymbol{))} \AgdaBound{y} \AgdaSymbol{=} \AgdaFunction{trans} \AgdaSymbol{(}\AgdaFunction{sym} \AgdaSymbol{(}\AgdaFunction{left-0} \AgdaFunction{⟦} \AgdaBound{y} \AgdaFunction{⟧₂}\AgdaSymbol{))} \AgdaSymbol{(}\AgdaFunction{cong} \AgdaSymbol{(λ} \AgdaBound{z} \AgdaSymbol{→} \AgdaBound{z} \AgdaFunction{+} \AgdaFunction{⟦} \AgdaBound{y} \AgdaFunction{⟧₂}\AgdaSymbol{)} \AgdaSymbol{(}\AgdaFunction{sym} \AgdaFunction{lemma₁}\AgdaSymbol{))}\<%
\\
\>[0]\AgdaIndent{2}{}\<[2]%
\>[2]\AgdaFunction{bplus-is-+} \AgdaSymbol{(}\AgdaInductiveConstructor{bn} \AgdaSymbol{\{}\AgdaNumber{0}\AgdaSymbol{\}} \AgdaSymbol{(}\AgdaInductiveConstructor{one} \AgdaInductiveConstructor{∷} \AgdaInductiveConstructor{[]}\AgdaSymbol{))} \AgdaBound{y} \AgdaSymbol{=} \AgdaFunction{trans} \AgdaSymbol{(}\AgdaFunction{+1-is-S} \AgdaBound{y}\AgdaSymbol{)} \AgdaSymbol{(}\AgdaFunction{sym} \AgdaSymbol{(}\AgdaFunction{add1-is-S} \AgdaFunction{⟦} \AgdaBound{y} \AgdaFunction{⟧₂}\AgdaSymbol{))}\<%
\\
\>[0]\AgdaIndent{2}{}\<[2]%
\>[2]\AgdaFunction{bplus-is-+} \AgdaSymbol{(}\AgdaInductiveConstructor{bn} \AgdaSymbol{\{}\AgdaInductiveConstructor{suc} \AgdaBound{n}\AgdaSymbol{\}} \AgdaSymbol{(}\AgdaBound{d₀} \AgdaInductiveConstructor{∷} \AgdaBound{l₀}\AgdaSymbol{))} \AgdaSymbol{(}\AgdaInductiveConstructor{bn} \AgdaSymbol{\{}\AgdaInductiveConstructor{ℕ.zero}\AgdaSymbol{\}} \AgdaSymbol{(}\AgdaInductiveConstructor{zero} \AgdaInductiveConstructor{∷} \AgdaInductiveConstructor{[]}\AgdaSymbol{))} \AgdaSymbol{=}\<%
\\
\>[2]\AgdaIndent{4}{}\<[4]%
\>[4]\AgdaFunction{trans} \AgdaSymbol{(}\AgdaFunction{sym} \AgdaSymbol{(}\AgdaFunction{right-0} \AgdaSymbol{\_))} \AgdaSymbol{(}\AgdaFunction{cong} \AgdaSymbol{(λ} \AgdaBound{x} \AgdaSymbol{→} \AgdaFunction{⟦} \AgdaInductiveConstructor{bn} \AgdaSymbol{(}\AgdaBound{d₀} \AgdaInductiveConstructor{∷} \AgdaBound{l₀}\AgdaSymbol{)} \AgdaFunction{⟧₂} \AgdaFunction{+} \AgdaBound{x}\AgdaSymbol{)} \AgdaSymbol{(}\AgdaFunction{sym} \AgdaFunction{lemma₁}\AgdaSymbol{))}\<%
\\
\>[0]\AgdaIndent{2}{}\<[2]%
\>[2]\AgdaFunction{bplus-is-+} \AgdaSymbol{(}\AgdaInductiveConstructor{bn} \AgdaSymbol{\{}\AgdaInductiveConstructor{suc} \AgdaBound{n}\AgdaSymbol{\}} \AgdaSymbol{(}\AgdaBound{d₀} \AgdaInductiveConstructor{∷} \AgdaBound{l₀}\AgdaSymbol{))} \AgdaSymbol{(}\AgdaInductiveConstructor{bn} \AgdaSymbol{\{}\AgdaInductiveConstructor{ℕ.zero}\AgdaSymbol{\}} \AgdaSymbol{(}\AgdaInductiveConstructor{one} \AgdaInductiveConstructor{∷} \AgdaInductiveConstructor{[]}\AgdaSymbol{))} \AgdaSymbol{=} \<[63]%
\>[63]\<%
\\
\>[2]\AgdaIndent{4}{}\<[4]%
\>[4]\AgdaKeyword{let} \AgdaBound{num} \AgdaSymbol{=} \AgdaInductiveConstructor{bn} \AgdaSymbol{(}\AgdaBound{d₀} \AgdaInductiveConstructor{∷} \AgdaBound{l₀}\AgdaSymbol{)} \AgdaKeyword{in}\<%
\\
\>[2]\AgdaIndent{4}{}\<[4]%
\>[4]\AgdaFunction{trans} \AgdaSymbol{(}\AgdaFunction{+1-is-S} \AgdaBound{num}\AgdaSymbol{)} \AgdaSymbol{(}\AgdaFunction{x+1} \AgdaFunction{⟦} \AgdaBound{num} \AgdaFunction{⟧₂}\AgdaSymbol{)} \<[39]%
\>[39]\<%
\\
\>[0]\AgdaIndent{2}{}\<[2]%
\>[2]\AgdaFunction{bplus-is-+} \AgdaSymbol{(}\AgdaInductiveConstructor{bn} \AgdaSymbol{\{}\AgdaInductiveConstructor{suc} \AgdaBound{n}\AgdaSymbol{\}} \AgdaSymbol{(}\AgdaInductiveConstructor{zero} \AgdaInductiveConstructor{∷} \AgdaBound{l₀}\AgdaSymbol{))} \AgdaSymbol{(}\AgdaInductiveConstructor{bn} \AgdaSymbol{\{}\AgdaInductiveConstructor{suc} \AgdaBound{n₁}\AgdaSymbol{\}} \AgdaSymbol{(}\AgdaInductiveConstructor{zero} \AgdaInductiveConstructor{∷} \AgdaBound{l₁}\AgdaSymbol{))} \AgdaSymbol{=} \<[66]%
\>[66]\<%
\\
\>[2]\AgdaIndent{4}{}\<[4]%
\>[4]\AgdaKeyword{let} \AgdaBound{n₁} \AgdaSymbol{=} \AgdaInductiveConstructor{bn} \AgdaBound{l₀}\<%
\\
\>[4]\AgdaIndent{8}{}\<[8]%
\>[8]\AgdaBound{n₂} \AgdaSymbol{=} \AgdaInductiveConstructor{bn} \AgdaBound{l₁}\<%
\\
\>[4]\AgdaIndent{8}{}\<[8]%
\>[8]\AgdaBound{num} \AgdaSymbol{=} \AgdaFunction{bplus} \AgdaBound{n₁} \AgdaBound{n₂}\<%
\\
\>[4]\AgdaIndent{8}{}\<[8]%
\>[8]\AgdaBound{v₁} \AgdaSymbol{=} \AgdaFunction{⟦} \AgdaBound{n₁} \AgdaFunction{⟧₂}\<%
\\
\>[4]\AgdaIndent{8}{}\<[8]%
\>[8]\AgdaBound{v₂} \AgdaSymbol{=} \AgdaFunction{⟦} \AgdaBound{n₂} \AgdaFunction{⟧₂} \AgdaKeyword{in}\<%
\\
\>[0]\AgdaIndent{4}{}\<[4]%
\>[4]\AgdaFunction{trans} \AgdaSymbol{(}\AgdaFunction{<<-is-*2} \AgdaBound{num}\AgdaSymbol{)} \AgdaSymbol{(}\AgdaFunction{trans} \AgdaSymbol{(}\AgdaFunction{cong₂} \AgdaFunction{\_+\_} \AgdaSymbol{(}\AgdaFunction{bplus-is-+} \AgdaBound{n₁} \AgdaBound{n₂}\AgdaSymbol{)} \AgdaSymbol{(}\AgdaFunction{bplus-is-+} \AgdaBound{n₁} \AgdaBound{n₂}\AgdaSymbol{))}\<%
\\
\>[0]\AgdaIndent{4}{}\<[4]%
\>[4]\AgdaSymbol{(}\AgdaFunction{trans} \AgdaSymbol{(}\AgdaFunction{cong} \AgdaSymbol{(λ} \AgdaBound{z} \AgdaSymbol{→} \AgdaBound{z} \AgdaFunction{+} \AgdaSymbol{(}\AgdaBound{v₁} \AgdaFunction{+} \AgdaBound{v₂}\AgdaSymbol{))} \AgdaSymbol{(}\AgdaField{comm-+} \AgdaBound{v₁} \AgdaBound{v₂}\AgdaSymbol{))}\<%
\\
\>[0]\AgdaIndent{4}{}\<[4]%
\>[4]\AgdaSymbol{(}\AgdaFunction{trans} \AgdaSymbol{(}\AgdaField{assoc-+} \AgdaBound{v₂} \AgdaBound{v₁} \AgdaSymbol{\_)} \AgdaSymbol{(}\<%
\\
\>[4]\AgdaIndent{5}{}\<[5]%
\>[5]\AgdaFunction{trans} \AgdaSymbol{(}\AgdaFunction{cong} \AgdaSymbol{(λ} \AgdaBound{z} \AgdaSymbol{→} \AgdaBound{v₂} \AgdaFunction{+} \AgdaBound{z}\AgdaSymbol{)} \AgdaSymbol{(}\AgdaFunction{trans} \AgdaSymbol{(}\AgdaFunction{sym} \AgdaSymbol{(}\AgdaField{assoc-+} \AgdaBound{v₁} \AgdaBound{v₁} \AgdaBound{v₂}\AgdaSymbol{))} \AgdaSymbol{(}\AgdaFunction{cong} \AgdaSymbol{(λ} \AgdaBound{z} \AgdaSymbol{→} \AgdaBound{z} \AgdaFunction{+} \AgdaBound{v₂}\AgdaSymbol{)} \AgdaSymbol{(}\AgdaFunction{sym} \AgdaSymbol{(}\AgdaFunction{<<-is-*2} \AgdaBound{n₁}\AgdaSymbol{)))))}\<%
\\
\>[0]\AgdaIndent{4}{}\<[4]%
\>[4]\AgdaSymbol{(}\AgdaFunction{trans} \AgdaSymbol{(}\AgdaField{comm-+} \AgdaBound{v₂} \AgdaSymbol{\_)} \AgdaSymbol{(}\<%
\\
\>[0]\AgdaIndent{4}{}\<[4]%
\>[4]\AgdaFunction{trans} \AgdaSymbol{(}\AgdaField{assoc-+} \AgdaSymbol{\_} \AgdaBound{v₂} \AgdaBound{v₂}\AgdaSymbol{)}\<%
\\
\>[4]\AgdaIndent{10}{}\<[10]%
\>[10]\AgdaSymbol{(}\AgdaFunction{cong} \AgdaSymbol{(λ} \AgdaBound{z} \AgdaSymbol{→} \AgdaFunction{⟦} \AgdaFunction{<<} \AgdaBound{n₁} \AgdaFunction{⟧₂} \AgdaFunction{+} \AgdaBound{z}\AgdaSymbol{)} \AgdaSymbol{(}\AgdaFunction{sym} \AgdaSymbol{(}\AgdaFunction{<<-is-*2} \AgdaBound{n₂}\AgdaSymbol{)))))))))}\<%
\\
\>[0]\AgdaIndent{2}{}\<[2]%
\>[2]\AgdaFunction{bplus-is-+} \AgdaSymbol{(}\AgdaInductiveConstructor{bn} \AgdaSymbol{\{}\AgdaInductiveConstructor{suc} \AgdaBound{n}\AgdaSymbol{\}} \AgdaSymbol{(}\AgdaInductiveConstructor{zero} \AgdaInductiveConstructor{∷} \AgdaBound{l₀}\AgdaSymbol{))} \AgdaSymbol{(}\AgdaInductiveConstructor{bn} \AgdaSymbol{\{}\AgdaInductiveConstructor{suc} \AgdaBound{n₁}\AgdaSymbol{\}} \AgdaSymbol{(}\AgdaInductiveConstructor{one} \AgdaInductiveConstructor{∷} \AgdaBound{l₁}\AgdaSymbol{))} \AgdaSymbol{=}\<%
\\
\>[2]\AgdaIndent{4}{}\<[4]%
\>[4]\AgdaKeyword{let} \AgdaBound{n₁} \AgdaSymbol{=} \AgdaInductiveConstructor{bn} \AgdaBound{l₀}\<%
\\
\>[4]\AgdaIndent{8}{}\<[8]%
\>[8]\AgdaBound{n₂} \AgdaSymbol{=} \AgdaInductiveConstructor{bn} \AgdaBound{l₁}\<%
\\
\>[4]\AgdaIndent{8}{}\<[8]%
\>[8]\AgdaBound{num} \AgdaSymbol{=} \AgdaFunction{bplus} \AgdaBound{n₁} \AgdaBound{n₂}\<%
\\
\>[4]\AgdaIndent{8}{}\<[8]%
\>[8]\AgdaBound{v₁} \AgdaSymbol{=} \AgdaFunction{⟦} \AgdaBound{n₁} \AgdaFunction{⟧₂}\<%
\\
\>[4]\AgdaIndent{8}{}\<[8]%
\>[8]\AgdaBound{v₂} \AgdaSymbol{=} \AgdaFunction{⟦} \AgdaBound{n₂} \AgdaFunction{⟧₂} \AgdaKeyword{in}\<%
\\
\>[0]\AgdaIndent{4}{}\<[4]%
\>[4]\AgdaFunction{trans} \AgdaSymbol{(}\AgdaFunction{+1-is-S} \AgdaSymbol{(}\AgdaFunction{<<} \AgdaBound{num}\AgdaSymbol{))} \AgdaSymbol{(}\<%
\\
\>[0]\AgdaIndent{4}{}\<[4]%
\>[4]\AgdaFunction{trans} \AgdaSymbol{(}\AgdaFunction{cong} \AgdaFunction{S} \AgdaSymbol{(}\AgdaFunction{trans} \AgdaSymbol{(}\AgdaFunction{<<-is-*2} \AgdaBound{num}\AgdaSymbol{)}\<%
\\
\>[4]\AgdaIndent{18}{}\<[18]%
\>[18]\AgdaSymbol{(}\AgdaFunction{trans} \AgdaSymbol{(}\AgdaFunction{cong₂} \AgdaFunction{\_+\_} \AgdaSymbol{(}\AgdaFunction{bplus-is-+} \AgdaBound{n₁} \AgdaBound{n₂}\AgdaSymbol{)} \AgdaSymbol{(}\AgdaFunction{bplus-is-+} \AgdaBound{n₁} \AgdaBound{n₂}\AgdaSymbol{))}\<%
\\
\>[18]\AgdaIndent{25}{}\<[25]%
\>[25]\AgdaSymbol{(}\AgdaFunction{shuffle} \AgdaBound{v₁} \AgdaBound{v₂}\AgdaSymbol{))))}\<%
\\
\>[0]\AgdaIndent{4}{}\<[4]%
\>[4]\AgdaSymbol{(}\AgdaFunction{trans} \AgdaSymbol{(}\AgdaFunction{sym} \AgdaSymbol{(}\AgdaFunction{x+Sy≡Sx+y} \AgdaSymbol{(}\AgdaBound{v₁} \AgdaFunction{+} \AgdaBound{v₁}\AgdaSymbol{)} \AgdaSymbol{(}\AgdaBound{v₂} \AgdaFunction{+} \AgdaBound{v₂}\AgdaSymbol{)))}\<%
\\
\>[4]\AgdaIndent{10}{}\<[10]%
\>[10]\AgdaSymbol{(}\AgdaFunction{cong₂} \AgdaFunction{\_+\_} \AgdaSymbol{(}\AgdaFunction{sym} \AgdaSymbol{(}\AgdaFunction{<<-is-*2} \AgdaBound{n₁}\AgdaSymbol{))}\<%
\\
\>[10]\AgdaIndent{21}{}\<[21]%
\>[21]\AgdaSymbol{(}\AgdaFunction{trans} \AgdaSymbol{(}\AgdaFunction{x+1} \AgdaSymbol{\_)}\<%
\\
\>[21]\AgdaIndent{28}{}\<[28]%
\>[28]\AgdaSymbol{(}\AgdaFunction{cong₂} \AgdaFunction{\_+\_} \AgdaSymbol{(}\AgdaFunction{trans} \AgdaSymbol{(}\AgdaFunction{sym} \AgdaSymbol{(}\AgdaFunction{<<-is-*2} \AgdaBound{n₂}\AgdaSymbol{))} \AgdaSymbol{(}\AgdaFunction{right-0} \AgdaSymbol{\_))} \AgdaFunction{lemma₂}\AgdaSymbol{)))))}\<%
\\
\>[0]\AgdaIndent{2}{}\<[2]%
\>[2]\AgdaFunction{bplus-is-+} \AgdaSymbol{(}\AgdaInductiveConstructor{bn} \AgdaSymbol{\{}\AgdaInductiveConstructor{suc} \AgdaBound{n}\AgdaSymbol{\}} \AgdaSymbol{(}\AgdaInductiveConstructor{one} \AgdaInductiveConstructor{∷} \AgdaBound{l₀}\AgdaSymbol{))} \AgdaSymbol{(}\AgdaInductiveConstructor{bn} \AgdaSymbol{\{}\AgdaInductiveConstructor{suc} \AgdaBound{n₁}\AgdaSymbol{\}} \AgdaSymbol{(}\AgdaInductiveConstructor{zero} \AgdaInductiveConstructor{∷} \AgdaBound{l₁}\AgdaSymbol{))} \AgdaSymbol{=}\<%
\\
\>[2]\AgdaIndent{4}{}\<[4]%
\>[4]\AgdaKeyword{let} \AgdaBound{n₁} \AgdaSymbol{=} \AgdaInductiveConstructor{bn} \AgdaBound{l₀}\<%
\\
\>[4]\AgdaIndent{8}{}\<[8]%
\>[8]\AgdaBound{n₂} \AgdaSymbol{=} \AgdaInductiveConstructor{bn} \AgdaBound{l₁}\<%
\\
\>[4]\AgdaIndent{8}{}\<[8]%
\>[8]\AgdaBound{num} \AgdaSymbol{=} \AgdaFunction{bplus} \AgdaBound{n₁} \AgdaBound{n₂}\<%
\\
\>[4]\AgdaIndent{8}{}\<[8]%
\>[8]\AgdaBound{v₁} \AgdaSymbol{=} \AgdaFunction{⟦} \AgdaBound{n₁} \AgdaFunction{⟧₂}\<%
\\
\>[4]\AgdaIndent{8}{}\<[8]%
\>[8]\AgdaBound{v₂} \AgdaSymbol{=} \AgdaFunction{⟦} \AgdaBound{n₂} \AgdaFunction{⟧₂} \AgdaKeyword{in}\<%
\\
\>[0]\AgdaIndent{5}{}\<[5]%
\>[5]\AgdaFunction{trans} \AgdaSymbol{(}\AgdaFunction{+1-is-S} \AgdaSymbol{(}\AgdaFunction{<<} \AgdaBound{num}\AgdaSymbol{))} \AgdaSymbol{(}\<%
\\
\>[0]\AgdaIndent{5}{}\<[5]%
\>[5]\AgdaFunction{trans} \AgdaSymbol{(}\AgdaFunction{cong} \AgdaFunction{S} \AgdaSymbol{(}\AgdaFunction{trans} \AgdaSymbol{(}\AgdaFunction{<<-is-*2} \AgdaBound{num}\AgdaSymbol{)}\<%
\\
\>[5]\AgdaIndent{19}{}\<[19]%
\>[19]\AgdaSymbol{(}\AgdaFunction{trans} \AgdaSymbol{(}\AgdaFunction{cong₂} \AgdaFunction{\_+\_} \AgdaSymbol{(}\AgdaFunction{bplus-is-+} \AgdaBound{n₁} \AgdaBound{n₂}\AgdaSymbol{)} \AgdaSymbol{(}\AgdaFunction{bplus-is-+} \AgdaBound{n₁} \AgdaBound{n₂}\AgdaSymbol{))}\<%
\\
\>[5]\AgdaIndent{19}{}\<[19]%
\>[19]\AgdaSymbol{(}\AgdaFunction{trans} \AgdaSymbol{(}\AgdaFunction{shuffle} \AgdaBound{v₁} \AgdaBound{v₂}\AgdaSymbol{)}\<%
\\
\>[19]\AgdaIndent{26}{}\<[26]%
\>[26]\AgdaSymbol{(}\AgdaField{comm-+} \AgdaSymbol{\_} \AgdaSymbol{\_)))))}\<%
\\
\>[0]\AgdaIndent{5}{}\<[5]%
\>[5]\AgdaSymbol{(}\AgdaFunction{trans} \AgdaSymbol{(}\AgdaFunction{sym} \AgdaSymbol{(}\AgdaFunction{x+Sy≡Sx+y} \AgdaSymbol{\_} \AgdaSymbol{\_))}\<%
\\
\>[0]\AgdaIndent{5}{}\<[5]%
\>[5]\AgdaSymbol{(}\AgdaFunction{trans} \AgdaSymbol{(}\AgdaField{comm-+} \AgdaSymbol{\_} \AgdaSymbol{\_)} \AgdaSymbol{(}\AgdaFunction{cong₂} \AgdaFunction{\_+\_} \AgdaSymbol{(}\AgdaFunction{trans} \AgdaSymbol{(}\AgdaFunction{x+1} \AgdaSymbol{\_)} \AgdaSymbol{(}\AgdaFunction{trans} \AgdaSymbol{(}\AgdaFunction{cong₂} \AgdaFunction{\_+\_} \AgdaSymbol{(}\AgdaFunction{sym} \AgdaSymbol{(}\AgdaFunction{<<-is-*2} \AgdaBound{n₁}\AgdaSymbol{))} \AgdaFunction{lemma₂}\AgdaSymbol{)} \AgdaSymbol{(}\AgdaFunction{cong} \AgdaSymbol{(λ} \AgdaBound{z} \AgdaSymbol{→} \AgdaBound{z} \AgdaFunction{+} \AgdaFunction{S} \AgdaFunction{Z}\AgdaSymbol{)} \AgdaSymbol{(}\AgdaFunction{right-0} \AgdaSymbol{\_))))}\<%
\\
\>[5]\AgdaIndent{36}{}\<[36]%
\>[36]\AgdaSymbol{(}\AgdaFunction{sym} \AgdaSymbol{(}\AgdaFunction{<<-is-*2} \AgdaBound{n₂}\AgdaSymbol{))))))}\<%
\\
\>[0]\AgdaIndent{2}{}\<[2]%
\>[2]\AgdaFunction{bplus-is-+} \AgdaSymbol{(}\AgdaInductiveConstructor{bn} \AgdaSymbol{\{}\AgdaInductiveConstructor{suc} \AgdaBound{n}\AgdaSymbol{\}} \AgdaSymbol{(}\AgdaInductiveConstructor{one} \AgdaInductiveConstructor{∷} \AgdaBound{l₀}\AgdaSymbol{))} \AgdaSymbol{(}\AgdaInductiveConstructor{bn} \AgdaSymbol{\{}\AgdaInductiveConstructor{suc} \AgdaBound{n₁}\AgdaSymbol{\}} \AgdaSymbol{(}\AgdaInductiveConstructor{one} \AgdaInductiveConstructor{∷} \AgdaBound{l₁}\AgdaSymbol{))} \AgdaSymbol{=}\<%
\\
\>[2]\AgdaIndent{4}{}\<[4]%
\>[4]\AgdaKeyword{let} \AgdaBound{n₁} \AgdaSymbol{=} \AgdaInductiveConstructor{bn} \AgdaBound{l₀}\<%
\\
\>[4]\AgdaIndent{8}{}\<[8]%
\>[8]\AgdaBound{n₂} \AgdaSymbol{=} \AgdaInductiveConstructor{bn} \AgdaBound{l₁}\<%
\\
\>[4]\AgdaIndent{8}{}\<[8]%
\>[8]\AgdaBound{num} \AgdaSymbol{=} \AgdaFunction{bplus} \AgdaBound{n₁} \AgdaBound{n₂}\<%
\\
\>[4]\AgdaIndent{8}{}\<[8]%
\>[8]\AgdaBound{v₁} \AgdaSymbol{=} \AgdaFunction{⟦} \AgdaBound{n₁} \AgdaFunction{⟧₂}\<%
\\
\>[4]\AgdaIndent{8}{}\<[8]%
\>[8]\AgdaBound{v₂} \AgdaSymbol{=} \AgdaFunction{⟦} \AgdaBound{n₂} \AgdaFunction{⟧₂} \AgdaKeyword{in}\<%
\\
\>[0]\AgdaIndent{5}{}\<[5]%
\>[5]\AgdaFunction{trans} \AgdaSymbol{(}\AgdaFunction{+1-is-S} \AgdaSymbol{(}\AgdaFunction{+1} \AgdaSymbol{(}\AgdaFunction{<<} \AgdaBound{num}\AgdaSymbol{)))}\<%
\\
\>[0]\AgdaIndent{5}{}\<[5]%
\>[5]\AgdaSymbol{(}\AgdaFunction{trans} \AgdaSymbol{(}\AgdaFunction{cong} \AgdaFunction{S} \AgdaSymbol{(}\AgdaFunction{+1-is-S} \AgdaSymbol{(}\AgdaFunction{<<} \AgdaBound{num}\AgdaSymbol{)))}\<%
\\
\>[0]\AgdaIndent{5}{}\<[5]%
\>[5]\AgdaSymbol{(}\AgdaFunction{trans} \AgdaSymbol{(}\AgdaFunction{cong} \AgdaSymbol{(λ} \AgdaBound{z} \AgdaSymbol{→} \AgdaFunction{S} \AgdaSymbol{(}\AgdaFunction{S} \AgdaBound{z}\AgdaSymbol{))} \AgdaSymbol{(}\AgdaFunction{trans} \AgdaSymbol{(}\AgdaFunction{<<-is-*2} \AgdaBound{num}\AgdaSymbol{)}\<%
\\
\>[5]\AgdaIndent{34}{}\<[34]%
\>[34]\AgdaSymbol{(}\AgdaFunction{trans} \AgdaSymbol{(}\AgdaFunction{cong₂} \AgdaFunction{\_+\_} \AgdaSymbol{(}\AgdaFunction{bplus-is-+} \AgdaBound{n₁} \AgdaBound{n₂}\AgdaSymbol{)} \AgdaSymbol{(}\AgdaFunction{bplus-is-+} \AgdaBound{n₁} \AgdaBound{n₂}\AgdaSymbol{))}\<%
\\
\>[5]\AgdaIndent{34}{}\<[34]%
\>[34]\AgdaSymbol{(}\AgdaFunction{shuffle} \AgdaBound{v₁} \AgdaBound{v₂}\AgdaSymbol{))))}\<%
\\
\>[0]\AgdaIndent{5}{}\<[5]%
\>[5]\AgdaSymbol{(}\AgdaFunction{trans} \AgdaSymbol{(}\AgdaFunction{cong} \AgdaFunction{S} \AgdaSymbol{(}\AgdaFunction{trans} \AgdaSymbol{(}\AgdaFunction{sym} \AgdaSymbol{(}\AgdaFunction{x+Sy≡Sx+y} \AgdaSymbol{\_} \AgdaSymbol{\_))} \AgdaSymbol{(}\AgdaField{comm-+} \AgdaSymbol{\_} \AgdaSymbol{\_)))}\<%
\\
\>[0]\AgdaIndent{5}{}\<[5]%
\>[5]\AgdaSymbol{(}\AgdaFunction{trans} \AgdaSymbol{(}\AgdaFunction{sym} \AgdaSymbol{(}\AgdaFunction{x+Sy≡Sx+y} \AgdaSymbol{\_} \AgdaSymbol{\_))}\<%
\\
\>[0]\AgdaIndent{5}{}\<[5]%
\>[5]\AgdaSymbol{(}\AgdaFunction{trans} \AgdaSymbol{(}\AgdaField{comm-+} \AgdaSymbol{\_} \AgdaSymbol{\_)}\<%
\\
\>[5]\AgdaIndent{12}{}\<[12]%
\>[12]\AgdaSymbol{(}\AgdaFunction{cong₂} \AgdaFunction{\_+\_} \AgdaSymbol{(}\AgdaFunction{trans} \AgdaSymbol{(}\AgdaFunction{cong} \AgdaFunction{S} \AgdaSymbol{(}\AgdaFunction{sym} \AgdaSymbol{(}\AgdaFunction{<<-is-*2} \AgdaBound{n₁}\AgdaSymbol{)))} \AgdaSymbol{(}\AgdaFunction{sym} \AgdaSymbol{(}\AgdaFunction{x+Sy≡Sx+y} \AgdaSymbol{\_} \AgdaSymbol{\_)))}\<%
\\
\>[12]\AgdaIndent{23}{}\<[23]%
\>[23]\AgdaSymbol{(}\AgdaFunction{trans} \AgdaSymbol{(}\AgdaFunction{cong} \AgdaFunction{S} \AgdaSymbol{(}\AgdaFunction{sym} \AgdaSymbol{(}\AgdaFunction{<<-is-*2} \AgdaBound{n₂}\AgdaSymbol{)))} \AgdaSymbol{(}\AgdaFunction{sym} \AgdaSymbol{(}\AgdaFunction{x+Sy≡Sx+y} \AgdaSymbol{\_} \AgdaSymbol{\_)))))))))}\<%
\end{code}
}

%% file: app-ctt.tex
\section{${\bf CTT}_{\bf uqe}$ Formalization}\label{app:cttuqe}

\setcounter{biformthy}{0}

{\churchqe}~\cite{FarmerArxiv16} is a version of simple type theory
with quotation and evaluation.  {\churchuqe}~\cite{FarmerArxiv17} is a
variant of {\churchqe} that admits undefined expressions, partial
functions, and multiple base types of individuals.  {\churchuqe} also
includes a definite description operator and conditional expression
operator.  This appendix presents a formalization of the biform theory
graph test case in {\churchuqe} (instead of {\churchqe}) since
{\churchuqe} contains a notion of theory morphism.  The reader is
expected to be familiar with the notation of {\churchuqe}.  The
following are additional notes to the reader:

\be

  \item $\set{o,\epsilon,\iota}$ is the set of base types for all
    eight biform theories given below.  The single nonlogical base
    type $\iota$ is used to represent the natural numbers.

  \item All constants that are not introduced as components of one of
    the biform theories listed below are logical constants of
    {\churchuqe}, either primitive or defined.
    $\mname{is-abs}_{\epsilon \tarrow o}$, $\mname{abs-body}_{\epsilon
      \tarrow \epsilon}$, and $\mname{is-closed}_{\epsilon \tarrow o}$
    are defined logical constants not in~\cite{FarmerArxiv16}.
    $\mname{is-abs}_{\epsilon \tarrow o} \, \textbf{A}_\epsilon$ holds
    iff $\textbf{A}_\epsilon$ represents an abstraction.  If
    $\textbf{A}_\epsilon$ represents an abstraction, then
    $\mname{abs-body}_{\epsilon \tarrow \epsilon} \,
    \textbf{A}_\epsilon$ represents the body of the abstraction.
    $\mname{is-closed}_{\epsilon \tarrow o} \, \textbf{A}_\epsilon$
    holds iff $\textbf{A}_\epsilon$ represents an expression that is
    closed (and eval-free).

  \item The type attached to a constant may be dropped when there is
    no loss of meaning.

\iffalse
  \item When it makes sense, the notation
    $\set{\textbf{A}_{\alpha}^{1}, \ldots, \textbf{A}_{\alpha}^{n}}$
    denotes the predicate
    \[\LambdaApp \textbf{x}_\alpha \mdot 
    (\textbf{x}_\alpha = \textbf{A}_{\alpha}^{1} \OR \cdots \OR
    \textbf{x}_\alpha = \textbf{A}_{\alpha}^{n}).\]
\fi

  \item A constant of a type of the form $\epsilon \tarrow \cdots
    \tarrow \epsilon$ that is intended to be implemented by a
    transformer has a name in upper case letters.

  \item Expressions of type $\epsilon$, i.e., expressions that denote
    constructions, are colored red to identify where reasoning about
    syntax occurs.

\ee

The following are the eight biform theories and the two noninclusive
theory morphisms in the test case:

\begin{biformthy}[BT\thebiformthy: Simple Theory of $0$ and $S$]\em
\bi

  \item[] 

  \item[] \textbf{Primitive Base Types}

  \be

    \item $\iota$ (type of natural numbers).

  \ee

  \item[] \textbf{Primitive Constants}

  \be

    \item $0_\iota$.

    \item $S_{\iota \tarrow \iota}$.

  \ee

  \item[] \textbf{Defined Constants (selected)}

  \be

    \item $1_\iota = S \, 0_\iota$.

    \item $\mname{IS-FO-BT1}_{\epsilon \tarrow \epsilon} = \LambdaApp
      x_\epsilon \mdot \textbf{B}_\epsilon$ {\sglsp} where
      $\textbf{B}_\epsilon$ is a complex expression such that
      $\syn{(\LambdaApp x_\epsilon \mdot \textbf{B}_\epsilon) \,
        \synbrack{\textbf{A}_\alpha}}$ equals $\syn{\synbrack{T_o}}$
      [$\syn{\synbrack{F_o}}$] if $\textbf{A}_\alpha$ is [not] a term
      or formula of first-order logic with equality whose variables
      are of type $\iota$ and whose nonlogical constants are members
      of $\set{0_\iota,S_{\iota \tarrow \iota}}$.

  \ee

  \item[] \textbf{Axioms}

  \be

    \item $S \, x_\iota \not= 0_\iota$.

    \item $S \, x_\iota = S \, y_\iota \Implies x_\iota =
      y_\iota$.

  \ee

  \item[] \textbf{Transformers}

  \be

    \item $\pi_1$ for $\mname{IS-FO-BT1}_{\epsilon \tarrow \epsilon}$
      is an efficient program that accesses the data stored in the
      data structures that represent expressions.

    \item $\pi_2$ for $\mname{IS-FO-BT1}_{\epsilon \tarrow \epsilon}$
      uses the definition of $\mname{IS-FO-BT1}_{\epsilon \tarrow
        \epsilon}$.

  \ee

\ei
\end{biformthy}

\begin{biformthy}[BT\thebiformthy: Simple Theory of $0$, $S$, and $+$]\em
\bi

  \item[]

  \item[] \textbf{Extended Theories} 

  \be

    \item BT1.

  \ee

  \item[] \textbf{Primitive Constants}

  \be

    \setcounter{enumi}{2}

    \item $+_{\iota \tarrow \iota \tarrow \iota}$ (infix).

    \item $\mname{BPLUS}_{\epsilon \tarrow \epsilon \tarrow \epsilon}$ (infix).

  \ee

  \item[] \textbf{Defined Constants (selected)}

  \be

    \setcounter{enumi}{2}

    \item $\mname{bnat}_{\iota \tarrow \iota \tarrow \iota} =
      \LambdaApp x_\iota \mdot \LambdaApp y_\iota \mdot ((x_\iota +
      x_\iota) + y_\iota)$.
      
    Notational definition:

    \bi

      \item[] $(0)_2 = \mname{bnat} \, 0_\iota \, 0_\iota$.
  
      \item[] $(1)_2 = \mname{bnat} \, 0_\iota \, 1_\iota$.
  
      \item[] $(a_1 \cdots a_n0)_2 = \mname{bnat} \, (a_1 \cdots
        a_n)_2 \, 0_\iota$ {\sglsp} where each $a_i$ is 0 or 1.
  
      \item[] $(a_1 \cdots a_n1)_2 = \mname{bnat} \, (a_1 \cdots
        a_n)_2 \, 1_\iota$ {\sglsp} where each $a_i$ is 0 or 1.
  
    \ei

    \item $\mname{is-bnum}_{\epsilon \tarrow o} = 
      \IotaApp f_{\epsilon \tarrow o} \mdot
      \ForallApp \syn{u_\epsilon} \mdot
      (f_{\epsilon \tarrow \epsilon} \, \syn{u_\epsilon} \Iff {}\\
      \hspace*{2ex}\ForsomeApp \syn{v_\epsilon} \mdot 
      \ForsomeApp \syn{w_\epsilon} \mdot
      (\syn{u_\epsilon} = \syn{\synbrack{\mname{bnat} \, 
      \commabrack{v_\epsilon} \, \commabrack{w_\epsilon}}} \And {}\\
      \hspace*{4ex}(\syn{v_\epsilon} = \syn{\synbrack{0_\iota}} \OR 
      f_{\epsilon \tarrow \epsilon} \, \syn{v_\epsilon}) \And
      (\syn{w_\epsilon} = \syn{\synbrack{0_\iota}} \OR 
      \syn{w_\epsilon} = \syn{\synbrack{1_\iota}})))$.

    \item $\mname{IS-FO-BT2}_{\epsilon \tarrow \epsilon} = \LambdaApp
      x_\epsilon \mdot \textbf{B}_\epsilon$ {\sglsp} where
      $\textbf{B}_\epsilon$ is a complex expression such that
      $\syn{(\LambdaApp x_\epsilon \mdot \textbf{B}_\epsilon) \,
        \synbrack{\textbf{A}_\alpha}}$ equals $\syn{\synbrack{T_o}}$
      [$\syn{\synbrack{F_o}}$] if $\textbf{A}_\alpha$ is [not] a term
      or formula of first-order logic with equality whose variables
      are of type $\iota$ and whose nonlogical constants are members
      of $\set{0_\iota,S_{\iota \tarrow \iota},+_{\iota \tarrow \iota
          \tarrow \iota}}$.
      
  \ee

  \item[] \textbf{Axioms}

  \be

    \setcounter{enumi}{2}

    \item $x_\iota + 0_\iota = x_\iota$.

    \item $x_\iota + S \, y_\iota = S \, (x_\iota + y_\iota)$.

    \item $\mname{is-bnum} \, \syn{u_\epsilon} \Implies
      \syn{u_\epsilon \; \mname{BPLUS} \; \synbrack{(0)_2}} =
      \syn{u_\epsilon}$.

    \item $\mname{is-bnum} \, \syn{u_\epsilon} \Implies
      \syn{\synbrack{(0)_2} \; \mname{BPLUS} \; u_\epsilon} =
      \syn{u_\epsilon}$.

    \item $\syn{\synbrack{(1)_2} \; \mname{BPLUS} \;
      \synbrack{(1)_2}} = \syn{\synbrack{(10)_2}}$.

    \item $\mname{is-bnum} \, \syn{u_\epsilon} \Implies {}\\
        \hspace*{2ex} \syn{\synbrack{\mname{bnat} \,
            \commabrack{u_\epsilon} \, 0_\iota} \; \mname{BPLUS} \;
          \synbrack{(1)_2}} = \syn{\synbrack{\mname{bnat} \,
            \commabrack{u_\epsilon} \, 1_\iota}}$.

    \item $\mname{is-bnum} \, \syn{u_\epsilon} \Implies {}\\
        \hspace*{2ex} \syn{\synbrack{\mname{bnat} \,
            \commabrack{u_\epsilon} \, 1_\iota} \; \mname{BPLUS} \;
          \synbrack{(1)_2}} = \syn{\synbrack{\mname{bnat} \,
            \commabrack{u_\epsilon \; \mname{BPLUS} \;
              \synbrack{(1)_2}} \, 0_\iota}}$.

    \item $\mname{is-bnum} \, \syn{u_\epsilon} \Implies {}\\
        \hspace*{2ex} \syn{\synbrack{(1)_2} \; \mname{BPLUS} \;
          \synbrack{\mname{bnat} \, \commabrack{u_\epsilon} \, 0_\iota}} =
        \syn{\synbrack{\mname{bnat} \, \commabrack{u_\epsilon} \,
            1_\iota}}$.

    \item $\mname{is-bnum} \, \syn{u_\epsilon} \Implies {}\\
        \hspace*{2ex} \syn{\synbrack{(1)_2} \; \mname{BPLUS} \;
          \synbrack{\mname{bnat} \, \commabrack{u_\epsilon} \, 0_\iota}} =
        \syn{\synbrack{\mname{bnat} \, \commabrack{u_\epsilon \;
              \mname{BPLUS} \; \synbrack{(1)_2}} \, 0_\iota}}$.

    \item $(\mname{is-bnum} \, \syn{u_\epsilon} \And \mname{is-bnum}
      \, \syn{v_\epsilon}) \Implies {}\\
        \hspace*{2ex} \syn{\synbrack{\mname{bnat} \,
            \commabrack{u_\epsilon} \, 0_\iota} \; \mname{BPLUS} \;
          \synbrack{\mname{bnat} \, \commabrack{v_\epsilon} 0_\iota}} =
        \syn{\synbrack{\mname{bnat} \, \commabrack{u_\epsilon \;
              \mname{BPLUS} \; v_\epsilon} \, 0_\iota}}$.

    \item $(\mname{is-bnum} \, \syn{u_\epsilon} \And \mname{is-bnum}
      \, \syn{v_\epsilon}) \Implies {}\\
        \hspace*{2ex} \syn{\synbrack{\mname{bnat} \,
            \commabrack{u_\epsilon} \, 0_\iota} \; \mname{BPLUS} \;
          \synbrack{\mname{bnat} \, \commabrack{v_\epsilon} \, 1_\iota}} =
        \syn{\synbrack{\mname{bnat} \, \commabrack{u_\epsilon \;
              \mname{BPLUS} \; v_\epsilon} \, 1_\iota}}$.

    \item $(\mname{is-bnum} \, \syn{u_\epsilon} \And \mname{is-bnum}
      \, \syn{v_\epsilon}) \Implies {}\\
        \hspace*{2ex} \syn{\synbrack{\mname{bnat} \,
            \commabrack{u_\epsilon} \, 1_\iota} \; \mname{BPLUS} \;
          \synbrack{\mname{bnat} \, \commabrack{v_\epsilon} \, 0_\iota}} =
        \syn{\synbrack{\mname{bnat} \, \commabrack{u_\epsilon \;
              \mname{BPLUS} \; v_\epsilon} \, 1_\iota}}$.

    \item $(\mname{is-bnum} \, \syn{u_\epsilon} \And \mname{is-bnum}
      \, \syn{v_\epsilon}) \Implies {}\\
        \hspace*{2ex} \syn{\synbrack{\mname{bnat} \,
            \commabrack{u_\epsilon} \, 1_\iota} \; \mname{BPLUS} \;
          \synbrack{\mname{bnat} \, \commabrack{v_\epsilon} \, 1_\iota}} = {}\\
        \hspace*{2ex}\syn{\synbrack{\mname{bnat} \,
            \commabrack{(u_\epsilon \; \mname{BPLUS} \; v_\epsilon)
              \; \mname{BPLUS} \; \synbrack{(1)_2}} \, 0_\iota}}$.

  \ee

  \item[] \textbf{Transformers}

  \be

      \setcounter{enumi}{2}

    \item $\pi_3$ for $\mname{BPLUS}_{\epsilon \tarrow \epsilon
      \tarrow \epsilon}$ is an efficient program that satisfies Axioms
      5--15.

    \item $\pi_4$ for $\mname{BPLUS}_{\epsilon \tarrow \epsilon
      \tarrow \epsilon}$ uses Axioms 5--15 as conditional rewrite
      rules.

    \item $\pi_5$ for $\mname{IS-FO-BT2}_{\epsilon \tarrow \epsilon}$
      is an efficient program that accesses the data stored in the
      data structures that represent expressions.

    \item $\pi_6$ for $\mname{IS-FO-BT2}_{\epsilon \tarrow \epsilon}$
      uses the definition of $\mname{IS-FO-BT2}_{\epsilon \tarrow
        \epsilon}$.

  \ee
    
  \item[] \textbf{Theorems (selected)}

    \be

      \setcounter{enumi}{0}

      \item Meaning formula schema for 
      $\mname{BPLUS}_{\epsilon \tarrow \epsilon \tarrow \epsilon}$

      $((\mname{is-bnum} \, \syn{\textbf{A}_\epsilon} \And 
      \mname{is-bnum} \, \syn{\textbf{B}_\epsilon}) 
      \Implies {}\\
      \hspace*{2ex}(\mname{is-bnum} \, 
      \syn{(\textbf{A}_\epsilon \; \mname{BPLUS} \; \textbf{B}_\epsilon)} \And {}\\
      \hspace*{2.8ex}(\sembrack{\syn{\textbf{A}_\epsilon \; \mname{BPLUS} \; 
      \textbf{B}_\epsilon}}_\iota = 
      \sembrack{\syn{\textbf{A}_\epsilon}}_\iota + 
      \sembrack{\syn{\textbf{B}_\epsilon}}_\iota)))$.

    \ee

\ei
\end{biformthy}

\begin{biformthy}[BT\thebiformthy: Simple Theory of $0$, $S$, $+$, and $*$]\em

\bi

  \item[]

  \item[] \textbf{Extended Theories} 

  \be

    \setcounter{enumi}{1}

    \item BT2.

  \ee

  \item[] \textbf{Primitive Constants}

  \be

    \setcounter{enumi}{4}

    \item $*_{\iota \tarrow \iota \tarrow \iota}$ (infix).

    \item $\mname{BTIMES}_{\epsilon \tarrow \epsilon \tarrow \epsilon}$ (infix).

  \ee

  \item[] \textbf{Defined Constants (selected)}

  \be

    \setcounter{enumi}{3}

    \item $\mname{IS-FO-BT3}_{\epsilon \tarrow \epsilon} = \LambdaApp
      x_\epsilon \mdot \textbf{B}_\epsilon$ {\sglsp} where
      $\textbf{B}_\epsilon$ is a complex expression such that
      $\syn{(\LambdaApp x_\epsilon \mdot \textbf{B}_\epsilon) \,
        \synbrack{\textbf{A}_\alpha}}$ equals $\syn{\synbrack{T_o}}$
      [$\syn{\synbrack{F_o}}$] if $\textbf{A}_\alpha$ is [not] a term
      or formula of first-order logic with equality whose variables
      are of type $\iota$ and whose nonlogical constants are members
      of $\set{0_\iota,S_{\iota \tarrow \iota},+_{\iota \tarrow \iota
          \tarrow \iota},*_{\iota \tarrow \iota \tarrow \iota}}$.

  \ee

  \item[] \textbf{Axioms}

  \be

    \setcounter{enumi}{15}

    \item $x_\iota * 0_\iota = 0_\iota$.

    \item $x_\iota * S \, y_\iota = (x_\iota * y_\iota) + x_\iota$.

    \item $\mname{is-bnum} \, \syn{u_\epsilon} \Implies
      \syn{u_\epsilon \; \mname{BTIMES} \; \synbrack{(0)_2}} =
      \syn{\synbrack{(0)_2}}$.

    \item $\mname{is-bnum} \, \syn{u_\epsilon} \Implies \syn{
      \synbrack{(0)_2} \; \mname{BTIMES} \; u_\epsilon} =
      \syn{\synbrack{(0)_2}}$.

    \item $\mname{is-bnum} \, \syn{u_\epsilon} \Implies
      \syn{u_\epsilon \; \mname{BTIMES} \; \synbrack{(1)_2}} =
      \syn{u_\epsilon}$.

    \item $\mname{is-bnum} \, \syn{u_\epsilon} \Implies
      \syn{\synbrack{(1)_2} \; \mname{BTIMES} \; u_\epsilon} =
      \syn{u_\epsilon}$.

    \item $(\mname{is-bnum} \, \syn{u_\epsilon} \And \mname{is-bnum}
      \, \syn{v_\epsilon}) \Implies {}\\
        \hspace*{2ex} \syn{\synbrack{\mname{bnat} \,
            \commabrack{u_\epsilon} \, 0} \; \mname{BTIMES} \;
          \syn{v_\epsilon}} = \syn{\synbrack{\mname{bnat} \,
            \commabrack{u_\epsilon \; \mname{BTIMES} \; v_\epsilon} \,
            0_\iota}}$.

    \item $(\mname{is-bnum} \, \syn{u_\epsilon} \And \mname{is-bnum}
      \, \syn{v_\epsilon}) \Implies {}\\
        \hspace*{2ex} \syn{\synbrack{\mname{bnat} \,
            \commabrack{u_\epsilon} \, 1_\iota} \; \mname{BTIMES} \;
          \syn{v_\epsilon}} = \syn{\synbrack{\mname{bnat} \,
            \commabrack{u_\epsilon \; \mname{BTIMES} \; v_\epsilon} \,
            0_\iota} \; \mname{badd} \; v_\epsilon}$.

    \item $(\mname{is-bnum} \, \syn{u_\epsilon} \And \mname{is-bnum}
      \, \syn{v_\epsilon}) \Implies {}\\
        \hspace*{2ex} \syn{\syn{v_\epsilon}\; \mname{BTIMES} \;
          \synbrack{\mname{bnat} \, \commabrack{u_\epsilon} \, 0_\iota}} =
        \syn{\synbrack{\mname{bnat} \, \commabrack{u_\epsilon \;
              \mname{BTIMES} \; v_\epsilon} \, 0_\iota}}$.

    \item $(\mname{is-bnum} \, \syn{u_\epsilon} \And 
        \mname{is-bnum} \, \syn{v_\epsilon}) \Implies {}\\
        \hspace*{2ex} \syn{\syn{v_\epsilon} \; \mname{BTIMES} \;
          \synbrack{\mname{bnat} \, \commabrack{u_\epsilon} \, 1_\iota}} =
        \syn{\synbrack{\mname{bnat} \, \commabrack{u_\epsilon \;
              \mname{BTIMES} \; v_\epsilon} \, 0_\iota} \; \mname{badd} \;
          v_\epsilon}$.

  \ee

  \item[] \textbf{Transformers}

  \be

      \setcounter{enumi}{6}

    \item $\pi_7$ for $\mname{BTIMES}_{\epsilon \tarrow \epsilon
      \tarrow \epsilon}$ is an efficient program that satisfies Axioms
      18--25.

    \item $\pi_8$ for $\mname{BTIMES}_{\epsilon \tarrow \epsilon
      \tarrow \epsilon}$ uses Axioms 18--25 as conditional rewrite
      rules.

    \item $\pi_9$ for $\mname{IS-FO-BT3}_{\epsilon \tarrow \epsilon}$
      is an efficient program that accesses the data stored in the
      data structures that represent expressions.

    \item $\pi_{10}$ for $\mname{IS-FO-BT3}_{\epsilon \tarrow
      \epsilon}$ uses the definition of $\mname{IS-FO-BT3}_{\epsilon
      \tarrow \epsilon}$.

  \ee

  \item[] \textbf{Theorems (selected)}

  \be

    \setcounter{enumi}{1}

    \item Meaning formula schema for
    $\mname{BTIMES}_{\epsilon \tarrow \epsilon \tarrow \epsilon}$

    $((\mname{is-bnum} \, \syn{\textbf{A}_\epsilon} \And 
      \mname{is-bnum} \, \syn{\textbf{B}_\epsilon}) 
      \Implies {}\\
      \hspace*{2ex}(\mname{is-bnum} \, 
      \syn{(\textbf{A}_\epsilon \; \mname{BTIMES} \; \textbf{B}_\epsilon)} \And {}\\
      \hspace*{2.8ex}(\sembrack{\syn{\textbf{A}_\epsilon \; \mname{BTIMES} \; 
      \textbf{B}_\epsilon}}_\iota = 
      \sembrack{\syn{\textbf{A}_\epsilon}}_\iota * 
      \sembrack{\syn{\textbf{B}_\epsilon}}_\iota)))$.

  \ee

\ei
\end{biformthy}

\begin{biformthy}[BT\thebiformthy: Robinson Arithmetic (Q)]\em
\bi

  \item[]

  \item[] \textbf{Extended Theories} 

  \be

    \setcounter{enumi}{2}

    \item BT3.

  \ee

  \item[] \textbf{Axioms}

  \be

    \setcounter{enumi}{25}

    \item $x_\iota = 0_\iota \OR \ForsomeApp y_\iota \mdot S \,
      y_\iota = x_\iota$.

  \ee

\ei
\end{biformthy}

\begin{biformthy}[BT\thebiformthy: Complete Theory of $0$ and $S$]\em

\bi

  \item[]

  \item[] \textbf{Extended Theories} 

  \be

    \setcounter{enumi}{0}

    \item BT1.

  \ee

  \item[] \textbf{Primitive Constants}

  \be

    \setcounter{enumi}{6}

    \item $\mname{BT5-DEC-PROC}_{\epsilon \tarrow \epsilon}$.

  \ee

  \item[] \textbf{Defined Constants (selected)}

  \be

    \setcounter{enumi}{5}

    \item $\mname{IS-FO-BT1-ABS}_{\epsilon \tarrow \epsilon} = {}\\
    \hspace*{2ex}\LambdaApp \syn{x_\epsilon} \mdot 
    (\If \; (\mname{is-abs}_{\epsilon \tarrow o} \, \syn{x_\epsilon}) \;
    \syn{(\mname{IS-FO-BT1}_{\epsilon \tarrow \epsilon} \,
    (\mname{abs-body}_{\epsilon \tarrow \epsilon} \, x_\epsilon))} \;
    \syn{\synbrack{F_o}})$.

  \ee

  \item[] \textbf{Axioms}

  \be

    \setcounter{enumi}{26}

    \item Induction Schema for $S$

    $\ForallApp \syn{f_\epsilon} \mdot 
    ((\mname{is-expr}_{\epsilon \tarrow o}^{\iota \tarrow o} \, \syn{f_\epsilon} \And
    \sembrack{\syn{\mname{IS-FO-BT1-ABS}_{\epsilon \tarrow \epsilon} \, 
    f_\epsilon}}_o) \Implies {} \\
    \hspace*{2ex}((\sembrack{\syn{f_\epsilon}}_{\iota \tarrow o} \, 0_\iota \And
    (\ForallApp x_\iota \mdot 
    \sembrack{\syn{f_\epsilon}}_{\iota \tarrow o} \, x_\iota \Implies
    \sembrack{\syn{f_\epsilon}}_{\iota \tarrow o} \, 
    (\mname{S} \, x_\iota))) \Implies 
    \ForallApp x_\iota \mdot 
    \sembrack{\syn{f_\epsilon}}_{\iota \tarrow o} \, x_\iota))$.

    \item Meaning Formula for $\mname{BT5-DEC-PROC}_{\epsilon \tarrow \epsilon}$

    $\ForallApp \syn{u_\epsilon} \mdot 
     ((\mname{is-expr}_{\epsilon \tarrow o}^{o} \, 
     \syn{u_\epsilon} \And
     \mname{is-closed}_{\epsilon \tarrow o} \, \syn{u_\epsilon} \And 
     \sembrack{\syn{\mname{IS-FO-BT1}_{\epsilon \tarrow \epsilon} \, 
     u_\epsilon}}_o) \Implies {}\\
     \hspace*{2ex}((\syn{\mname{BT5-DEC-PROC}_{\epsilon \tarrow \epsilon} \, u_\epsilon} = 
     \syn{\synbrack{T_o}} \OR 
     \syn{\mname{BT5-DEC-PROC}_{\epsilon \tarrow \epsilon} \, u_\epsilon} = 
     \syn{\synbrack{F_o}}) \And {}\\
     \hspace*{3.1ex}
     \sembrack{\syn{\mname{BT5-DEC-PROC}_{\epsilon \tarrow \epsilon} \, u_\epsilon}}_o =
     \sembrack{\syn{u_\epsilon}}_o))$.

  \ee

  \item[] \textbf{Transformers}

  \be

    \setcounter{enumi}{10}

    \item $\pi_{11}$ for $\mname{BT5-DEC-PROC}_{\epsilon \tarrow
      \epsilon \tarrow \epsilon}$ is an efficient decision procedure
      that satisfies Axiom 28.

    \item $\pi_{12}$ for $\mname{IS-FO-BT1-ABS}_{\epsilon \tarrow
      \epsilon}$ is an efficient program that accesses the data stored
      in the data structures that represent expressions.

    \item $\pi_{13}$ for $\mname{IS-FO-BT1-ABS}_{\epsilon \tarrow
      \epsilon}$ uses the definition of
      $\mname{IS-FO-BT1-ABS}_{\epsilon \tarrow \epsilon}$.

  \ee

\ei
\end{biformthy}

\begin{biformthy}[BT\thebiformthy: Presburger Arithmetic]\em

\bi

  \item[]

  \item[] \textbf{Extended Theories} 

  \be

    \item[1.] BT2.

    \item[5.] BT5.

  \ee
    
  \item[] \textbf{Primitive Constants}

  \be

    \setcounter{enumi}{7}

    \item $\mname{BT6-DEC-PROC}_{\epsilon \tarrow \epsilon}$.

  \ee

  \item[] \textbf{Defined Constants (selected)}

  \be

    \setcounter{enumi}{6}

    \item $\mname{IS-FO-BT2-ABS}_{\epsilon \tarrow \epsilon} = {}\\
    \hspace*{2ex}\LambdaApp \syn{x_\epsilon} \mdot 
    (\If \; (\mname{is-abs}_{\epsilon \tarrow o} \, \syn{x_\epsilon}) \;
    \syn{(\mname{IS-FO-BT2}_{\epsilon \tarrow \epsilon} \,
    (\mname{abs-body}_{\epsilon \tarrow \epsilon} \, x_\epsilon))} \;
    \syn{\synbrack{F_o}})$.

  \ee

  \item[] \textbf{Axioms}

  \be

    \setcounter{enumi}{28}

    \item Induction Schema for $S$ and $+$

    $\ForallApp \syn{f_\epsilon} \mdot 
    ((\mname{is-expr}_{\epsilon \tarrow o}^{\iota \tarrow o} \, \syn{f_\epsilon} \And
    \sembrack{\syn{\mname{IS-FO-BT2-ABS}_{\epsilon \tarrow \epsilon} \, 
    f_\epsilon}}_o) \Implies {} \\
    \hspace*{2ex}((\sembrack{\syn{f_\epsilon}}_{\iota \tarrow o} \, 0_\iota \And
    (\ForallApp x_\iota \mdot 
    \sembrack{\syn{f_\epsilon}}_{\iota \tarrow o} \, x_\iota \Implies
    \sembrack{\syn{f_\epsilon}}_{\iota \tarrow o} \, 
    (\mname{S} \, x_\iota))) \Implies 
    \ForallApp x_\iota \mdot 
    \sembrack{\syn{f_\epsilon}}_{\iota \tarrow o} \, x_\iota))$.

    \item Meaning formula for $\mname{BT6-DEC-PROC}_{\epsilon \tarrow \epsilon}$.

    $\ForallApp \syn{u_\epsilon} \mdot 
    ((\mname{is-expr}_{\epsilon \tarrow o}^{o} \, 
    \syn{u_\epsilon}  \And
    \mname{is-closed}_{\epsilon \tarrow o} \, \syn{u_\epsilon} \And 
    \sembrack{\syn{\mname{IS-FO-BT2}_{\epsilon \tarrow \epsilon} \, 
    u_\epsilon}}_o) \Implies {}\\
    \hspace*{2ex}((\syn{\mname{BT6-DEC-PROC}_{\epsilon \tarrow \epsilon} \, u_\epsilon} = 
    \syn{\synbrack{T_o}} \OR 
    \syn{\mname{BT6-DEC-PROC}_{\epsilon \tarrow \epsilon} \, u_\epsilon} = 
    \syn{\synbrack{F_o}}) \And {}\\
    \hspace*{3.1ex}
    \sembrack{\syn{\mname{BT6-DEC-PROC}_{\epsilon \tarrow \epsilon} \, u_\epsilon}}_o =
    \sembrack{\syn{u_\epsilon}}_o))$.

  \ee

  \item[] \textbf{Transformers}

  \be

    \setcounter{enumi}{13}

    \item $\pi_{14}$ for $\mname{BT6-DEC-PROC}_{\epsilon \tarrow
      \epsilon \tarrow \epsilon}$ is an efficient decision procedure
      that satisfies Axiom 30.

    \item $\pi_{15}$ for $\mname{IS-FO-BT2-ABS}_{\epsilon \tarrow
      \epsilon}$ is an efficient program that accesses the data
      stored in the data structures that represent expressions.

    \item $\pi_{16}$ for $\mname{IS-FO-BT2-ABS}_{\epsilon \tarrow
      \epsilon}$ uses the definition of
      $\mname{IS-FO-BT2-ABS}_{\epsilon \tarrow \epsilon}$.

  \ee
    
  \item[] \textbf{Theorems (selected)}

    \be

      \setcounter{enumi}{2}

      \item Meaning formula for 
      $\mname{BPLUS}_{\epsilon \tarrow \epsilon \tarrow \epsilon}$

      $\ForallApp \syn{u_\epsilon} \mdot \ForallApp \syn{v_\epsilon} \mdot
      ((\mname{is-bnum} \, \syn{u_\epsilon} \And \mname{is-bnum} \, \syn{v_\epsilon}) 
      \Implies {}\\
      \hspace*{2ex}(\mname{is-bnum} \, 
      \syn{(u_\epsilon \; \mname{BPLUS} \; v_\epsilon)} \And {}\\
      \hspace*{2.8ex}(\sembrack{\syn{u_\epsilon \; \mname{BPLUS} \; 
      v_\epsilon}}_\iota = 
      \sembrack{\syn{u_\epsilon}}_\iota + \sembrack{\syn{v_\epsilon}}_\iota)))$.

    \ee

\ei
\end{biformthy}

\begin{biformthy}[BT\thebiformthy: First-Order Peano Arithmetic]\em

\bi

  \item[]

  \item[] \textbf{Extended Theories} 

  \be

    \item[3.] BT3.

    \item[6.] BT6.

  \ee

  \item[] \textbf{Defined Constants (selected)}

  \be

    \setcounter{enumi}{7}

    \item $\mname{IS-FO-BT3-ABS}_{\epsilon \tarrow \epsilon} = {}\\
    \hspace*{2ex}\LambdaApp \syn{x_\epsilon} \mdot 
    (\If \; (\mname{is-abs}_{\epsilon \tarrow o} \, \syn{x_\epsilon}) \;
    \syn{(\mname{IS-FO-BT3}_{\epsilon \tarrow \epsilon} \,
    (\mname{abs-body}_{\epsilon \tarrow \epsilon} \, x_\epsilon))} \;
    \syn{\synbrack{F_o}})$.

  \ee

  \item[] \textbf{Axioms}

  \be

    \setcounter{enumi}{30}

    \item Induction Schema for $S$, $+$, and $*$

    $\ForallApp \syn{f_\epsilon} \mdot 
    ((\mname{is-expr}_{\epsilon \tarrow o}^{\iota \tarrow o} \, \syn{f_\epsilon} \And
    \sembrack{\syn{\mname{IS-FO-BT3-ABS}_{\epsilon \tarrow \epsilon} \, 
    f_\epsilon}}_o) \Implies {} \\
    \hspace*{2ex}((\sembrack{\syn{f_\epsilon}}_{\iota \tarrow o} \, 0_\iota \And
    (\ForallApp x_\iota \mdot 
    \sembrack{\syn{f_\epsilon}}_{\iota \tarrow o} \, x_\iota \Implies
    \sembrack{\syn{f_\epsilon}}_{\iota \tarrow o} \, 
    (\mname{S} \, x_\iota))) \Implies 
    \ForallApp x_\iota \mdot 
    \sembrack{\syn{f_\epsilon}}_{\iota \tarrow o} \, x_\iota))$.

  \ee

  \item[] \textbf{Transformers}

  \be

    \setcounter{enumi}{16}

    \item $\pi_{17}$ for $\mname{IS-FO-BT3-ABS}_{\epsilon \tarrow
      \epsilon}$ is an efficient program that accesses the data stored
      in the data structures that represent expressions.

    \item $\pi_{18}$ for $\mname{IS-FO-BT3-ABS}_{\epsilon \tarrow
      \epsilon}$ uses the definition of
      $\mname{IS-FO-BT3-ABS}_{\epsilon \tarrow \epsilon}$.

  \ee

  \item[] \textbf{Theorems (selected)}

  \be

    \setcounter{enumi}{3}

    \item Axiom 26.

    \item Meaning formula
    $\mname{BTIMES}_{\epsilon \tarrow \epsilon \tarrow \epsilon}$

    $\ForallApp \syn{u_\epsilon} \mdot \ForallApp \syn{v_\epsilon} \mdot
    ((\mname{is-bnum} \, \syn{u_\epsilon} \And \mname{is-bnum} \, \syn{v_\epsilon}) 
    \Implies {}\\
    \hspace*{2ex}(\mname{is-bnum} \, 
    \syn{(u_\epsilon \; \mname{BTIMES} \; v_\epsilon)} \And {}\\
    \hspace*{2.8ex}(\sembrack{\syn{u_\epsilon \; \mname{BTIMES} \; 
    v_\epsilon}}_\iota = 
    \sembrack{\syn{u_\epsilon}}_\iota * \sembrack{\syn{v_\epsilon}}_\iota)))$.

  \ee

\ei
\end{biformthy}

\begin{biformthy}[BT\thebiformthy: Higher-Order Peano Arithmetic]\em

\bi

  \item[]

  \item[] \textbf{Extended Theories} 

  \be

    \setcounter{enumi}{0}

    \item BT1.

  \ee

  \item[] \textbf{Defined Constants (selected)}

  \be

    \setcounter{enumi}{8}

    \item $+_{\iota \tarrow \iota \tarrow \iota} =
    \IotaApp f_{\iota \tarrow \iota \tarrow \iota} \mdot
    \ForallApp x_\iota \mdot \ForallApp y_\iota \mdot {}\\
    \hspace*{2ex} (f_{\iota \tarrow \iota \tarrow \iota} \, 
    x_\iota \, 0_\iota = x_\iota \And {}\\
    \hspace*{2.8ex}f_{\iota \tarrow \iota \tarrow \iota} \, 
    x_\iota \, (S \, y_\iota) = S \, 
    (f_{\iota \tarrow \iota \tarrow \iota} \, x_\iota \,
    y_\iota))$.

    \item $*_{\iota \tarrow \iota \tarrow \iota} =
    \IotaApp f_{\iota \tarrow \iota \tarrow \iota} \mdot
    \ForallApp x_\iota \mdot \ForallApp y_\iota \mdot {}\\
    \hspace*{2ex} (f_{\iota \tarrow \iota \tarrow \iota} \, 
    x_\iota \, 0_\iota = 0_\iota \And {}\\
    \hspace*{2.8ex}f_{\iota \tarrow \iota \tarrow \iota} \, 
    x_\iota \, (S \, y_\iota) =  
    (f_{\iota \tarrow \iota \tarrow \iota} \, x_\iota \,
    y_\iota) + x_\iota)$.

  \ee

  \item[] \textbf{Axioms}

  \be

    \setcounter{enumi}{31}

    \item Induction Axiom for the Natural Numbers

    $\ForallApp p_{\iota \tarrow o} \mdot ((p_{\iota \tarrow o} \, 0_\iota \And 
    (\ForallApp x_\iota \mdot (p_{\iota \tarrow o} \, x_\iota \Implies 
    p_{\iota \tarrow o} \, (S \, x_\iota)))) \Implies {}\\
    \hspace*{2ex}\ForallApp x_\iota \mdot 
    p_{\iota \tarrow o} \, x_\iota)$.

  \ee

  \item[] \textbf{Theorems (selected)}

  \be

    \setcounter{enumi}{5}

    \item Axiom 27 (Induction Schema for $S$).

    \item Axiom 39 (Induction Schema for $S$ and $+$).

    \item Axiom 31 (Induction Schema for $S$, $+$, and $*$).

  \ee

\ei
\end{biformthy}

\begin{thymorphism}[BT4-to-BT7]\em
\be

  \item[]

  \item[] \textbf{Source Theory} BT4.

  \item[] \textbf{Target Theory} BT7.

  \item[] \textbf{Translation} 

  \bi

    \item[] $\mu$ is defined as follows:

    \bi

      \item[] $\mu(o) = \LambdaApp x_o \mdot T_o$.

      \item[] $\mu(\epsilon) = \LambdaApp x_\epsilon \mdot T_o$.

      \item[] $\mu(\iota) = \LambdaApp x_\iota \mdot T_o$.

    \ei

    $\nu$ is the identity on the constants of BT4.

  \ei

  \item[] \textbf{Nontrivial Obligations} 

  \bi

    \item[] Axiom 26.

  \ei  

\ee
\end{thymorphism}

\begin{thymorphism}[BT7-to-BT8]\em
\be

  \item[]

  \item[] \textbf{Source Theory} BT7.

  \item[] \textbf{Target Theory} BT8.

  \item[] \textbf{Translation} 

  \bi

    \item[] $\mu$ is defined as follows:

    \bi

      \item[] $\mu(o) = \LambdaApp x_o \mdot T_o$.

      \item[] $\mu(\epsilon) = \LambdaApp x_\epsilon \mdot T_o$.

      \item[] $\mu(\iota) = \LambdaApp x_\iota \mdot T_o$.

    \ei

    $\nu$ is the identity on the constants of BT7.

  \ei

  \item[] \textbf{Nontrivial Obligations} 

  \bi

    \item[] Axioms 3--25, 27--31.

  \ei  

\ee
\end{thymorphism}

%% file: DefiniteDescr.tex
This defines the tools needed to do the equivalent of definite description in
MLTT.  

\begin{code}%
\>\AgdaKeyword{module} \AgdaModule{DefiniteDescr} \AgdaKeyword{where}\<%
\\
\>\AgdaKeyword{open} \AgdaKeyword{import} \AgdaModule{Relation.Binary.PropositionalEquality} \AgdaKeyword{using} \AgdaSymbol{(}\AgdaDatatype{\_≡\_}\AgdaSymbol{)}\<%
\\
\>\AgdaKeyword{open} \AgdaKeyword{import} \AgdaModule{Data.Product} \AgdaKeyword{using} \AgdaSymbol{(}\AgdaRecord{Σ}\AgdaSymbol{;}\AgdaFunction{\_×\_}\AgdaSymbol{)}\<%
\end{code}
  
\noindent Normal contractability of a type

\begin{code}%
\>\AgdaFunction{isContr} \AgdaSymbol{:} \AgdaPrimitiveType{Set₀} \AgdaSymbol{→} \AgdaPrimitiveType{Set₀}\<%
\\
\>\AgdaFunction{isContr} \AgdaBound{A} \AgdaSymbol{=} \AgdaRecord{Σ} \AgdaBound{A} \AgdaSymbol{(λ} \AgdaBound{a} \AgdaSymbol{→} \AgdaSymbol{∀} \AgdaBound{b} \AgdaSymbol{→} \AgdaBound{a} \AgdaDatatype{≡} \AgdaBound{b}\AgdaSymbol{)}\<%
\end{code}

We now "expand out" that definition when A is a Σ-type
with the first type that of a 2-argument function, and the
second is a predicate (and thus automatically contractible).
We could elide the second part via proof irrelevance.

\begin{code}%
\>\AgdaFunction{isContr₂} \AgdaSymbol{:} \AgdaSymbol{(}\AgdaBound{B} \AgdaSymbol{:} \AgdaPrimitiveType{Set₀}\AgdaSymbol{)} \AgdaSymbol{→} \AgdaKeyword{let} \AgdaBound{bin} \AgdaSymbol{=} \AgdaBound{B} \AgdaSymbol{→} \AgdaBound{B} \AgdaSymbol{→} \AgdaBound{B} \AgdaKeyword{in} \AgdaSymbol{(}\AgdaBound{bin} \AgdaSymbol{→} \AgdaPrimitiveType{Set₀}\AgdaSymbol{)} \AgdaSymbol{→} \AgdaPrimitiveType{Set₀}\<%
\\
\>\AgdaFunction{isContr₂} \AgdaBound{A} \AgdaBound{P} \AgdaSymbol{=}\<%
\\
\>[0]\AgdaIndent{2}{}\<[2]%
\>[2]\AgdaKeyword{let} \AgdaBound{bin} \AgdaSymbol{=} \AgdaBound{A} \AgdaSymbol{→} \AgdaBound{A} \AgdaSymbol{→} \AgdaBound{A} \AgdaKeyword{in}\<%
\\
\>[0]\AgdaIndent{2}{}\<[2]%
\>[2]\AgdaRecord{Σ} \AgdaBound{bin} \AgdaSymbol{(λ} \AgdaBound{f} \AgdaSymbol{→} \AgdaBound{P} \AgdaBound{f} \AgdaFunction{×}\<%
\\
\>[2]\AgdaIndent{15}{}\<[15]%
\>[15]\AgdaSymbol{(∀} \AgdaSymbol{(}\AgdaBound{g} \AgdaSymbol{:} \AgdaBound{A} \AgdaSymbol{→} \AgdaBound{A} \AgdaSymbol{→} \AgdaBound{A}\AgdaSymbol{)} \AgdaSymbol{→} \AgdaBound{P} \AgdaBound{g} \AgdaSymbol{→} \AgdaSymbol{∀} \AgdaBound{a} \AgdaBound{b} \AgdaSymbol{→} \AgdaBound{f} \AgdaBound{a} \AgdaBound{b} \AgdaDatatype{≡} \AgdaBound{g} \AgdaBound{a} \AgdaBound{b}\AgdaSymbol{)} \AgdaFunction{×}\<%
\\
\>[2]\AgdaIndent{15}{}\<[15]%
\>[15]\AgdaSymbol{(∀} \AgdaSymbol{(}\AgdaBound{pf₁} \AgdaBound{pf₂} \AgdaSymbol{:} \AgdaBound{P} \AgdaBound{f}\AgdaSymbol{)} \AgdaSymbol{→} \AgdaBound{pf₁} \AgdaDatatype{≡} \AgdaBound{pf₂}\AgdaSymbol{))}\<%
\end{code}

%% file: Equiv.tex
\begin{code}%
\>\AgdaSymbol{\{-\#} \AgdaKeyword{OPTIONS} --without-K \AgdaSymbol{\#-\}}\<%
\\
\\
\>\AgdaKeyword{module} \AgdaModule{Equiv} \AgdaKeyword{where}\<%
\\
\\
\>\AgdaKeyword{open} \AgdaKeyword{import} \AgdaModule{Level} \AgdaKeyword{using} \AgdaSymbol{(}\AgdaPrimitive{\_⊔\_}\AgdaSymbol{)}\<%
\\
\>\AgdaKeyword{open} \AgdaKeyword{import} \AgdaModule{Function} \AgdaKeyword{using} \AgdaSymbol{(}\AgdaFunction{\_∘\_}\AgdaSymbol{;} \AgdaFunction{id}\AgdaSymbol{;} \AgdaFunction{flip}\AgdaSymbol{)}\<%
\\
\>\AgdaKeyword{open} \AgdaKeyword{import} \AgdaModule{Relation.Binary} \AgdaKeyword{using} \AgdaSymbol{(}\AgdaRecord{IsEquivalence}\AgdaSymbol{)}\<%
\\
\>\AgdaKeyword{open} \AgdaKeyword{import} \AgdaModule{Relation.Binary.PropositionalEquality}\<%
\\
\>[0]\AgdaIndent{2}{}\<[2]%
\>[2]\AgdaKeyword{using} \AgdaSymbol{(}\AgdaDatatype{\_≡\_}\AgdaSymbol{;} \AgdaInductiveConstructor{refl}\AgdaSymbol{;} \AgdaFunction{sym}\AgdaSymbol{;} \AgdaFunction{trans}\AgdaSymbol{;} \AgdaFunction{cong}\AgdaSymbol{;} \AgdaFunction{cong₂}\AgdaSymbol{;} \AgdaKeyword{module} \AgdaModule{≡-Reasoning}\AgdaSymbol{)}\<%
\\
\\
\>\AgdaKeyword{infix} \AgdaNumber{4} \AgdaFunction{\_∼\_}\<%
\\
\>\AgdaKeyword{infix} \AgdaNumber{4} \AgdaRecord{\_≃\_}\<%
\\
\>\AgdaKeyword{infixr} \AgdaNumber{5} \AgdaFunction{\_●\_}\<%
\\
\\
\>\AgdaComment{------------------------------------------------------------------------------}\<%
\\
\>\AgdaComment{-- Extensional equivalence of (unary) functions}\<%
\\
\\
\>\AgdaFunction{\_∼\_} \AgdaSymbol{:} \AgdaSymbol{∀} \AgdaSymbol{\{}\AgdaBound{ℓ} \AgdaBound{ℓ'}\AgdaSymbol{\}} \AgdaSymbol{→} \AgdaSymbol{\{}\AgdaBound{A} \AgdaSymbol{:} \AgdaPrimitiveType{Set} \AgdaBound{ℓ}\AgdaSymbol{\}} \AgdaSymbol{\{}\AgdaBound{B} \AgdaSymbol{:} \AgdaPrimitiveType{Set} \AgdaBound{ℓ'}\AgdaSymbol{\}} \AgdaSymbol{→} \AgdaSymbol{(}\AgdaBound{f} \AgdaBound{g} \AgdaSymbol{:} \AgdaBound{A} \AgdaSymbol{→} \AgdaBound{B}\AgdaSymbol{)} \AgdaSymbol{→} \AgdaPrimitiveType{Set} \AgdaSymbol{(}\AgdaBound{ℓ} \AgdaPrimitive{⊔} \AgdaBound{ℓ'}\AgdaSymbol{)}\<%
\\
\>\AgdaFunction{\_∼\_} \AgdaSymbol{\{}\AgdaArgument{A} \AgdaSymbol{=} \AgdaBound{A}\AgdaSymbol{\}} \AgdaBound{f} \AgdaBound{g} \AgdaSymbol{=} \AgdaSymbol{(}\AgdaBound{x} \AgdaSymbol{:} \AgdaBound{A}\AgdaSymbol{)} \AgdaSymbol{→} \AgdaBound{f} \AgdaBound{x} \AgdaDatatype{≡} \AgdaBound{g} \AgdaBound{x}\<%
\\
\\
\>\AgdaFunction{refl∼} \AgdaSymbol{:} \AgdaSymbol{∀} \AgdaSymbol{\{}\AgdaBound{ℓ} \AgdaBound{ℓ'}\AgdaSymbol{\}} \AgdaSymbol{\{}\AgdaBound{A} \AgdaSymbol{:} \AgdaPrimitiveType{Set} \AgdaBound{ℓ}\AgdaSymbol{\}} \AgdaSymbol{\{}\AgdaBound{B} \AgdaSymbol{:} \AgdaPrimitiveType{Set} \AgdaBound{ℓ'}\AgdaSymbol{\}} \AgdaSymbol{\{}\AgdaBound{f} \AgdaSymbol{:} \AgdaBound{A} \AgdaSymbol{→} \AgdaBound{B}\AgdaSymbol{\}} \AgdaSymbol{→} \AgdaSymbol{(}\AgdaBound{f} \AgdaFunction{∼} \AgdaBound{f}\AgdaSymbol{)}\<%
\\
\>\AgdaFunction{refl∼} \AgdaSymbol{\_} \AgdaSymbol{=} \AgdaInductiveConstructor{refl}\<%
\\
\\
\>\AgdaFunction{sym∼} \AgdaSymbol{:} \AgdaSymbol{∀} \AgdaSymbol{\{}\AgdaBound{ℓ} \AgdaBound{ℓ'}\AgdaSymbol{\}} \AgdaSymbol{\{}\AgdaBound{A} \AgdaSymbol{:} \AgdaPrimitiveType{Set} \AgdaBound{ℓ}\AgdaSymbol{\}} \AgdaSymbol{\{}\AgdaBound{B} \AgdaSymbol{:} \AgdaPrimitiveType{Set} \AgdaBound{ℓ'}\AgdaSymbol{\}} \AgdaSymbol{\{}\AgdaBound{f} \AgdaBound{g} \AgdaSymbol{:} \AgdaBound{A} \AgdaSymbol{→} \AgdaBound{B}\AgdaSymbol{\}} \AgdaSymbol{→} \AgdaSymbol{(}\AgdaBound{f} \AgdaFunction{∼} \AgdaBound{g}\AgdaSymbol{)} \AgdaSymbol{→} \AgdaSymbol{(}\AgdaBound{g} \AgdaFunction{∼} \AgdaBound{f}\AgdaSymbol{)}\<%
\\
\>\AgdaFunction{sym∼} \AgdaBound{H} \AgdaBound{x} \AgdaSymbol{=} \AgdaFunction{sym} \AgdaSymbol{(}\AgdaBound{H} \AgdaBound{x}\AgdaSymbol{)}\<%
\\
\\
\>\AgdaFunction{trans∼} \AgdaSymbol{:} \AgdaSymbol{∀} \AgdaSymbol{\{}\AgdaBound{ℓ} \AgdaBound{ℓ'}\AgdaSymbol{\}} \AgdaSymbol{\{}\AgdaBound{A} \AgdaSymbol{:} \AgdaPrimitiveType{Set} \AgdaBound{ℓ}\AgdaSymbol{\}} \AgdaSymbol{\{}\AgdaBound{B} \AgdaSymbol{:} \AgdaPrimitiveType{Set} \AgdaBound{ℓ'}\AgdaSymbol{\}} \AgdaSymbol{\{}\AgdaBound{f} \AgdaBound{g} \AgdaBound{h} \AgdaSymbol{:} \AgdaBound{A} \AgdaSymbol{→} \AgdaBound{B}\AgdaSymbol{\}} \AgdaSymbol{→} \AgdaSymbol{(}\AgdaBound{f} \AgdaFunction{∼} \AgdaBound{g}\AgdaSymbol{)} \AgdaSymbol{→} \AgdaSymbol{(}\AgdaBound{g} \AgdaFunction{∼} \AgdaBound{h}\AgdaSymbol{)} \AgdaSymbol{→} \AgdaSymbol{(}\AgdaBound{f} \AgdaFunction{∼} \AgdaBound{h}\AgdaSymbol{)}\<%
\\
\>\AgdaFunction{trans∼} \AgdaBound{H} \AgdaBound{G} \AgdaBound{x} \AgdaSymbol{=} \AgdaFunction{trans} \AgdaSymbol{(}\AgdaBound{H} \AgdaBound{x}\AgdaSymbol{)} \<[28]%
\>[28]\AgdaSymbol{(}\AgdaBound{G} \AgdaBound{x}\AgdaSymbol{)}\<%
\\
\\
\>\AgdaFunction{∘-resp-∼} \AgdaSymbol{:} \AgdaSymbol{∀} \AgdaSymbol{\{}\AgdaBound{ℓA} \AgdaBound{ℓB} \AgdaBound{ℓC}\AgdaSymbol{\}} \AgdaSymbol{\{}\AgdaBound{A} \AgdaSymbol{:} \AgdaPrimitiveType{Set} \AgdaBound{ℓA}\AgdaSymbol{\}} \AgdaSymbol{\{}\AgdaBound{B} \AgdaSymbol{:} \AgdaPrimitiveType{Set} \AgdaBound{ℓB}\AgdaSymbol{\}} \AgdaSymbol{\{}\AgdaBound{C} \AgdaSymbol{:} \AgdaPrimitiveType{Set} \AgdaBound{ℓC}\AgdaSymbol{\}} \AgdaSymbol{\{}\AgdaBound{f} \AgdaBound{h} \AgdaSymbol{:} \AgdaBound{B} \AgdaSymbol{→} \AgdaBound{C}\AgdaSymbol{\}} \AgdaSymbol{\{}\AgdaBound{g} \AgdaBound{i} \AgdaSymbol{:} \AgdaBound{A} \AgdaSymbol{→} \AgdaBound{B}\AgdaSymbol{\}} \AgdaSymbol{→}\<%
\\
\>[0]\AgdaIndent{2}{}\<[2]%
\>[2]\AgdaSymbol{(}\AgdaBound{f} \AgdaFunction{∼} \AgdaBound{h}\AgdaSymbol{)} \AgdaSymbol{→} \AgdaSymbol{(}\AgdaBound{g} \AgdaFunction{∼} \AgdaBound{i}\AgdaSymbol{)} \AgdaSymbol{→} \AgdaBound{f} \AgdaFunction{∘} \AgdaBound{g} \AgdaFunction{∼} \AgdaBound{h} \AgdaFunction{∘} \AgdaBound{i}\<%
\\
\>\AgdaFunction{∘-resp-∼} \AgdaSymbol{\{}\AgdaArgument{f} \AgdaSymbol{=} \AgdaBound{f}\AgdaSymbol{\}} \AgdaSymbol{\{}\AgdaArgument{i} \AgdaSymbol{=} \AgdaBound{i}\AgdaSymbol{\}} \AgdaBound{f∼h} \AgdaBound{g∼i} \AgdaBound{x} \AgdaSymbol{=} \AgdaFunction{trans} \AgdaSymbol{(}\AgdaFunction{cong} \AgdaBound{f} \AgdaSymbol{(}\AgdaBound{g∼i} \AgdaBound{x}\AgdaSymbol{))} \AgdaSymbol{(}\AgdaBound{f∼h} \AgdaSymbol{(}\AgdaBound{i} \AgdaBound{x}\AgdaSymbol{))}\<%
\\
\\
\>\AgdaFunction{isEquivalence∼} \AgdaSymbol{:} \AgdaSymbol{∀} \AgdaSymbol{\{}\AgdaBound{ℓ} \AgdaBound{ℓ′}\AgdaSymbol{\}} \AgdaSymbol{\{}\AgdaBound{A} \AgdaSymbol{:} \AgdaPrimitiveType{Set} \AgdaBound{ℓ}\AgdaSymbol{\}} \AgdaSymbol{\{}\AgdaBound{B} \AgdaSymbol{:} \AgdaPrimitiveType{Set} \AgdaBound{ℓ′}\AgdaSymbol{\}} \AgdaSymbol{→} \AgdaRecord{IsEquivalence} \AgdaSymbol{(}\AgdaFunction{\_∼\_} \AgdaSymbol{\{}\AgdaBound{ℓ}\AgdaSymbol{\}} \AgdaSymbol{\{}\AgdaBound{ℓ′}\AgdaSymbol{\}} \AgdaSymbol{\{}\AgdaBound{A}\AgdaSymbol{\}} \AgdaSymbol{\{}\AgdaBound{B}\AgdaSymbol{\})}\<%
\\
\>\AgdaFunction{isEquivalence∼} \AgdaSymbol{=} \AgdaKeyword{record} \AgdaSymbol{\{} \AgdaField{refl} \AgdaSymbol{=} \AgdaFunction{refl∼} \AgdaSymbol{;} \AgdaField{sym} \AgdaSymbol{=} \AgdaFunction{sym∼} \AgdaSymbol{;} \AgdaField{trans} \AgdaSymbol{=} \AgdaFunction{trans∼} \AgdaSymbol{\}}\<%
\\
\\
\>\AgdaComment{------------------------------------------------------------------------------}\<%
\\
\>\AgdaComment{-- Quasi-equivalences, in a more useful packaging}\<%
\\
\\
\>\AgdaKeyword{record} \AgdaRecord{\_≃\_} \AgdaSymbol{\{}\AgdaBound{ℓ} \AgdaBound{ℓ'}\AgdaSymbol{\}} \AgdaSymbol{(}\AgdaBound{A} \AgdaSymbol{:} \AgdaPrimitiveType{Set} \AgdaBound{ℓ}\AgdaSymbol{)} \AgdaSymbol{(}\AgdaBound{B} \AgdaSymbol{:} \AgdaPrimitiveType{Set} \AgdaBound{ℓ'}\AgdaSymbol{)} \AgdaSymbol{:} \AgdaPrimitiveType{Set} \AgdaSymbol{(}\AgdaBound{ℓ} \AgdaPrimitive{⊔} \AgdaBound{ℓ'}\AgdaSymbol{)} \AgdaKeyword{where}\<%
\\
\>[0]\AgdaIndent{2}{}\<[2]%
\>[2]\AgdaKeyword{constructor} \AgdaInductiveConstructor{qeq}\<%
\\
\>[0]\AgdaIndent{2}{}\<[2]%
\>[2]\AgdaKeyword{field}\<%
\\
\>[2]\AgdaIndent{4}{}\<[4]%
\>[4]\AgdaField{f} \AgdaSymbol{:} \AgdaBound{A} \AgdaSymbol{→} \AgdaBound{B}\<%
\\
\>[2]\AgdaIndent{4}{}\<[4]%
\>[4]\AgdaField{g} \AgdaSymbol{:} \AgdaBound{B} \AgdaSymbol{→} \AgdaBound{A}\<%
\\
\>[2]\AgdaIndent{4}{}\<[4]%
\>[4]\AgdaField{α} \AgdaSymbol{:} \AgdaSymbol{(}\AgdaField{f} \AgdaFunction{∘} \AgdaField{g}\AgdaSymbol{)} \AgdaFunction{∼} \AgdaFunction{id}\<%
\\
\>[2]\AgdaIndent{4}{}\<[4]%
\>[4]\AgdaField{β} \AgdaSymbol{:} \AgdaSymbol{(}\AgdaField{g} \AgdaFunction{∘} \AgdaField{f}\AgdaSymbol{)} \AgdaFunction{∼} \AgdaFunction{id}\<%
\\
\>[2]\AgdaIndent{4}{}\<[4]%
\>[4]\AgdaComment{-- to make it contractible, could add}\<%
\\
\>[2]\AgdaIndent{4}{}\<[4]%
\>[4]\AgdaComment{-- τ : ∀ x → cong f (β x) P.≡ α (f x)}\<%
\\
\\
\>\AgdaFunction{id≃} \AgdaSymbol{:} \AgdaSymbol{∀} \AgdaSymbol{\{}\AgdaBound{ℓ}\AgdaSymbol{\}} \AgdaSymbol{\{}\AgdaBound{A} \AgdaSymbol{:} \AgdaPrimitiveType{Set} \AgdaBound{ℓ}\AgdaSymbol{\}} \AgdaSymbol{→} \AgdaBound{A} \AgdaRecord{≃} \AgdaBound{A}\<%
\\
\>\AgdaFunction{id≃} \AgdaSymbol{=} \AgdaInductiveConstructor{qeq} \AgdaFunction{id} \AgdaFunction{id} \AgdaSymbol{(λ} \AgdaBound{\_} \AgdaSymbol{→} \AgdaInductiveConstructor{refl}\AgdaSymbol{)} \AgdaSymbol{(λ} \AgdaBound{\_} \AgdaSymbol{→} \AgdaInductiveConstructor{refl}\AgdaSymbol{)}\<%
\\
\\
\>\AgdaFunction{sym≃} \AgdaSymbol{:} \AgdaSymbol{∀} \AgdaSymbol{\{}\AgdaBound{ℓ} \AgdaBound{ℓ′}\AgdaSymbol{\}} \AgdaSymbol{\{}\AgdaBound{A} \AgdaSymbol{:} \AgdaPrimitiveType{Set} \AgdaBound{ℓ}\AgdaSymbol{\}} \AgdaSymbol{\{}\AgdaBound{B} \AgdaSymbol{:} \AgdaPrimitiveType{Set} \AgdaBound{ℓ′}\AgdaSymbol{\}} \AgdaSymbol{→} \AgdaSymbol{(}\AgdaBound{A} \AgdaRecord{≃} \AgdaBound{B}\AgdaSymbol{)} \AgdaSymbol{→} \AgdaBound{B} \AgdaRecord{≃} \AgdaBound{A}\<%
\\
\>\AgdaFunction{sym≃} \AgdaSymbol{(}\AgdaInductiveConstructor{qeq} \AgdaBound{f} \AgdaBound{g} \AgdaBound{α} \AgdaBound{β}\AgdaSymbol{)} \AgdaSymbol{=} \AgdaInductiveConstructor{qeq} \AgdaBound{g} \AgdaBound{f} \AgdaBound{β} \AgdaBound{α}\<%
\\
\\
\>\AgdaFunction{trans≃} \AgdaSymbol{:} \<[10]%
\>[10]\AgdaSymbol{∀} \AgdaSymbol{\{}\AgdaBound{ℓ} \AgdaBound{ℓ′} \AgdaBound{ℓ″}\AgdaSymbol{\}} \AgdaSymbol{\{}\AgdaBound{A} \AgdaSymbol{:} \AgdaPrimitiveType{Set} \AgdaBound{ℓ}\AgdaSymbol{\}} \AgdaSymbol{\{}\AgdaBound{B} \AgdaSymbol{:} \AgdaPrimitiveType{Set} \AgdaBound{ℓ′}\AgdaSymbol{\}} \AgdaSymbol{\{}\AgdaBound{C} \AgdaSymbol{:} \AgdaPrimitiveType{Set} \AgdaBound{ℓ″}\AgdaSymbol{\}} \AgdaSymbol{→}\<%
\\
\>[0]\AgdaIndent{2}{}\<[2]%
\>[2]\AgdaBound{A} \AgdaRecord{≃} \AgdaBound{B} \AgdaSymbol{→} \AgdaBound{B} \AgdaRecord{≃} \AgdaBound{C} \AgdaSymbol{→} \AgdaBound{A} \AgdaRecord{≃} \AgdaBound{C}\<%
\\
\>\AgdaFunction{trans≃} \AgdaSymbol{\{}\AgdaArgument{A} \AgdaSymbol{=} \AgdaBound{A}\AgdaSymbol{\}} \AgdaSymbol{\{}\AgdaBound{B}\AgdaSymbol{\}} \AgdaSymbol{\{}\AgdaBound{C}\AgdaSymbol{\}} \AgdaSymbol{(}\AgdaInductiveConstructor{qeq} \AgdaBound{f} \AgdaBound{f⁻¹} \AgdaBound{fα} \AgdaBound{fβ}\AgdaSymbol{)} \AgdaSymbol{(}\AgdaInductiveConstructor{qeq} \AgdaBound{g} \AgdaBound{g⁻¹} \AgdaBound{gα} \AgdaBound{gβ}\AgdaSymbol{)} \AgdaSymbol{=}\<%
\\
\>[0]\AgdaIndent{2}{}\<[2]%
\>[2]\AgdaInductiveConstructor{qeq} \AgdaSymbol{(}\AgdaBound{g} \AgdaFunction{∘} \AgdaBound{f}\AgdaSymbol{)} \AgdaSymbol{(}\AgdaBound{f⁻¹} \AgdaFunction{∘} \AgdaBound{g⁻¹}\AgdaSymbol{)} \AgdaSymbol{(λ} \AgdaBound{x} \AgdaSymbol{→} \AgdaFunction{trans} \AgdaSymbol{(}\AgdaFunction{cong} \AgdaBound{g} \AgdaSymbol{(}\AgdaBound{fα} \AgdaSymbol{(}\AgdaBound{g⁻¹} \AgdaBound{x}\AgdaSymbol{)))} \AgdaSymbol{(}\AgdaBound{gα} \AgdaBound{x}\AgdaSymbol{))}\<%
\\
\>[2]\AgdaIndent{26}{}\<[26]%
\>[26]\AgdaSymbol{(λ} \AgdaBound{x} \AgdaSymbol{→} \AgdaFunction{trans} \AgdaSymbol{(}\AgdaFunction{cong} \AgdaBound{f⁻¹} \AgdaSymbol{(}\AgdaBound{gβ} \AgdaSymbol{(}\AgdaBound{f} \AgdaBound{x}\AgdaSymbol{)))} \AgdaSymbol{(}\AgdaBound{fβ} \AgdaBound{x}\AgdaSymbol{))}\<%
\\
\\
\>\AgdaComment{-- more convenient infix version, flipped}\<%
\\
\\
\>\AgdaFunction{\_●\_} \AgdaSymbol{:} \AgdaSymbol{∀} \AgdaSymbol{\{}\AgdaBound{ℓ} \AgdaBound{ℓ′} \AgdaBound{ℓ″}\AgdaSymbol{\}} \AgdaSymbol{\{}\AgdaBound{A} \AgdaSymbol{:} \AgdaPrimitiveType{Set} \AgdaBound{ℓ}\AgdaSymbol{\}} \AgdaSymbol{\{}\AgdaBound{B} \AgdaSymbol{:} \AgdaPrimitiveType{Set} \AgdaBound{ℓ′}\AgdaSymbol{\}} \AgdaSymbol{\{}\AgdaBound{C} \AgdaSymbol{:} \AgdaPrimitiveType{Set} \AgdaBound{ℓ″}\AgdaSymbol{\}} \AgdaSymbol{→}\<%
\\
\>[0]\AgdaIndent{2}{}\<[2]%
\>[2]\AgdaBound{B} \AgdaRecord{≃} \AgdaBound{C} \AgdaSymbol{→} \AgdaBound{A} \AgdaRecord{≃} \AgdaBound{B} \AgdaSymbol{→} \AgdaBound{A} \AgdaRecord{≃} \AgdaBound{C}\<%
\\
\>\AgdaFunction{\_●\_} \AgdaSymbol{=} \AgdaFunction{flip} \AgdaFunction{trans≃}\<%
\\
\\
\>\AgdaFunction{≃IsEquiv} \AgdaSymbol{:} \AgdaRecord{IsEquivalence} \AgdaSymbol{\{}\AgdaPrimitive{Level.suc} \AgdaPrimitive{Level.zero}\AgdaSymbol{\}} \AgdaSymbol{\{}\AgdaPrimitive{Level.zero}\AgdaSymbol{\}} \AgdaSymbol{\{}\AgdaPrimitiveType{Set}\AgdaSymbol{\}} \AgdaRecord{\_≃\_}\<%
\\
\>\AgdaFunction{≃IsEquiv} \AgdaSymbol{=} \AgdaKeyword{record} \AgdaSymbol{\{} \AgdaField{refl} \AgdaSymbol{=} \AgdaFunction{id≃} \AgdaSymbol{;} \AgdaField{sym} \AgdaSymbol{=} \AgdaFunction{sym≃} \AgdaSymbol{;} \AgdaField{trans} \AgdaSymbol{=} \AgdaFunction{trans≃} \AgdaSymbol{\}}\<%
\\
\\
\>\AgdaComment{-- equivalences are injective}\<%
\\
\\
\>\AgdaFunction{inj≃} \AgdaSymbol{:} \AgdaSymbol{∀} \AgdaSymbol{\{}\AgdaBound{ℓ} \AgdaBound{ℓ'}\AgdaSymbol{\}} \AgdaSymbol{\{}\AgdaBound{A} \AgdaSymbol{:} \AgdaPrimitiveType{Set} \AgdaBound{ℓ}\AgdaSymbol{\}} \AgdaSymbol{\{}\AgdaBound{B} \AgdaSymbol{:} \AgdaPrimitiveType{Set} \AgdaBound{ℓ'}\AgdaSymbol{\}} \AgdaSymbol{→}\<%
\\
\>[0]\AgdaIndent{2}{}\<[2]%
\>[2]\AgdaSymbol{(}\AgdaBound{eq} \AgdaSymbol{:} \AgdaBound{A} \AgdaRecord{≃} \AgdaBound{B}\AgdaSymbol{)} \AgdaSymbol{→} \AgdaSymbol{(}\AgdaBound{x} \AgdaBound{y} \AgdaSymbol{:} \AgdaBound{A}\AgdaSymbol{)} \AgdaSymbol{→} \AgdaSymbol{(}\AgdaField{\_≃\_.f} \AgdaBound{eq} \AgdaBound{x} \AgdaDatatype{≡} \AgdaField{\_≃\_.f} \AgdaBound{eq} \AgdaBound{y} \AgdaSymbol{→} \AgdaBound{x} \AgdaDatatype{≡} \AgdaBound{y}\AgdaSymbol{)}\<%
\\
\>\AgdaFunction{inj≃} \AgdaSymbol{(}\AgdaInductiveConstructor{qeq} \AgdaBound{f} \AgdaBound{g} \AgdaBound{α} \AgdaBound{β}\AgdaSymbol{)} \AgdaBound{x} \AgdaBound{y} \AgdaBound{p} \AgdaSymbol{=} \AgdaFunction{trans}\<%
\\
\>[0]\AgdaIndent{2}{}\<[2]%
\>[2]\AgdaSymbol{(}\AgdaFunction{sym} \AgdaSymbol{(}\AgdaBound{β} \AgdaBound{x}\AgdaSymbol{))} \AgdaSymbol{(}\AgdaFunction{trans}\<%
\\
\>[0]\AgdaIndent{2}{}\<[2]%
\>[2]\AgdaSymbol{(}\AgdaFunction{cong} \AgdaBound{g} \AgdaBound{p}\AgdaSymbol{)} \AgdaSymbol{(}\<%
\\
\>[0]\AgdaIndent{2}{}\<[2]%
\>[2]\AgdaBound{β} \AgdaBound{y}\AgdaSymbol{))}\<%
\end{code}

%% file: NumPlus.tex
\AgdaHide{
\begin{code}%
\>\AgdaKeyword{module} \AgdaModule{NumPlus} \AgdaKeyword{where}\<%
\\
\\
\>\AgdaKeyword{data} \AgdaDatatype{ℕ+} \AgdaSymbol{:} \AgdaPrimitiveType{Set₀} \AgdaKeyword{where}\<%
\\
\>[0]\AgdaIndent{2}{}\<[2]%
\>[2]\AgdaInductiveConstructor{z+} \AgdaSymbol{:} \AgdaDatatype{ℕ+}\<%
\\
\>[0]\AgdaIndent{2}{}\<[2]%
\>[2]\AgdaInductiveConstructor{s+} \AgdaSymbol{:} \AgdaDatatype{ℕ+} \AgdaSymbol{→} \AgdaDatatype{ℕ+}\<%
\\
\>[0]\AgdaIndent{2}{}\<[2]%
\>[2]\AgdaInductiveConstructor{\_`+\_} \AgdaSymbol{:} \AgdaDatatype{ℕ+} \AgdaSymbol{→} \AgdaDatatype{ℕ+} \AgdaSymbol{→} \AgdaDatatype{ℕ+}\<%
\\
\\
\>\AgdaKeyword{data} \AgdaDatatype{ℕ+X} \AgdaSymbol{(}\AgdaBound{V} \AgdaSymbol{:} \AgdaPrimitiveType{Set₀}\AgdaSymbol{)} \AgdaSymbol{:} \AgdaPrimitiveType{Set₀} \AgdaKeyword{where}\<%
\\
\>[0]\AgdaIndent{2}{}\<[2]%
\>[2]\AgdaInductiveConstructor{z} \AgdaSymbol{:} \AgdaDatatype{ℕ+X} \AgdaBound{V}\<%
\\
\>[0]\AgdaIndent{2}{}\<[2]%
\>[2]\AgdaInductiveConstructor{s} \AgdaSymbol{:} \AgdaDatatype{ℕ+X} \AgdaBound{V} \AgdaSymbol{→} \AgdaDatatype{ℕ+X} \AgdaBound{V}\<%
\\
\>[0]\AgdaIndent{2}{}\<[2]%
\>[2]\AgdaInductiveConstructor{\_`+\_} \AgdaSymbol{:} \AgdaDatatype{ℕ+X} \AgdaBound{V} \AgdaSymbol{→} \AgdaDatatype{ℕ+X} \AgdaBound{V} \AgdaSymbol{→} \AgdaDatatype{ℕ+X} \AgdaBound{V}\<%
\\
\>[0]\AgdaIndent{2}{}\<[2]%
\>[2]\AgdaInductiveConstructor{v} \AgdaSymbol{:} \AgdaBound{V} \AgdaSymbol{→} \AgdaDatatype{ℕ+X} \AgdaBound{V}\<%
\end{code}
}

%% file: NumPlusTimes.tex
\begin{code}%
\>\AgdaKeyword{module} \AgdaModule{NumPlusTimes} \AgdaKeyword{where}\<%
\\
\\
\>\AgdaKeyword{data} \AgdaDatatype{ℕ*} \AgdaSymbol{(}\AgdaBound{V} \AgdaSymbol{:} \AgdaPrimitiveType{Set₀}\AgdaSymbol{)} \AgdaSymbol{:} \AgdaPrimitiveType{Set₀} \AgdaKeyword{where}\<%
\\
\>[0]\AgdaIndent{2}{}\<[2]%
\>[2]\AgdaInductiveConstructor{z} \AgdaSymbol{:} \AgdaDatatype{ℕ*} \AgdaBound{V}\<%
\\
\>[0]\AgdaIndent{2}{}\<[2]%
\>[2]\AgdaInductiveConstructor{s} \AgdaSymbol{:} \AgdaDatatype{ℕ*} \AgdaBound{V} \AgdaSymbol{→} \AgdaDatatype{ℕ*} \AgdaBound{V}\<%
\\
\>[0]\AgdaIndent{2}{}\<[2]%
\>[2]\AgdaInductiveConstructor{\_`+\_} \AgdaSymbol{:} \AgdaDatatype{ℕ*} \AgdaBound{V} \AgdaSymbol{→} \AgdaDatatype{ℕ*} \AgdaBound{V} \AgdaSymbol{→} \AgdaDatatype{ℕ*} \AgdaBound{V}\<%
\\
\>[0]\AgdaIndent{2}{}\<[2]%
\>[2]\AgdaInductiveConstructor{\_`*\_} \AgdaSymbol{:} \AgdaDatatype{ℕ*} \AgdaBound{V} \AgdaSymbol{→} \AgdaDatatype{ℕ*} \AgdaBound{V} \AgdaSymbol{→} \AgdaDatatype{ℕ*} \AgdaBound{V}\<%
\\
\>[0]\AgdaIndent{2}{}\<[2]%
\>[2]\AgdaInductiveConstructor{v} \AgdaSymbol{:} \AgdaBound{V} \AgdaSymbol{→} \AgdaDatatype{ℕ*} \AgdaBound{V}\<%
\end{code}

%% file: Variables.tex
\begin{code}%
\>\AgdaKeyword{module} \AgdaModule{Variables} \AgdaKeyword{where}\<%
\\
\\
\>\AgdaKeyword{open} \AgdaKeyword{import} \AgdaModule{Level} \AgdaKeyword{using} \AgdaSymbol{(}\AgdaPostulate{Level}\AgdaSymbol{;} \AgdaPrimitive{zero}\AgdaSymbol{;} \AgdaPrimitive{suc}\AgdaSymbol{)}\<%
\\
\>\AgdaKeyword{open} \AgdaKeyword{import} \AgdaModule{Relation.Binary} \AgdaKeyword{using} \AgdaSymbol{(}\AgdaRecord{DecSetoid}\AgdaSymbol{)}\<%
\\
\>\AgdaKeyword{open} \AgdaKeyword{import} \AgdaModule{Relation.Nullary} \AgdaKeyword{using} \AgdaSymbol{(}\AgdaDatatype{Dec}\AgdaSymbol{;} \AgdaInductiveConstructor{yes}\AgdaSymbol{;} \AgdaInductiveConstructor{no}\AgdaSymbol{)}\<%
\\
\>\AgdaKeyword{open} \AgdaKeyword{import} \AgdaModule{Data.Bool} \AgdaKeyword{using} \AgdaSymbol{(}\AgdaDatatype{Bool}\AgdaSymbol{)} \AgdaKeyword{renaming} \AgdaSymbol{(}\AgdaFunction{\_≟\_} \AgdaSymbol{to} \AgdaFunction{\_=𝔹\_}\AgdaSymbol{)}\<%
\\
\>\AgdaKeyword{open} \AgdaKeyword{import} \AgdaModule{Relation.Binary.PropositionalEquality}\<%
\\
\>[0]\AgdaIndent{2}{}\<[2]%
\>[2]\AgdaKeyword{using} \AgdaSymbol{(}\AgdaDatatype{\_≡\_}\AgdaSymbol{;} \AgdaInductiveConstructor{refl}\AgdaSymbol{;} \AgdaFunction{sym}\AgdaSymbol{;} \AgdaFunction{trans}\AgdaSymbol{)}\<%
\\
\>\AgdaKeyword{open} \AgdaKeyword{import} \AgdaModule{Data.Empty} \AgdaKeyword{using} \AgdaSymbol{(}\AgdaDatatype{⊥}\AgdaSymbol{)}\<%
\\
\>\AgdaKeyword{open} \AgdaKeyword{import} \AgdaModule{Data.Unit} \AgdaKeyword{using} \AgdaSymbol{(}\AgdaRecord{⊤}\AgdaSymbol{;} \AgdaInductiveConstructor{tt}\AgdaSymbol{)}\<%
\\
\\
\>\AgdaKeyword{private}\<%
\\
\>[0]\AgdaIndent{2}{}\<[2]%
\>[2]\AgdaFunction{DT} \AgdaSymbol{:} \AgdaPrimitiveType{Set} \AgdaSymbol{(}\AgdaPrimitive{suc} \AgdaPrimitive{zero}\AgdaSymbol{)}\<%
\\
\>[0]\AgdaIndent{2}{}\<[2]%
\>[2]\AgdaFunction{DT} \AgdaSymbol{=} \AgdaRecord{DecSetoid} \AgdaPrimitive{zero} \AgdaPrimitive{zero}\<%
\\
\\
\>\AgdaFunction{NoVars} \AgdaSymbol{:} \AgdaFunction{DT}\<%
\\
\>\AgdaFunction{NoVars} \AgdaSymbol{=} \AgdaKeyword{record}\<%
\\
\>[0]\AgdaIndent{2}{}\<[2]%
\>[2]\AgdaSymbol{\{} \AgdaField{Carrier} \AgdaSymbol{=} \AgdaDatatype{⊥}\<%
\\
\>[0]\AgdaIndent{2}{}\<[2]%
\>[2]\AgdaSymbol{;} \AgdaField{\_≈\_} \AgdaSymbol{=} \AgdaSymbol{λ} \AgdaBound{\_} \AgdaBound{\_} \AgdaSymbol{→} \AgdaRecord{⊤}\<%
\\
\>[0]\AgdaIndent{2}{}\<[2]%
\>[2]\AgdaSymbol{;} \AgdaField{isDecEquivalence} \AgdaSymbol{=} \AgdaKeyword{record}\<%
\\
\>[2]\AgdaIndent{4}{}\<[4]%
\>[4]\AgdaSymbol{\{} \AgdaField{isEquivalence} \AgdaSymbol{=} \AgdaKeyword{record} \AgdaSymbol{\{} \AgdaField{refl} \AgdaSymbol{=} \AgdaInductiveConstructor{tt} \AgdaSymbol{;} \AgdaField{sym} \AgdaSymbol{=} \AgdaSymbol{λ} \AgdaBound{\_} \AgdaSymbol{→} \AgdaInductiveConstructor{tt} \AgdaSymbol{;} \AgdaField{trans} \AgdaSymbol{=} \AgdaSymbol{λ} \AgdaBound{\_} \AgdaBound{\_} \AgdaSymbol{→} \AgdaInductiveConstructor{tt} \AgdaSymbol{\}}\<%
\\
\>[2]\AgdaIndent{4}{}\<[4]%
\>[4]\AgdaSymbol{;} \AgdaField{\_≟\_} \AgdaSymbol{=} \AgdaSymbol{λ} \AgdaSymbol{()} \AgdaSymbol{\}} \AgdaSymbol{\}}\<%
\\
\\
\>\AgdaFunction{DBool} \AgdaSymbol{:} \AgdaFunction{DT}\<%
\\
\>\AgdaFunction{DBool} \AgdaSymbol{=} \AgdaKeyword{record}\<%
\\
\>[0]\AgdaIndent{2}{}\<[2]%
\>[2]\AgdaSymbol{\{} \AgdaField{Carrier} \AgdaSymbol{=} \AgdaDatatype{Bool}\<%
\\
\>[0]\AgdaIndent{2}{}\<[2]%
\>[2]\AgdaSymbol{;} \AgdaField{\_≈\_} \AgdaSymbol{=} \AgdaDatatype{\_≡\_}\<%
\\
\>[0]\AgdaIndent{2}{}\<[2]%
\>[2]\AgdaSymbol{;} \AgdaField{isDecEquivalence} \AgdaSymbol{=} \AgdaKeyword{record}\<%
\\
\>[2]\AgdaIndent{4}{}\<[4]%
\>[4]\AgdaSymbol{\{} \AgdaField{isEquivalence} \AgdaSymbol{=} \AgdaKeyword{record} \AgdaSymbol{\{} \AgdaField{refl} \AgdaSymbol{=} \AgdaInductiveConstructor{refl} \AgdaSymbol{;} \AgdaField{sym} \AgdaSymbol{=} \AgdaFunction{sym} \AgdaSymbol{;} \AgdaField{trans} \AgdaSymbol{=} \AgdaFunction{trans} \AgdaSymbol{\}}\<%
\\
\>[2]\AgdaIndent{4}{}\<[4]%
\>[4]\AgdaSymbol{;} \AgdaField{\_≟\_} \AgdaSymbol{=} \AgdaFunction{\_=𝔹\_} \AgdaSymbol{\}} \AgdaSymbol{\}} \<[22]%
\>[22]\<%
\\
\>[2]\AgdaIndent{4}{}\<[4]%
\>[4]\<%
\\
\>\AgdaComment{-- For convenience, some simple "languages of variables"}\<%
\\
\>\AgdaKeyword{module} \AgdaModule{VarLangs} \AgdaKeyword{where}\<%
\\
\>[0]\AgdaIndent{2}{}\<[2]%
\>[2]\AgdaKeyword{data} \AgdaDatatype{X} \AgdaSymbol{:} \AgdaPrimitiveType{Set₀} \AgdaKeyword{where} \AgdaInductiveConstructor{x} \AgdaSymbol{:} \AgdaDatatype{X}\<%
\\
\>[0]\AgdaIndent{2}{}\<[2]%
\>[2]\AgdaFunction{XV} \AgdaSymbol{:} \AgdaFunction{DT}\<%
\\
\>[0]\AgdaIndent{2}{}\<[2]%
\>[2]\AgdaFunction{XV} \AgdaSymbol{=} \AgdaKeyword{record}\<%
\\
\>[2]\AgdaIndent{4}{}\<[4]%
\>[4]\AgdaSymbol{\{} \AgdaField{Carrier} \AgdaSymbol{=} \AgdaDatatype{X}\<%
\\
\>[2]\AgdaIndent{4}{}\<[4]%
\>[4]\AgdaSymbol{;} \AgdaField{\_≈\_} \AgdaSymbol{=} \AgdaSymbol{λ} \AgdaBound{\_} \AgdaBound{\_} \AgdaSymbol{→} \AgdaRecord{⊤}\<%
\\
\>[2]\AgdaIndent{4}{}\<[4]%
\>[4]\AgdaSymbol{;} \AgdaField{isDecEquivalence} \AgdaSymbol{=} \AgdaKeyword{record}\<%
\\
\>[4]\AgdaIndent{6}{}\<[6]%
\>[6]\AgdaSymbol{\{} \AgdaField{isEquivalence} \AgdaSymbol{=} \AgdaKeyword{record}\<%
\\
\>[6]\AgdaIndent{8}{}\<[8]%
\>[8]\AgdaSymbol{\{} \AgdaField{refl} \AgdaSymbol{=} \AgdaInductiveConstructor{tt}\<%
\\
\>[6]\AgdaIndent{8}{}\<[8]%
\>[8]\AgdaSymbol{;} \AgdaField{sym} \AgdaSymbol{=} \AgdaSymbol{λ} \AgdaBound{\_} \AgdaSymbol{→} \AgdaInductiveConstructor{tt}\<%
\\
\>[6]\AgdaIndent{8}{}\<[8]%
\>[8]\AgdaSymbol{;} \AgdaField{trans} \AgdaSymbol{=} \AgdaSymbol{λ} \AgdaBound{\_} \AgdaBound{\_} \AgdaSymbol{→} \AgdaInductiveConstructor{tt} \AgdaSymbol{\}}\<%
\\
\>[0]\AgdaIndent{6}{}\<[6]%
\>[6]\AgdaSymbol{;} \AgdaField{\_≟\_} \AgdaSymbol{=} \AgdaSymbol{λ} \AgdaBound{\_} \AgdaBound{\_} \AgdaSymbol{→} \AgdaInductiveConstructor{yes} \AgdaInductiveConstructor{tt} \AgdaSymbol{\}} \AgdaSymbol{\}}\<%
\end{code}

%% file: T2.tex
\AgdaHide{
\begin{code}%
\>\AgdaKeyword{module} \AgdaModule{T2} \AgdaKeyword{where}\<%
\\
\>\AgdaKeyword{open} \AgdaKeyword{import} \AgdaModule{T1} \AgdaKeyword{using} \AgdaSymbol{(}\AgdaRecord{BT₁}\AgdaSymbol{)}\<%
\\
\>\AgdaKeyword{open} \AgdaKeyword{import} \AgdaModule{Numerals}\<%
\\
\>\AgdaKeyword{open} \AgdaKeyword{import} \AgdaModule{NumPlus}\<%
\\
\>\AgdaKeyword{open} \AgdaKeyword{import} \AgdaModule{NatVar} \AgdaKeyword{using} \AgdaSymbol{(}\AgdaDatatype{ℕX}\AgdaSymbol{)}\<%
\\
\\
\>\AgdaKeyword{open} \AgdaKeyword{import} \AgdaModule{Relation.Binary} \AgdaKeyword{using} \AgdaSymbol{(}\AgdaRecord{DecSetoid}\AgdaSymbol{)}\<%
\\
\>\AgdaKeyword{open} \AgdaModule{DecSetoid} \AgdaKeyword{using} \AgdaSymbol{(}\AgdaField{Carrier}\AgdaSymbol{)}\<%
\\
\>\AgdaKeyword{open} \AgdaKeyword{import} \AgdaModule{Level} \AgdaKeyword{using} \AgdaSymbol{()} \AgdaKeyword{renaming} \AgdaSymbol{(}\AgdaPrimitive{zero} \AgdaSymbol{to} \AgdaPrimitive{lzero}\AgdaSymbol{)}\<%
\\
\\
\>\AgdaKeyword{open} \AgdaKeyword{import} \AgdaModule{Relation.Binary.PropositionalEquality}\<%
\\
\>[0]\AgdaIndent{2}{}\<[2]%
\>[2]\AgdaKeyword{using} \AgdaSymbol{(}\AgdaDatatype{\_≡\_}\AgdaSymbol{;} \AgdaInductiveConstructor{refl}\AgdaSymbol{;} \AgdaFunction{trans}\AgdaSymbol{;} \AgdaFunction{cong}\AgdaSymbol{;} \AgdaFunction{sym}\AgdaSymbol{;} \AgdaFunction{cong₂}\AgdaSymbol{)}\<%
\\
\>\AgdaKeyword{open} \AgdaKeyword{import} \AgdaModule{Data.Nat} \AgdaKeyword{using} \AgdaSymbol{(}\AgdaDatatype{ℕ}\AgdaSymbol{;} \AgdaInductiveConstructor{suc}\AgdaSymbol{)} \AgdaComment{-- instead of defining our own}\<%
\\
\>[0]\AgdaIndent{2}{}\<[2]%
\>[2]\AgdaComment{-- isomorphic copy}\<%
\\
\>\AgdaKeyword{open} \AgdaKeyword{import} \AgdaModule{Data.Vec} \AgdaKeyword{using} \AgdaSymbol{(}\AgdaInductiveConstructor{\_∷\_}\AgdaSymbol{;} \AgdaInductiveConstructor{[]}\AgdaSymbol{;} \AgdaDatatype{Vec}\AgdaSymbol{)}\<%
\\
\>\AgdaKeyword{open} \AgdaKeyword{import} \AgdaModule{Language} \AgdaKeyword{using} \AgdaSymbol{(}\AgdaRecord{GroundLanguage}\AgdaSymbol{;} \AgdaKeyword{module} \AgdaModule{FOL}\AgdaSymbol{)}\<%
\\
\>\AgdaKeyword{open} \AgdaModule{GroundLanguage} \AgdaKeyword{using} \AgdaSymbol{(}\AgdaField{value}\AgdaSymbol{)}\<%
\\
\\
\>\AgdaKeyword{private}\<%
\\
\>[0]\AgdaIndent{2}{}\<[2]%
\>[2]\AgdaFunction{DT} \AgdaSymbol{=} \AgdaRecord{DecSetoid} \AgdaPrimitive{lzero} \AgdaPrimitive{lzero}\<%
\\
\>\AgdaComment{-------------------------------------------------------------------}\<%
\\
\>[2]\AgdaIndent{4}{}\<[4]%
\>[4]\<%
\\
\>\AgdaKeyword{record} \AgdaRecord{BT₂} \AgdaSymbol{(}\AgdaBound{t1} \AgdaSymbol{:} \AgdaRecord{BT₁}\AgdaSymbol{)} \AgdaSymbol{:} \AgdaPrimitiveType{Set} \AgdaKeyword{where}\<%
\\
\>[0]\AgdaIndent{2}{}\<[2]%
\>[2]\AgdaKeyword{open} \AgdaModule{BT₁} \AgdaBound{t1} \AgdaKeyword{public}\<%
\\
\>[0]\AgdaIndent{2}{}\<[2]%
\>[2]\AgdaKeyword{field}\<%
\\
\>[2]\AgdaIndent{4}{}\<[4]%
\>[4]\AgdaField{\_+\_} \AgdaSymbol{:} \AgdaFunction{nat} \AgdaSymbol{→} \AgdaFunction{nat} \AgdaSymbol{→} \AgdaFunction{nat}\<%
\\
\>[2]\AgdaIndent{4}{}\<[4]%
\>[4]\AgdaField{right-0} \AgdaSymbol{:} \AgdaSymbol{∀} \AgdaBound{x} \AgdaSymbol{→} \AgdaBound{x} \AgdaField{+} \AgdaFunction{Z} \AgdaDatatype{≡} \AgdaBound{x}\<%
\\
\>[2]\AgdaIndent{4}{}\<[4]%
\>[4]\AgdaField{x+Sy≡Sx+y} \AgdaSymbol{:} \AgdaSymbol{∀} \AgdaBound{x} \AgdaBound{y} \AgdaSymbol{→} \AgdaBound{x} \AgdaField{+} \AgdaFunction{S} \AgdaBound{y} \AgdaDatatype{≡} \AgdaFunction{S} \AgdaSymbol{(}\AgdaBound{x} \AgdaField{+} \AgdaBound{y}\AgdaSymbol{)}\<%
\\
\>[2]\AgdaIndent{4}{}\<[4]%
\>[4]\AgdaComment{-- Wikipedia lists}\<%
\\
\>[2]\AgdaIndent{4}{}\<[4]%
\>[4]\AgdaComment{-- y ≡ 0 ⊎ Σ ℕ (λ x → S x ≡ y)}\<%
\\
\>[2]\AgdaIndent{4}{}\<[4]%
\>[4]\AgdaComment{-- as an additional axiom.  It allows addition}\<%
\\
\>[2]\AgdaIndent{4}{}\<[4]%
\>[4]\AgdaComment{-- to be defined recursively.}\<%
\\
\\
\>[0]\AgdaIndent{2}{}\<[2]%
\>[2]\AgdaFunction{bnat} \AgdaSymbol{:} \AgdaFunction{nat} \AgdaSymbol{→} \AgdaFunction{nat} \AgdaSymbol{→} \AgdaFunction{nat}\<%
\\
\>[0]\AgdaIndent{2}{}\<[2]%
\>[2]\AgdaFunction{bnat} \AgdaBound{x} \AgdaBound{y} \AgdaSymbol{=} \AgdaSymbol{(}\AgdaBound{x} \AgdaField{+} \AgdaBound{x}\AgdaSymbol{)} \AgdaField{+} \AgdaBound{y}\<%
\\
\\
\>[0]\AgdaIndent{2}{}\<[2]%
\>[2]\AgdaComment{-- the following two functions are not (unfortunately)}\<%
\\
\>[0]\AgdaIndent{2}{}\<[2]%
\>[2]\AgdaComment{-- private, as T2a will need to prove things about them.}\<%
\\
\>[0]\AgdaIndent{2}{}\<[2]%
\>[2]\AgdaFunction{dig-to-nat} \AgdaSymbol{:} \AgdaDatatype{BinDigit} \AgdaSymbol{→} \AgdaFunction{nat}\<%
\\
\>[0]\AgdaIndent{2}{}\<[2]%
\>[2]\AgdaFunction{dig-to-nat} \AgdaInductiveConstructor{zero} \AgdaSymbol{=} \AgdaFunction{Z}\<%
\\
\>[0]\AgdaIndent{2}{}\<[2]%
\>[2]\AgdaFunction{dig-to-nat} \AgdaInductiveConstructor{one} \AgdaSymbol{=} \AgdaFunction{S} \AgdaFunction{Z}\<%
\\
\\
\>[0]\AgdaIndent{2}{}\<[2]%
\>[2]\AgdaFunction{unroll} \AgdaSymbol{:} \AgdaSymbol{\{}\AgdaBound{n} \AgdaSymbol{:} \AgdaDatatype{ℕ}\AgdaSymbol{\}} \AgdaSymbol{→} \AgdaDatatype{Vec} \AgdaDatatype{BinDigit} \AgdaBound{n} \AgdaSymbol{→} \AgdaFunction{nat}\<%
\\
\>[0]\AgdaIndent{2}{}\<[2]%
\>[2]\AgdaFunction{unroll} \AgdaInductiveConstructor{[]} \AgdaSymbol{=} \AgdaFunction{Z}\<%
\\
\>[0]\AgdaIndent{2}{}\<[2]%
\>[2]\AgdaFunction{unroll} \AgdaSymbol{(}\AgdaBound{x} \AgdaInductiveConstructor{∷} \AgdaBound{l}\AgdaSymbol{)} \AgdaSymbol{=} \AgdaFunction{bnat} \AgdaSymbol{(}\AgdaFunction{unroll} \AgdaBound{l}\AgdaSymbol{)} \AgdaSymbol{(}\AgdaFunction{dig-to-nat} \AgdaBound{x}\AgdaSymbol{)}\<%
\\
\\
\>[0]\AgdaIndent{2}{}\<[2]%
\>[2]\AgdaFunction{⟦\_⟧₂} \AgdaSymbol{:} \AgdaDatatype{BNum} \AgdaSymbol{→} \AgdaFunction{nat}\<%
\\
\>[0]\AgdaIndent{2}{}\<[2]%
\>[2]\AgdaFunction{⟦} \AgdaInductiveConstructor{bn} \AgdaSymbol{(}\AgdaBound{x} \AgdaInductiveConstructor{∷} \AgdaBound{l}\AgdaSymbol{)} \AgdaFunction{⟧₂} \AgdaSymbol{=} \AgdaFunction{bnat} \AgdaSymbol{(}\AgdaFunction{unroll} \AgdaBound{l}\AgdaSymbol{)} \AgdaSymbol{(}\AgdaFunction{dig-to-nat} \AgdaBound{x}\AgdaSymbol{)}\<%
\\
\\
\>[0]\AgdaIndent{2}{}\<[2]%
\>[2]\AgdaComment{-- just to make sure we've done things right.}\<%
\\
\>[0]\AgdaIndent{2}{}\<[2]%
\>[2]\AgdaFunction{lemma₁} \AgdaSymbol{:} \AgdaFunction{⟦} \AgdaFunction{0b} \AgdaFunction{⟧₂} \AgdaDatatype{≡} \AgdaFunction{Z}\<%
\\
\>[0]\AgdaIndent{2}{}\<[2]%
\>[2]\AgdaFunction{lemma₁} \AgdaSymbol{=} \AgdaFunction{trans} \AgdaSymbol{(}\AgdaField{right-0} \AgdaSymbol{\_)} \AgdaSymbol{(}\AgdaField{right-0} \AgdaFunction{Z}\AgdaSymbol{)}\<%
\\
\\
\>[0]\AgdaIndent{2}{}\<[2]%
\>[2]\AgdaFunction{lemma₂} \AgdaSymbol{:} \AgdaFunction{⟦} \AgdaFunction{1b} \AgdaFunction{⟧₂} \AgdaDatatype{≡} \AgdaFunction{S} \AgdaFunction{Z}\<%
\\
\>[0]\AgdaIndent{2}{}\<[2]%
\>[2]\AgdaFunction{lemma₂} \AgdaSymbol{=} \AgdaFunction{trans} \AgdaSymbol{(}\AgdaFunction{cong} \AgdaSymbol{(λ} \AgdaBound{z} \AgdaSymbol{→} \AgdaBound{z} \AgdaField{+} \AgdaFunction{S} \AgdaFunction{Z}\AgdaSymbol{)} \AgdaSymbol{(}\AgdaField{right-0} \AgdaFunction{Z}\AgdaSymbol{))} \AgdaSymbol{(}\<%
\\
\>[2]\AgdaIndent{11}{}\<[11]%
\>[11]\AgdaFunction{trans} \AgdaSymbol{(}\AgdaField{x+Sy≡Sx+y} \AgdaFunction{Z} \AgdaFunction{Z}\AgdaSymbol{)}\<%
\\
\>[11]\AgdaIndent{17}{}\<[17]%
\>[17]\AgdaSymbol{(}\AgdaFunction{cong} \AgdaFunction{S} \AgdaSymbol{(}\AgdaField{right-0} \AgdaFunction{Z}\AgdaSymbol{)))}\<%
\\
\\
\>[0]\AgdaIndent{2}{}\<[2]%
\>[2]\AgdaComment{-- two coherence theorems are provable here (<< x is x + x and + 1 on the right is S)}\<%
\\
\>[0]\AgdaIndent{2}{}\<[2]%
\>[2]\AgdaFunction{<<-is-*2} \AgdaSymbol{:} \AgdaSymbol{∀} \AgdaBound{x} \AgdaSymbol{→} \AgdaFunction{⟦} \AgdaFunction{<<} \AgdaBound{x} \AgdaFunction{⟧₂} \AgdaDatatype{≡} \AgdaFunction{⟦} \AgdaBound{x} \AgdaFunction{⟧₂} \AgdaField{+} \AgdaFunction{⟦} \AgdaBound{x} \AgdaFunction{⟧₂}\<%
\\
\>[0]\AgdaIndent{2}{}\<[2]%
\>[2]\AgdaFunction{<<-is-*2} \AgdaSymbol{(}\AgdaInductiveConstructor{bn} \AgdaSymbol{(}\AgdaBound{x} \AgdaInductiveConstructor{∷} \AgdaBound{x₁}\AgdaSymbol{))} \AgdaSymbol{=} \AgdaKeyword{let} \AgdaBound{num} \AgdaSymbol{=} \AgdaFunction{⟦} \AgdaInductiveConstructor{bn} \AgdaSymbol{(}\AgdaBound{x} \AgdaInductiveConstructor{∷} \AgdaBound{x₁}\AgdaSymbol{)} \AgdaFunction{⟧₂} \AgdaKeyword{in} \AgdaField{right-0} \AgdaSymbol{(}\AgdaBound{num} \AgdaField{+} \AgdaBound{num}\AgdaSymbol{)}\<%
\\
\\
\>[0]\AgdaIndent{2}{}\<[2]%
\>[2]\AgdaFunction{x+1} \AgdaSymbol{:} \AgdaSymbol{∀} \AgdaBound{x} \AgdaSymbol{→} \AgdaFunction{S} \AgdaBound{x} \AgdaDatatype{≡} \AgdaBound{x} \AgdaField{+} \AgdaFunction{⟦} \AgdaFunction{1b} \AgdaFunction{⟧₂}\<%
\\
\>[0]\AgdaIndent{2}{}\<[2]%
\>[2]\AgdaFunction{x+1} \AgdaBound{x} \AgdaSymbol{=} \AgdaFunction{sym} \AgdaSymbol{(}\AgdaFunction{trans} \AgdaSymbol{(}\AgdaFunction{cong} \AgdaSymbol{(λ} \AgdaBound{z} \AgdaSymbol{→} \AgdaBound{x} \AgdaField{+} \AgdaBound{z}\AgdaSymbol{)} \AgdaFunction{lemma₂}\AgdaSymbol{)} \AgdaSymbol{(}\<%
\\
\>[2]\AgdaIndent{15}{}\<[15]%
\>[15]\AgdaFunction{trans} \AgdaSymbol{(}\AgdaField{x+Sy≡Sx+y} \AgdaBound{x} \AgdaFunction{Z}\AgdaSymbol{)} \AgdaSymbol{(}\<%
\\
\>[2]\AgdaIndent{15}{}\<[15]%
\>[15]\AgdaFunction{cong} \AgdaFunction{S} \AgdaSymbol{(}\AgdaField{right-0} \AgdaBound{x}\AgdaSymbol{))))}\<%
\\
\\
\>[0]\AgdaIndent{2}{}\<[2]%
\>[2]\AgdaFunction{nat+-lang} \AgdaSymbol{:} \AgdaRecord{GroundLanguage} \AgdaFunction{nat}\<%
\\
\>[0]\AgdaIndent{2}{}\<[2]%
\>[2]\AgdaFunction{nat+-lang} \AgdaSymbol{=} \AgdaKeyword{record} \AgdaSymbol{\{} \AgdaField{Lang} \AgdaSymbol{=} \AgdaSymbol{λ} \AgdaBound{X} \AgdaSymbol{→} \AgdaDatatype{ℕ+X} \AgdaSymbol{(}\AgdaField{Carrier} \AgdaBound{X}\AgdaSymbol{)}\<%
\\
\>[2]\AgdaIndent{21}{}\<[21]%
\>[21]\AgdaSymbol{;} \AgdaField{value} \AgdaSymbol{=} \AgdaSymbol{λ} \AgdaSymbol{\{}\AgdaBound{V}\AgdaSymbol{\}} \AgdaSymbol{→} \AgdaFunction{val} \AgdaSymbol{\{}\AgdaBound{V}\AgdaSymbol{\}} \AgdaSymbol{\}}\<%
\\
\>[0]\AgdaIndent{3}{}\<[3]%
\>[3]\AgdaKeyword{where}\<%
\\
\>[3]\AgdaIndent{4}{}\<[4]%
\>[4]\AgdaFunction{val} \AgdaSymbol{:} \AgdaSymbol{\{}\AgdaBound{V} \AgdaSymbol{:} \AgdaFunction{DT}\AgdaSymbol{\}} \AgdaSymbol{→} \AgdaDatatype{ℕ+X} \AgdaSymbol{(}\AgdaField{Carrier} \AgdaBound{V}\AgdaSymbol{)} \AgdaSymbol{→} \AgdaSymbol{(}\AgdaField{Carrier} \AgdaBound{V} \AgdaSymbol{→} \AgdaFunction{nat}\AgdaSymbol{)} \AgdaSymbol{→} \AgdaFunction{nat}\<%
\\
\>[3]\AgdaIndent{4}{}\<[4]%
\>[4]\AgdaFunction{val} \<[12]%
\>[12]\AgdaInductiveConstructor{z} \<[21]%
\>[21]\AgdaBound{env} \AgdaSymbol{=} \AgdaFunction{Z}\<%
\\
\>[3]\AgdaIndent{4}{}\<[4]%
\>[4]\AgdaFunction{val} \AgdaSymbol{\{}\AgdaBound{V}\AgdaSymbol{\}} \AgdaSymbol{(}\AgdaInductiveConstructor{s} \AgdaBound{n}\AgdaSymbol{)} \<[21]%
\>[21]\AgdaBound{env} \AgdaSymbol{=} \AgdaFunction{S} \AgdaSymbol{(}\AgdaFunction{val} \AgdaSymbol{\{}\AgdaBound{V}\AgdaSymbol{\}} \AgdaBound{n} \AgdaBound{env}\AgdaSymbol{)}\<%
\\
\>[3]\AgdaIndent{4}{}\<[4]%
\>[4]\AgdaFunction{val} \AgdaSymbol{\{}\AgdaBound{V}\AgdaSymbol{\}} \AgdaSymbol{(}\AgdaBound{e} \AgdaInductiveConstructor{`+} \AgdaBound{f}\AgdaSymbol{)} \AgdaBound{env} \AgdaSymbol{=} \AgdaFunction{val} \AgdaSymbol{\{}\AgdaBound{V}\AgdaSymbol{\}} \AgdaBound{e} \AgdaBound{env} \AgdaField{+} \AgdaFunction{val} \AgdaSymbol{\{}\AgdaBound{V}\AgdaSymbol{\}} \AgdaBound{f} \AgdaBound{env}\<%
\\
\>[3]\AgdaIndent{4}{}\<[4]%
\>[4]\AgdaFunction{val} \<[12]%
\>[12]\AgdaSymbol{(}\AgdaInductiveConstructor{v} \AgdaBound{x}\AgdaSymbol{)} \<[21]%
\>[21]\AgdaBound{env} \AgdaSymbol{=} \AgdaBound{env} \AgdaBound{x}\<%
\\
\\
\>[0]\AgdaIndent{2}{}\<[2]%
\>[2]\AgdaKeyword{module} \AgdaModule{fo₂} \AgdaSymbol{=} \AgdaModule{FOL} \AgdaFunction{nat+-lang}\<%
\\
\\
\>[0]\AgdaIndent{2}{}\<[2]%
\>[2]\AgdaComment{-- we can inject ℕX into ℕ+X}\<%
\\
\>[0]\AgdaIndent{2}{}\<[2]%
\>[2]\AgdaFunction{inject1⇒2} \AgdaSymbol{:} \AgdaSymbol{∀\{}\AgdaBound{V}\AgdaSymbol{\}} \AgdaSymbol{→} \AgdaDatatype{ℕX} \AgdaBound{V} \AgdaSymbol{→} \AgdaDatatype{ℕ+X} \AgdaBound{V}\<%
\\
\>[0]\AgdaIndent{2}{}\<[2]%
\>[2]\AgdaFunction{inject1⇒2} \AgdaInductiveConstructor{ℕX.z} \<[21]%
\>[21]\AgdaSymbol{=} \AgdaInductiveConstructor{z}\<%
\\
\>[0]\AgdaIndent{2}{}\<[2]%
\>[2]\AgdaFunction{inject1⇒2} \AgdaSymbol{(}\AgdaInductiveConstructor{ℕX.s} \AgdaBound{t}\AgdaSymbol{)} \AgdaSymbol{=} \AgdaInductiveConstructor{s} \AgdaSymbol{(}\AgdaFunction{inject1⇒2} \AgdaBound{t}\AgdaSymbol{)}\<%
\\
\>[0]\AgdaIndent{2}{}\<[2]%
\>[2]\AgdaFunction{inject1⇒2} \AgdaSymbol{(}\AgdaInductiveConstructor{ℕX.v} \AgdaBound{x}\AgdaSymbol{)} \AgdaSymbol{=} \AgdaInductiveConstructor{v} \AgdaBound{x}\<%
\end{code}
}

%% file: T3.tex
\begin{code}%
\>\AgdaKeyword{module} \AgdaModule{T3} \AgdaKeyword{where}\<%
\\
\>\AgdaKeyword{open} \AgdaKeyword{import} \AgdaModule{T1} \AgdaKeyword{using} \AgdaSymbol{(}\AgdaRecord{BT₁}\AgdaSymbol{)}\<%
\\
\>\AgdaKeyword{open} \AgdaKeyword{import} \AgdaModule{T2} \AgdaKeyword{using} \AgdaSymbol{(}\AgdaRecord{BT₂}\AgdaSymbol{)}\<%
\\
\>\AgdaKeyword{open} \AgdaKeyword{import} \AgdaModule{Numerals} \AgdaKeyword{using} \AgdaSymbol{(}\AgdaFunction{\_btimes\_}\AgdaSymbol{)}\<%
\\
\>\AgdaKeyword{open} \AgdaKeyword{import} \AgdaModule{NumPlusTimes}\<%
\\
\>\AgdaKeyword{open} \AgdaKeyword{import} \AgdaModule{Language} \AgdaKeyword{using} \AgdaSymbol{(}\AgdaRecord{GroundLanguage}\AgdaSymbol{;} \AgdaKeyword{module} \AgdaModule{FOL}\AgdaSymbol{)}\<%
\\
\\
\>\AgdaKeyword{open} \AgdaKeyword{import} \AgdaModule{Level} \AgdaKeyword{renaming} \AgdaSymbol{(}\AgdaPrimitive{zero} \AgdaSymbol{to} \AgdaPrimitive{lzero}\AgdaSymbol{)}\<%
\\
\>\AgdaKeyword{open} \AgdaKeyword{import} \AgdaModule{Relation.Binary} \AgdaKeyword{using} \AgdaSymbol{(}\AgdaRecord{DecSetoid}\AgdaSymbol{)}\<%
\\
\>\AgdaKeyword{open} \AgdaModule{DecSetoid} \AgdaKeyword{using} \AgdaSymbol{(}\AgdaField{Carrier}\AgdaSymbol{)}\<%
\\
\>\AgdaKeyword{open} \AgdaKeyword{import} \AgdaModule{Relation.Binary.PropositionalEquality} \AgdaKeyword{using} \AgdaSymbol{(}\AgdaDatatype{\_≡\_}\AgdaSymbol{)}\<%
\\
\>\AgdaKeyword{private}\<%
\\
\>[0]\AgdaIndent{2}{}\<[2]%
\>[2]\AgdaFunction{DT} \AgdaSymbol{=} \AgdaRecord{DecSetoid} \AgdaPrimitive{lzero} \AgdaPrimitive{lzero}\<%
\\
\\
\>\AgdaKeyword{record} \AgdaRecord{BT₃} \AgdaSymbol{(}\AgdaBound{t₁} \AgdaSymbol{:} \AgdaRecord{BT₁}\AgdaSymbol{)} \AgdaSymbol{(}\AgdaBound{t₂} \AgdaSymbol{:} \AgdaRecord{BT₂} \AgdaBound{t₁}\AgdaSymbol{)} \AgdaSymbol{:} \AgdaPrimitiveType{Set₀} \AgdaKeyword{where}\<%
\\
\>[0]\AgdaIndent{2}{}\<[2]%
\>[2]\AgdaKeyword{open} \AgdaModule{BT₂} \AgdaBound{t₂} \AgdaKeyword{public}\<%
\\
\\
\>[0]\AgdaIndent{2}{}\<[2]%
\>[2]\AgdaKeyword{field}\<%
\\
\>[2]\AgdaIndent{4}{}\<[4]%
\>[4]\AgdaField{\_*\_} \AgdaSymbol{:} \AgdaFunction{nat} \AgdaSymbol{→} \AgdaFunction{nat} \AgdaSymbol{→} \AgdaFunction{nat}\<%
\\
\>[2]\AgdaIndent{4}{}\<[4]%
\>[4]\AgdaField{right-zero} \AgdaSymbol{:} \AgdaSymbol{∀} \AgdaBound{x} \AgdaSymbol{→} \AgdaBound{x} \AgdaField{*} \AgdaFunction{Z} \AgdaDatatype{≡} \AgdaFunction{Z}\<%
\\
\>[2]\AgdaIndent{4}{}\<[4]%
\>[4]\AgdaField{S*} \AgdaSymbol{:} \AgdaSymbol{∀} \AgdaBound{x} \AgdaBound{y} \AgdaSymbol{→} \AgdaBound{x} \AgdaField{*} \AgdaFunction{S} \AgdaBound{y} \AgdaDatatype{≡} \AgdaSymbol{(}\AgdaBound{x} \AgdaField{*} \AgdaBound{y}\AgdaSymbol{)} \AgdaFunction{+} \AgdaBound{x}\<%
\\
\>[2]\AgdaIndent{4}{}\<[4]%
\>[4]\AgdaField{btimes-is-*} \AgdaSymbol{:} \AgdaSymbol{∀} \AgdaBound{a} \AgdaBound{b} \AgdaSymbol{→} \AgdaFunction{⟦} \AgdaBound{a} \AgdaFunction{btimes} \AgdaBound{b} \AgdaFunction{⟧₂} \AgdaDatatype{≡} \AgdaFunction{⟦} \AgdaBound{a} \AgdaFunction{⟧₂} \AgdaField{*} \AgdaFunction{⟦} \AgdaBound{b} \AgdaFunction{⟧₂}\<%
\\
\\
\>[0]\AgdaIndent{2}{}\<[2]%
\>[2]\AgdaFunction{nat*-lang} \AgdaSymbol{:} \AgdaRecord{GroundLanguage} \AgdaFunction{nat}\<%
\\
\>[0]\AgdaIndent{2}{}\<[2]%
\>[2]\AgdaFunction{nat*-lang} \AgdaSymbol{=} \AgdaKeyword{record} \AgdaSymbol{\{} \AgdaField{Lang} \AgdaSymbol{=} \AgdaSymbol{λ} \AgdaBound{X} \AgdaSymbol{→} \AgdaDatatype{ℕ*} \AgdaSymbol{(}\AgdaField{Carrier} \AgdaBound{X}\AgdaSymbol{)}\<%
\\
\>[2]\AgdaIndent{21}{}\<[21]%
\>[21]\AgdaSymbol{;} \AgdaField{value} \AgdaSymbol{=} \AgdaSymbol{λ} \AgdaSymbol{\{}\AgdaBound{V}\AgdaSymbol{\}} \AgdaSymbol{→} \AgdaFunction{val} \AgdaSymbol{\{}\AgdaBound{V}\AgdaSymbol{\}} \AgdaSymbol{\}}\<%
\\
\>[0]\AgdaIndent{3}{}\<[3]%
\>[3]\AgdaKeyword{where}\<%
\\
\>[3]\AgdaIndent{4}{}\<[4]%
\>[4]\AgdaFunction{val} \AgdaSymbol{:} \AgdaSymbol{\{}\AgdaBound{V} \AgdaSymbol{:} \AgdaFunction{DT}\AgdaSymbol{\}} \AgdaSymbol{→} \AgdaDatatype{ℕ*} \AgdaSymbol{(}\AgdaField{Carrier} \AgdaBound{V}\AgdaSymbol{)} \AgdaSymbol{→} \AgdaSymbol{(}\AgdaField{Carrier} \AgdaBound{V} \AgdaSymbol{→} \AgdaFunction{nat}\AgdaSymbol{)} \AgdaSymbol{→} \AgdaFunction{nat}\<%
\\
\>[3]\AgdaIndent{4}{}\<[4]%
\>[4]\AgdaFunction{val} \<[12]%
\>[12]\AgdaInductiveConstructor{z} \<[21]%
\>[21]\AgdaBound{env} \AgdaSymbol{=} \AgdaFunction{Z}\<%
\\
\>[3]\AgdaIndent{4}{}\<[4]%
\>[4]\AgdaFunction{val} \AgdaSymbol{\{}\AgdaBound{V}\AgdaSymbol{\}} \AgdaSymbol{(}\AgdaInductiveConstructor{s} \AgdaBound{n}\AgdaSymbol{)} \<[21]%
\>[21]\AgdaBound{env} \AgdaSymbol{=} \AgdaFunction{S} \AgdaSymbol{(}\AgdaFunction{val} \AgdaSymbol{\{}\AgdaBound{V}\AgdaSymbol{\}} \AgdaBound{n} \AgdaBound{env}\AgdaSymbol{)}\<%
\\
\>[3]\AgdaIndent{4}{}\<[4]%
\>[4]\AgdaFunction{val} \AgdaSymbol{\{}\AgdaBound{V}\AgdaSymbol{\}} \AgdaSymbol{(}\AgdaBound{e} \AgdaInductiveConstructor{`+} \AgdaBound{f}\AgdaSymbol{)} \AgdaBound{env} \AgdaSymbol{=} \AgdaFunction{val} \AgdaSymbol{\{}\AgdaBound{V}\AgdaSymbol{\}} \AgdaBound{e} \AgdaBound{env} \AgdaFunction{+} \AgdaFunction{val} \AgdaSymbol{\{}\AgdaBound{V}\AgdaSymbol{\}} \AgdaBound{f} \AgdaBound{env}\<%
\\
\>[3]\AgdaIndent{4}{}\<[4]%
\>[4]\AgdaFunction{val} \AgdaSymbol{\{}\AgdaBound{V}\AgdaSymbol{\}} \AgdaSymbol{(}\AgdaBound{e} \AgdaInductiveConstructor{`*} \AgdaBound{f}\AgdaSymbol{)} \AgdaBound{env} \AgdaSymbol{=} \AgdaFunction{val} \AgdaSymbol{\{}\AgdaBound{V}\AgdaSymbol{\}} \AgdaBound{e} \AgdaBound{env} \AgdaFunction{+} \AgdaFunction{val} \AgdaSymbol{\{}\AgdaBound{V}\AgdaSymbol{\}} \AgdaBound{f} \AgdaBound{env}\<%
\\
\>[3]\AgdaIndent{4}{}\<[4]%
\>[4]\AgdaFunction{val} \<[12]%
\>[12]\AgdaSymbol{(}\AgdaInductiveConstructor{v} \AgdaBound{x}\AgdaSymbol{)} \<[21]%
\>[21]\AgdaBound{env} \AgdaSymbol{=} \AgdaBound{env} \AgdaBound{x}\<%
\\
\\
\>[0]\AgdaIndent{2}{}\<[2]%
\>[2]\AgdaKeyword{module} \AgdaModule{fo₃} \AgdaSymbol{=} \AgdaModule{FOL} \AgdaFunction{nat*-lang}\<%
\end{code}

%% file: T4.tex
\begin{code}%
\>\AgdaKeyword{module} \AgdaModule{T4} \AgdaKeyword{where}\<%
\\
\>\AgdaKeyword{open} \AgdaKeyword{import} \AgdaModule{T1} \AgdaKeyword{using} \AgdaSymbol{(}\AgdaRecord{BT₁}\AgdaSymbol{)}\<%
\\
\>\AgdaKeyword{open} \AgdaKeyword{import} \AgdaModule{T2} \AgdaKeyword{using} \AgdaSymbol{(}\AgdaRecord{BT₂}\AgdaSymbol{)}\<%
\\
\>\AgdaKeyword{open} \AgdaKeyword{import} \AgdaModule{T3} \AgdaKeyword{using} \AgdaSymbol{(}\AgdaRecord{BT₃}\AgdaSymbol{)}\<%
\\
\>\AgdaKeyword{open} \AgdaKeyword{import} \AgdaModule{Numerals}\<%
\\
\\
\>\AgdaKeyword{open} \AgdaKeyword{import} \AgdaModule{Relation.Binary.PropositionalEquality} \AgdaKeyword{using} \AgdaSymbol{(}\AgdaDatatype{\_≡\_}\AgdaSymbol{)}\<%
\\
\>\AgdaKeyword{open} \AgdaKeyword{import} \AgdaModule{Data.Sum} \AgdaKeyword{using} \AgdaSymbol{(}\AgdaDatatype{\_⊎\_}\AgdaSymbol{)}\<%
\\
\>\AgdaKeyword{open} \AgdaKeyword{import} \AgdaModule{Data.Product} \AgdaKeyword{using} \AgdaSymbol{(}\AgdaRecord{Σ}\AgdaSymbol{)}\<%
\\
\\
\>\AgdaKeyword{record} \AgdaRecord{BT₄} \AgdaSymbol{(}\AgdaBound{t1} \AgdaSymbol{:} \AgdaRecord{BT₁}\AgdaSymbol{)} \AgdaSymbol{(}\AgdaBound{t2} \AgdaSymbol{:} \AgdaRecord{BT₂} \AgdaBound{t1}\AgdaSymbol{)} \AgdaSymbol{(}\AgdaBound{t3} \AgdaSymbol{:} \AgdaRecord{BT₃} \AgdaBound{t1} \AgdaBound{t2}\AgdaSymbol{)} \AgdaSymbol{:} \AgdaPrimitiveType{Set₀} \AgdaKeyword{where}\<%
\\
\>[0]\AgdaIndent{2}{}\<[2]%
\>[2]\AgdaKeyword{open} \AgdaModule{BT₃} \AgdaBound{t3} \AgdaKeyword{public}\<%
\\
\>[0]\AgdaIndent{2}{}\<[2]%
\>[2]\AgdaKeyword{field}\<%
\\
\>[2]\AgdaIndent{4}{}\<[4]%
\>[4]\AgdaField{no-junk} \AgdaSymbol{:} \AgdaSymbol{∀} \AgdaBound{x} \AgdaSymbol{→} \AgdaBound{x} \AgdaDatatype{≡} \AgdaFunction{Z} \AgdaDatatype{⊎} \AgdaRecord{Σ} \AgdaFunction{nat} \AgdaSymbol{(λ} \AgdaBound{y} \AgdaSymbol{→} \AgdaFunction{S} \AgdaBound{y} \AgdaDatatype{≡} \AgdaBound{x}\AgdaSymbol{)}\<%
\end{code}

%% file: T5.tex
\begin{code}%
\>\AgdaKeyword{module} \AgdaModule{T5} \AgdaKeyword{where}\<%
\\
\>\AgdaKeyword{open} \AgdaKeyword{import} \AgdaModule{Relation.Binary} \AgdaKeyword{using} \AgdaSymbol{(}\AgdaRecord{DecSetoid}\AgdaSymbol{)}\<%
\\
\>\AgdaKeyword{open} \AgdaKeyword{import} \AgdaModule{Level} \AgdaKeyword{using} \AgdaSymbol{()} \AgdaKeyword{renaming} \AgdaSymbol{(}\AgdaPrimitive{zero} \AgdaSymbol{to} \AgdaPrimitive{lzero}\AgdaSymbol{)}\<%
\\
\\
\>\AgdaFunction{DT} \AgdaSymbol{:} \AgdaPrimitiveType{Set₁}\<%
\\
\>\AgdaFunction{DT} \AgdaSymbol{=} \AgdaRecord{DecSetoid} \AgdaPrimitive{lzero} \AgdaPrimitive{lzero}\<%
\\
\\
\>\AgdaKeyword{open} \AgdaKeyword{import} \AgdaModule{T1} \AgdaKeyword{using} \AgdaSymbol{(}\AgdaRecord{BT₁}\AgdaSymbol{)}\<%
\\
\\
\>\AgdaKeyword{open} \AgdaKeyword{import} \AgdaModule{Relation.Binary.PropositionalEquality} \AgdaKeyword{using} \AgdaSymbol{(}\AgdaDatatype{\_≡\_}\AgdaSymbol{)}\<%
\\
\>\AgdaKeyword{open} \AgdaKeyword{import} \AgdaModule{Data.Empty} \AgdaKeyword{using} \AgdaSymbol{(}\AgdaDatatype{⊥}\AgdaSymbol{)}\<%
\\
\>\AgdaKeyword{open} \AgdaKeyword{import} \AgdaModule{Data.Sum} \AgdaKeyword{using} \AgdaSymbol{(}\AgdaDatatype{\_⊎\_}\AgdaSymbol{)}\<%
\\
\>\AgdaKeyword{open} \AgdaKeyword{import} \AgdaModule{Data.Product} \AgdaKeyword{using} \AgdaSymbol{(}\AgdaRecord{Σ}\AgdaSymbol{;}\AgdaFunction{\_×\_}\AgdaSymbol{;}\AgdaInductiveConstructor{\_,\_}\AgdaSymbol{)}\<%
\\
\>\AgdaKeyword{open} \AgdaKeyword{import} \AgdaModule{Data.Bool} \AgdaKeyword{using} \AgdaSymbol{(}\AgdaDatatype{Bool}\AgdaSymbol{)}\<%
\\
\>\AgdaKeyword{open} \AgdaKeyword{import} \AgdaModule{Equiv} \AgdaKeyword{using} \AgdaSymbol{(}\AgdaRecord{\_≃\_}\AgdaSymbol{)}\<%
\\
\\
\>\AgdaKeyword{open} \AgdaKeyword{import} \AgdaModule{Variables} \AgdaKeyword{using} \AgdaSymbol{(}\AgdaKeyword{module} \AgdaModule{VarLangs}\AgdaSymbol{;} \AgdaFunction{NoVars}\AgdaSymbol{;} \AgdaFunction{DBool}\AgdaSymbol{)}\<%
\\
\>\AgdaKeyword{open} \AgdaKeyword{import} \AgdaModule{Language}\<%
\\
\>[0]\AgdaIndent{2}{}\<[2]%
\>[2]\AgdaKeyword{using} \AgdaSymbol{(}\AgdaKeyword{module} \AgdaModule{FOL}\AgdaSymbol{;}\<%
\\
\>[2]\AgdaIndent{9}{}\<[9]%
\>[9]\AgdaKeyword{module} \AgdaModule{LogicOverL}\AgdaSymbol{)}\<%
\\
\\
\>\AgdaKeyword{record} \AgdaRecord{BT₅} \AgdaSymbol{(}\AgdaBound{t₁} \AgdaSymbol{:} \AgdaRecord{BT₁}\AgdaSymbol{)} \AgdaSymbol{:} \AgdaPrimitiveType{Set₁} \AgdaKeyword{where}\<%
\\
\>[0]\AgdaIndent{2}{}\<[2]%
\>[2]\AgdaKeyword{open} \AgdaModule{BT₁} \AgdaBound{t₁} \AgdaKeyword{public}\<%
\\
\>[0]\AgdaIndent{2}{}\<[2]%
\>[2]\AgdaKeyword{open} \AgdaModule{VarLangs} \AgdaKeyword{using} \AgdaSymbol{(}\AgdaFunction{XV}\AgdaSymbol{;} \AgdaInductiveConstructor{x}\AgdaSymbol{)}\<%
\\
\>[0]\AgdaIndent{2}{}\<[2]%
\>[2]\AgdaKeyword{open} \AgdaModule{DecSetoid} \AgdaKeyword{using} \AgdaSymbol{(}\AgdaField{Carrier}\AgdaSymbol{)}\<%
\\
\>[0]\AgdaIndent{2}{}\<[2]%
\>[2]\AgdaKeyword{open} \AgdaModule{DecSetoid} \AgdaFunction{DBool} \AgdaKeyword{using} \AgdaSymbol{(}\_≈\_\AgdaSymbol{)}\<%
\\
\>[0]\AgdaIndent{2}{}\<[2]%
\>[2]\AgdaKeyword{open} \AgdaModule{LogicOverL} \AgdaFunction{fo₁.LoL-FOL}\<%
\\
\>[0]\AgdaIndent{2}{}\<[2]%
\>[2]\AgdaKeyword{open} \AgdaModule{fo₁} \AgdaKeyword{using} \AgdaSymbol{(}\AgdaDatatype{FOL}\AgdaSymbol{;} \AgdaInductiveConstructor{tt}\AgdaSymbol{;} \AgdaInductiveConstructor{ff}\AgdaSymbol{)}\<%
\\
\>[2]\AgdaIndent{4}{}\<[4]%
\>[4]\<%
\\
\>[0]\AgdaIndent{2}{}\<[2]%
\>[2]\AgdaKeyword{field}\<%
\\
\>[2]\AgdaIndent{4}{}\<[4]%
\>[4]\AgdaField{induct} \AgdaSymbol{:} \AgdaSymbol{(}\AgdaBound{e} \AgdaSymbol{:} \AgdaDatatype{FOL} \AgdaFunction{XV}\AgdaSymbol{)} \AgdaSymbol{→}\<%
\\
\>[4]\AgdaIndent{6}{}\<[6]%
\>[6]\AgdaFunction{⟦} \AgdaBound{e} \AgdaFunction{⟧} \AgdaSymbol{(λ} \AgdaSymbol{\{} \AgdaInductiveConstructor{x} \AgdaSymbol{→} \AgdaFunction{Z} \AgdaSymbol{\})} \AgdaSymbol{→}\<%
\\
\>[4]\AgdaIndent{6}{}\<[6]%
\>[6]\AgdaSymbol{(∀} \AgdaSymbol{(}\AgdaBound{y} \AgdaSymbol{:} \AgdaFunction{nat}\AgdaSymbol{)} \AgdaSymbol{→} \AgdaFunction{⟦} \AgdaBound{e} \AgdaFunction{⟧} \AgdaSymbol{(λ} \AgdaSymbol{\{}\AgdaInductiveConstructor{x} \AgdaSymbol{→} \AgdaBound{y}\AgdaSymbol{\})} \AgdaSymbol{→} \AgdaFunction{⟦} \AgdaBound{e} \AgdaFunction{⟧} \AgdaSymbol{(λ} \AgdaSymbol{\{}\AgdaInductiveConstructor{x} \AgdaSymbol{→} \AgdaFunction{S} \AgdaBound{y}\AgdaSymbol{\}))} \AgdaSymbol{→}\<%
\\
\>[4]\AgdaIndent{6}{}\<[6]%
\>[6]\AgdaSymbol{∀} \AgdaSymbol{(}\AgdaBound{y} \AgdaSymbol{:} \AgdaFunction{nat}\AgdaSymbol{)} \AgdaSymbol{→} \AgdaFunction{⟦} \AgdaBound{e} \AgdaFunction{⟧} \AgdaSymbol{(λ} \AgdaSymbol{\{}\AgdaInductiveConstructor{x} \AgdaSymbol{→} \AgdaBound{y}\AgdaSymbol{\})}\<%
\\
\>[0]\AgdaIndent{2}{}\<[2]%
\>[2]\AgdaKeyword{postulate}\<%
\\
\>[2]\AgdaIndent{4}{}\<[4]%
\>[4]\AgdaPostulate{decide} \AgdaSymbol{:} \AgdaSymbol{∀} \AgdaSymbol{\{}\AgdaBound{W}\AgdaSymbol{\}} \AgdaSymbol{→} \AgdaSymbol{(}\AgdaField{Carrier} \AgdaBound{W} \AgdaSymbol{→} \AgdaFunction{nat}\AgdaSymbol{)} \AgdaSymbol{→} \AgdaDatatype{FOL} \AgdaBound{W} \AgdaSymbol{→} \AgdaDatatype{FOL} \AgdaFunction{NoVars} \AgdaComment{-- T5-dec-proc}\<%
\\
\>[2]\AgdaIndent{4}{}\<[4]%
\>[4]\AgdaPostulate{meaning-decide} \AgdaSymbol{:} \AgdaSymbol{\{}\AgdaBound{W} \AgdaSymbol{:} \AgdaFunction{DT}\AgdaSymbol{\}} \AgdaSymbol{(}\AgdaBound{env} \AgdaSymbol{:} \AgdaField{Carrier} \AgdaBound{W} \AgdaSymbol{→} \AgdaFunction{nat}\AgdaSymbol{)} \AgdaSymbol{→} \AgdaSymbol{(}\AgdaBound{env′} \AgdaSymbol{:} \AgdaDatatype{⊥} \AgdaSymbol{→} \AgdaFunction{nat}\AgdaSymbol{)} \AgdaSymbol{→}\<%
\\
\>[4]\AgdaIndent{6}{}\<[6]%
\>[6]\AgdaSymbol{(}\AgdaBound{e} \AgdaSymbol{:} \AgdaDatatype{FOL} \AgdaBound{W}\AgdaSymbol{)} \AgdaSymbol{→}\<%
\\
\>[4]\AgdaIndent{6}{}\<[6]%
\>[6]\AgdaKeyword{let} \AgdaBound{res} \AgdaSymbol{=} \AgdaPostulate{decide} \AgdaBound{env} \AgdaBound{e} \AgdaKeyword{in}\<%
\\
\>[4]\AgdaIndent{6}{}\<[6]%
\>[6]\AgdaSymbol{(}\AgdaBound{res} \AgdaDatatype{≡} \AgdaInductiveConstructor{tt} \AgdaDatatype{⊎} \AgdaBound{res} \AgdaDatatype{≡} \AgdaInductiveConstructor{ff}\AgdaSymbol{)} \AgdaFunction{×} \AgdaSymbol{(}\AgdaFunction{⟦} \AgdaBound{e} \AgdaFunction{⟧} \AgdaBound{env}\AgdaSymbol{)} \AgdaRecord{≃} \AgdaSymbol{(}\AgdaFunction{⟦} \AgdaBound{res} \AgdaFunction{⟧} \AgdaBound{env′}\AgdaSymbol{)}\<%
\end{code}

%% file: T7.tex
\AgdaHide{
\begin{code}%
\>\AgdaKeyword{module} \AgdaModule{T7} \AgdaKeyword{where}\<%
\\
\>\AgdaKeyword{open} \AgdaKeyword{import} \AgdaModule{Relation.Binary} \AgdaKeyword{using} \AgdaSymbol{(}\AgdaRecord{DecSetoid}\AgdaSymbol{)}\<%
\\
\>\AgdaKeyword{open} \AgdaKeyword{import} \AgdaModule{Level} \AgdaKeyword{using} \AgdaSymbol{()} \AgdaKeyword{renaming} \AgdaSymbol{(}\AgdaPrimitive{zero} \AgdaSymbol{to} \AgdaPrimitive{lzero}\AgdaSymbol{)}\<%
\\
\\
\>\AgdaFunction{DT} \AgdaSymbol{:} \AgdaPrimitiveType{Set₁}\<%
\\
\>\AgdaFunction{DT} \AgdaSymbol{=} \AgdaRecord{DecSetoid} \AgdaPrimitive{lzero} \AgdaPrimitive{lzero}\<%
\\
\\
\>\AgdaKeyword{open} \AgdaKeyword{import} \AgdaModule{T1} \AgdaKeyword{using} \AgdaSymbol{(}\AgdaRecord{BT₁}\AgdaSymbol{)}\<%
\\
\>\AgdaKeyword{open} \AgdaKeyword{import} \AgdaModule{T2} \AgdaKeyword{using} \AgdaSymbol{(}\AgdaRecord{BT₂}\AgdaSymbol{)}\<%
\\
\>\AgdaKeyword{open} \AgdaKeyword{import} \AgdaModule{T3} \AgdaKeyword{using} \AgdaSymbol{(}\AgdaRecord{BT₃}\AgdaSymbol{)}\<%
\\
\>\AgdaKeyword{open} \AgdaKeyword{import} \AgdaModule{T5} \AgdaKeyword{using} \AgdaSymbol{(}\AgdaRecord{BT₅}\AgdaSymbol{)}\<%
\\
\>\AgdaKeyword{open} \AgdaKeyword{import} \AgdaModule{T6} \AgdaKeyword{using} \AgdaSymbol{(}\AgdaRecord{BT₆}\AgdaSymbol{)}\<%
\\
\\
\>\AgdaKeyword{open} \AgdaKeyword{import} \AgdaModule{Relation.Binary.PropositionalEquality} \AgdaKeyword{using} \AgdaSymbol{(}\AgdaDatatype{\_≡\_}\AgdaSymbol{)}\<%
\\
\>\AgdaKeyword{open} \AgdaKeyword{import} \AgdaModule{Data.Empty} \AgdaKeyword{using} \AgdaSymbol{(}\AgdaDatatype{⊥}\AgdaSymbol{)}\<%
\\
\>\AgdaKeyword{open} \AgdaKeyword{import} \AgdaModule{Data.Sum} \AgdaKeyword{using} \AgdaSymbol{(}\AgdaDatatype{\_⊎\_}\AgdaSymbol{)}\<%
\\
\>\AgdaKeyword{open} \AgdaKeyword{import} \AgdaModule{Data.Product} \AgdaKeyword{using} \AgdaSymbol{(}\AgdaRecord{Σ}\AgdaSymbol{;}\AgdaFunction{\_×\_}\AgdaSymbol{;}\AgdaInductiveConstructor{\_,\_}\AgdaSymbol{)}\<%
\\
\>\AgdaKeyword{open} \AgdaKeyword{import} \AgdaModule{Data.Bool} \AgdaKeyword{using} \AgdaSymbol{(}\AgdaDatatype{Bool}\AgdaSymbol{)}\<%
\\
\>\AgdaKeyword{open} \AgdaKeyword{import} \AgdaModule{Equiv} \AgdaKeyword{using} \AgdaSymbol{(}\AgdaRecord{\_≃\_}\AgdaSymbol{)}\<%
\\
\\
\>\AgdaKeyword{open} \AgdaKeyword{import} \AgdaModule{Variables} \AgdaKeyword{using} \AgdaSymbol{(}\AgdaKeyword{module} \AgdaModule{VarLangs}\AgdaSymbol{;} \AgdaFunction{NoVars}\AgdaSymbol{;} \AgdaFunction{DBool}\AgdaSymbol{)}\<%
\\
\>\AgdaKeyword{open} \AgdaKeyword{import} \AgdaModule{Language}\<%
\\
\>[0]\AgdaIndent{2}{}\<[2]%
\>[2]\AgdaKeyword{using} \AgdaSymbol{(}\AgdaKeyword{module} \AgdaModule{FOL}\AgdaSymbol{;}\<%
\\
\>[2]\AgdaIndent{9}{}\<[9]%
\>[9]\AgdaKeyword{module} \AgdaModule{LogicOverL}\AgdaSymbol{)}\<%
\end{code}
}

\begin{code}%
\>\AgdaKeyword{record} \AgdaRecord{BT₇} \AgdaSymbol{\{}\AgdaBound{t₁} \AgdaSymbol{:} \AgdaRecord{BT₁}\AgdaSymbol{\}} \AgdaSymbol{\{}\AgdaBound{t₂} \AgdaSymbol{:} \AgdaRecord{BT₂} \AgdaBound{t₁}\AgdaSymbol{\}} \AgdaSymbol{(}\AgdaBound{t₃} \AgdaSymbol{:} \AgdaRecord{BT₃} \AgdaBound{t₁} \AgdaBound{t₂}\AgdaSymbol{)} \AgdaSymbol{(}\AgdaBound{t₅} \AgdaSymbol{:} \AgdaRecord{BT₅} \AgdaBound{t₁}\AgdaSymbol{)} \AgdaSymbol{(}\AgdaBound{t₆} \AgdaSymbol{:} \AgdaRecord{BT₆} \AgdaBound{t₂} \AgdaBound{t₅}\AgdaSymbol{)} \AgdaSymbol{:} \AgdaPrimitiveType{Set₁} \AgdaKeyword{where}\<%
\\
\>[0]\AgdaIndent{2}{}\<[2]%
\>[2]\AgdaKeyword{open} \AgdaModule{VarLangs} \AgdaKeyword{using} \AgdaSymbol{(}\AgdaFunction{XV}\AgdaSymbol{;} \AgdaInductiveConstructor{x}\AgdaSymbol{)}\<%
\\
\>[0]\AgdaIndent{2}{}\<[2]%
\>[2]\AgdaKeyword{open} \AgdaModule{DecSetoid} \AgdaKeyword{using} \AgdaSymbol{(}\AgdaField{Carrier}\AgdaSymbol{)}\<%
\\
\>[0]\AgdaIndent{2}{}\<[2]%
\>[2]\AgdaKeyword{open} \AgdaModule{BT₃} \AgdaBound{t₃} \AgdaKeyword{public}\<%
\\
\>[0]\AgdaIndent{2}{}\<[2]%
\>[2]\AgdaKeyword{open} \AgdaModule{fo₃} \AgdaKeyword{using} \AgdaSymbol{(}\AgdaDatatype{FOL}\AgdaSymbol{;} \AgdaInductiveConstructor{tt}\AgdaSymbol{;} \AgdaInductiveConstructor{ff}\AgdaSymbol{;} \AgdaFunction{LoL-FOL}\AgdaSymbol{;} \AgdaInductiveConstructor{\_and\_}\AgdaSymbol{;} \AgdaInductiveConstructor{all}\AgdaSymbol{)}\<%
\\
\>[0]\AgdaIndent{2}{}\<[2]%
\>[2]\AgdaKeyword{open} \AgdaModule{LogicOverL} \AgdaFunction{LoL-FOL}\<%
\\
\\
\>[0]\AgdaIndent{2}{}\<[2]%
\>[2]\AgdaKeyword{field}\<%
\\
\>[2]\AgdaIndent{4}{}\<[4]%
\>[4]\AgdaField{induct} \AgdaSymbol{:} \AgdaSymbol{(}\AgdaBound{e} \AgdaSymbol{:} \AgdaDatatype{FOL} \AgdaFunction{XV}\AgdaSymbol{)} \AgdaSymbol{→}\<%
\\
\>[4]\AgdaIndent{6}{}\<[6]%
\>[6]\AgdaFunction{⟦} \AgdaBound{e} \AgdaFunction{⟧} \AgdaSymbol{(λ} \AgdaSymbol{\{} \AgdaInductiveConstructor{x} \AgdaSymbol{→} \AgdaFunction{⟦} \AgdaNumber{0} \AgdaFunction{⟧₁} \AgdaSymbol{\})} \AgdaSymbol{→}\<%
\\
\>[4]\AgdaIndent{6}{}\<[6]%
\>[6]\AgdaSymbol{(∀} \AgdaBound{y} \AgdaSymbol{→} \AgdaFunction{⟦} \AgdaBound{e} \AgdaFunction{⟧} \AgdaSymbol{(λ} \AgdaSymbol{\{}\AgdaInductiveConstructor{x} \AgdaSymbol{→} \AgdaBound{y}\AgdaSymbol{\})} \AgdaSymbol{→} \AgdaFunction{⟦} \AgdaBound{e} \AgdaFunction{⟧} \AgdaSymbol{(λ} \AgdaSymbol{\{}\AgdaInductiveConstructor{x} \AgdaSymbol{→} \AgdaFunction{S} \AgdaBound{y}\AgdaSymbol{\}))} \AgdaSymbol{→}\<%
\\
\>[4]\AgdaIndent{6}{}\<[6]%
\>[6]\AgdaSymbol{∀} \AgdaBound{y} \AgdaSymbol{→} \AgdaFunction{⟦} \AgdaBound{e} \AgdaFunction{⟧} \AgdaSymbol{(λ} \AgdaSymbol{\{}\AgdaInductiveConstructor{x} \AgdaSymbol{→} \AgdaBound{y}\AgdaSymbol{\})}\<%
\\
\>[0]\AgdaIndent{2}{}\<[2]%
\>[2]\AgdaKeyword{postulate}\<%
\\
\>[2]\AgdaIndent{4}{}\<[4]%
\>[4]\AgdaPostulate{decide} \AgdaSymbol{:} \AgdaSymbol{∀} \AgdaSymbol{\{}\AgdaBound{W}\AgdaSymbol{\}} \AgdaSymbol{→} \AgdaSymbol{(}\AgdaField{Carrier} \AgdaBound{W} \AgdaSymbol{→} \AgdaFunction{nat}\AgdaSymbol{)} \AgdaSymbol{→} \AgdaDatatype{FOL} \AgdaBound{W} \AgdaSymbol{→} \AgdaDatatype{FOL} \AgdaFunction{NoVars}\<%
\\
\>[2]\AgdaIndent{4}{}\<[4]%
\>[4]\AgdaPostulate{meaning-decide} \AgdaSymbol{:} \AgdaSymbol{\{}\AgdaBound{W} \AgdaSymbol{:} \AgdaFunction{DT}\AgdaSymbol{\}} \AgdaSymbol{(}\AgdaBound{env} \AgdaSymbol{:} \AgdaField{Carrier} \AgdaBound{W} \AgdaSymbol{→} \AgdaFunction{nat}\AgdaSymbol{)} \AgdaSymbol{→} \AgdaSymbol{(}\AgdaBound{env′} \AgdaSymbol{:} \AgdaDatatype{⊥} \AgdaSymbol{→} \AgdaFunction{nat}\AgdaSymbol{)} \AgdaSymbol{→}\<%
\\
\>[4]\AgdaIndent{6}{}\<[6]%
\>[6]\AgdaSymbol{(}\AgdaBound{e} \AgdaSymbol{:} \AgdaDatatype{FOL} \AgdaBound{W}\AgdaSymbol{)} \AgdaSymbol{→}\<%
\\
\>[4]\AgdaIndent{6}{}\<[6]%
\>[6]\AgdaKeyword{let} \AgdaBound{res} \AgdaSymbol{=} \AgdaPostulate{decide} \AgdaBound{env} \AgdaBound{e} \AgdaKeyword{in}\<%
\\
\>[4]\AgdaIndent{6}{}\<[6]%
\>[6]\AgdaSymbol{(}\AgdaBound{res} \AgdaDatatype{≡} \AgdaInductiveConstructor{tt} \AgdaDatatype{⊎} \AgdaBound{res} \AgdaDatatype{≡} \AgdaInductiveConstructor{ff}\AgdaSymbol{)} \AgdaFunction{×} \AgdaSymbol{(}\AgdaFunction{⟦} \AgdaBound{e} \AgdaFunction{⟧} \AgdaBound{env}\AgdaSymbol{)} \AgdaRecord{≃} \AgdaSymbol{(}\AgdaFunction{⟦} \AgdaBound{res} \AgdaFunction{⟧} \AgdaBound{env′}\AgdaSymbol{)}\<%
\end{code}

%% file: T8.tex
\begin{code}%
\>\AgdaKeyword{module} \AgdaModule{T8} \AgdaKeyword{where}\<%
\\
\>\AgdaKeyword{open} \AgdaKeyword{import} \AgdaModule{DefiniteDescr} \AgdaKeyword{using} \AgdaSymbol{(}\AgdaFunction{isContr₂}\AgdaSymbol{)}\<%
\\
\\
\>\AgdaKeyword{open} \AgdaKeyword{import} \AgdaModule{Relation.Nullary} \AgdaKeyword{using} \AgdaSymbol{(}\AgdaFunction{¬\_}\AgdaSymbol{)}\<%
\\
\>\AgdaKeyword{open} \AgdaKeyword{import} \AgdaModule{Relation.Binary.PropositionalEquality} \AgdaKeyword{using} \AgdaSymbol{(}\AgdaDatatype{\_≡\_}\AgdaSymbol{)}\<%
\\
\>\AgdaKeyword{open} \AgdaKeyword{import} \AgdaModule{Data.Product} \AgdaKeyword{using} \AgdaSymbol{(}\AgdaRecord{Σ}\AgdaSymbol{;} \AgdaField{proj₁}\AgdaSymbol{;} \AgdaFunction{\_×\_}\AgdaSymbol{)}\<%
\\
\\
\>\AgdaKeyword{record} \AgdaRecord{BT₈} \AgdaSymbol{:} \AgdaPrimitiveType{Set₁} \AgdaKeyword{where}\<%
\\
\>[0]\AgdaIndent{2}{}\<[2]%
\>[2]\AgdaKeyword{field}\<%
\\
\>[2]\AgdaIndent{4}{}\<[4]%
\>[4]\AgdaField{ι} \AgdaSymbol{:} \AgdaPrimitiveType{Set₀}\<%
\\
\>[2]\AgdaIndent{4}{}\<[4]%
\>[4]\AgdaField{ze} \AgdaSymbol{:} \AgdaField{ι}\<%
\\
\>[2]\AgdaIndent{4}{}\<[4]%
\>[4]\AgdaField{S} \AgdaSymbol{:} \AgdaField{ι} \AgdaSymbol{→} \AgdaField{ι}\<%
\\
\>[2]\AgdaIndent{4}{}\<[4]%
\>[4]\AgdaField{S≠Z} \AgdaSymbol{:} \AgdaSymbol{∀} \AgdaBound{x} \AgdaSymbol{→} \AgdaFunction{¬} \AgdaSymbol{(}\AgdaField{S} \AgdaBound{x} \AgdaDatatype{≡} \AgdaField{ze}\AgdaSymbol{)}\<%
\\
\>[2]\AgdaIndent{4}{}\<[4]%
\>[4]\AgdaField{inj} \AgdaSymbol{:} \AgdaSymbol{∀} \AgdaBound{x} \AgdaBound{y} \AgdaSymbol{→} \AgdaField{S} \AgdaBound{x} \AgdaDatatype{≡} \AgdaField{S} \AgdaBound{y} \AgdaSymbol{→} \AgdaBound{x} \AgdaDatatype{≡} \AgdaBound{y}\<%
\\
\>[2]\AgdaIndent{4}{}\<[4]%
\>[4]\AgdaField{induct} \AgdaSymbol{:} \AgdaSymbol{(}\AgdaBound{p} \AgdaSymbol{:} \AgdaField{ι} \AgdaSymbol{→} \AgdaPrimitiveType{Set₀}\AgdaSymbol{)} \AgdaSymbol{→} \AgdaBound{p} \AgdaField{ze} \AgdaSymbol{→} \AgdaSymbol{(∀} \AgdaBound{x} \AgdaSymbol{→} \AgdaBound{p} \AgdaBound{x} \AgdaSymbol{→} \AgdaBound{p} \AgdaSymbol{(}\AgdaField{S} \AgdaBound{x}\AgdaSymbol{))} \AgdaSymbol{→} \AgdaSymbol{(∀} \AgdaBound{y} \AgdaSymbol{→} \AgdaBound{p} \AgdaBound{y}\AgdaSymbol{)}\<%
\\
\\
\>[0]\AgdaIndent{2}{}\<[2]%
\>[2]\AgdaFunction{bin} \AgdaSymbol{:} \AgdaPrimitiveType{Set₀}\<%
\\
\>[0]\AgdaIndent{2}{}\<[2]%
\>[2]\AgdaFunction{bin} \AgdaSymbol{=} \AgdaField{ι} \AgdaSymbol{→} \AgdaField{ι} \AgdaSymbol{→} \AgdaField{ι}\<%
\\
\\
\>[0]\AgdaIndent{2}{}\<[2]%
\>[2]\AgdaFunction{+-pred} \AgdaSymbol{:} \AgdaFunction{bin} \AgdaSymbol{→} \AgdaPrimitiveType{Set₀}\<%
\\
\>[0]\AgdaIndent{2}{}\<[2]%
\>[2]\AgdaFunction{+-pred} \AgdaBound{f} \AgdaSymbol{=} \AgdaSymbol{(∀} \AgdaBound{x} \AgdaSymbol{→} \AgdaBound{f} \AgdaBound{x} \AgdaField{ze} \AgdaDatatype{≡} \AgdaBound{x}\AgdaSymbol{)} \AgdaFunction{×}\<%
\\
\>[2]\AgdaIndent{13}{}\<[13]%
\>[13]\AgdaSymbol{(∀} \AgdaBound{x} \AgdaBound{y} \AgdaSymbol{→} \AgdaBound{f} \AgdaBound{x} \AgdaSymbol{(}\AgdaField{S} \AgdaBound{y}\AgdaSymbol{)} \AgdaDatatype{≡} \AgdaField{S} \AgdaSymbol{(}\AgdaBound{f} \AgdaBound{x} \AgdaBound{y}\AgdaSymbol{))}\<%
\\
\\
\>[0]\AgdaIndent{2}{}\<[2]%
\>[2]\AgdaKeyword{field}\<%
\\
\>[2]\AgdaIndent{4}{}\<[4]%
\>[4]\AgdaField{+-uniq} \AgdaSymbol{:} \AgdaFunction{isContr₂} \AgdaField{ι} \AgdaFunction{+-pred}\<%
\\
\\
\>[0]\AgdaIndent{2}{}\<[2]%
\>[2]\AgdaFunction{\_+\_} \AgdaSymbol{:} \AgdaFunction{bin}\<%
\\
\>[0]\AgdaIndent{2}{}\<[2]%
\>[2]\AgdaFunction{\_+\_} \AgdaSymbol{=} \AgdaField{proj₁} \AgdaField{+-uniq}\<%
\\
\\
\>[0]\AgdaIndent{2}{}\<[2]%
\>[2]\AgdaFunction{*-pred} \AgdaSymbol{:} \AgdaFunction{bin} \AgdaSymbol{→} \AgdaPrimitiveType{Set₀}\<%
\\
\>[0]\AgdaIndent{2}{}\<[2]%
\>[2]\AgdaFunction{*-pred} \AgdaBound{f} \AgdaSymbol{=} \AgdaSymbol{(∀} \AgdaBound{x} \AgdaSymbol{→} \AgdaBound{f} \AgdaBound{x} \AgdaField{ze} \AgdaDatatype{≡} \AgdaField{ze}\AgdaSymbol{)} \AgdaFunction{×}\<%
\\
\>[2]\AgdaIndent{13}{}\<[13]%
\>[13]\AgdaSymbol{(∀} \AgdaBound{x} \AgdaBound{y} \AgdaSymbol{→} \AgdaBound{f} \AgdaBound{x} \AgdaSymbol{(}\AgdaField{S} \AgdaBound{y}\AgdaSymbol{)} \AgdaDatatype{≡} \AgdaBound{f} \AgdaBound{x} \AgdaBound{y} \AgdaFunction{+} \AgdaBound{x}\AgdaSymbol{)}\<%
\\
\\
\>[0]\AgdaIndent{2}{}\<[2]%
\>[2]\AgdaKeyword{field}\<%
\\
\>[2]\AgdaIndent{4}{}\<[4]%
\>[4]\AgdaField{*-uniq} \AgdaSymbol{:} \AgdaFunction{isContr₂} \AgdaField{ι} \AgdaFunction{*-pred}\<%
\\
\\
\>[0]\AgdaIndent{2}{}\<[2]%
\>[2]\AgdaFunction{\_*\_} \AgdaSymbol{:} \AgdaFunction{bin}\<%
\\
\>[0]\AgdaIndent{2}{}\<[2]%
\>[2]\AgdaFunction{\_*\_} \AgdaSymbol{=} \AgdaField{proj₁} \AgdaField{*-uniq}\<%
\end{code}